\documentclass{naturep}
\usepackage[utf8]{inputenc}
\usepackage{lineno}
\usepackage{setspace}
\usepackage{color}
\usepackage{graphicx}
\usepackage{booktabs}
\usepackage{amssymb}
\usepackage{amsmath}
\usepackage{hyperref}
\usepackage{adjustbox}
\usepackage{subcaption}
\usepackage{tablefootnote}
\usepackage[margin=0.6in]{geometry}
\usepackage{verbatim}
\usepackage{caption}
\usepackage{paralist}
\usepackage{cleveref}
\usepackage{multirow}
\usepackage{tabularx} 
\usepackage{array}

\newcommand{\ours}{\textsc{TITAN}}
\newcommand{\oursv}{\ours_{\text{V}}}
\newcommand{\ourscr}{\ours_{\text{L}}}
\newcommand{\oursvc}{\ours}
\newcommand{\oursvr}{\ours}
\newcommand{\oursvl}{\ours}
\newcommand{\oursvcr}{\ours}

\newcommand\Heading[1]{
  \noindent\textbf{\Large{#1}}
}

\newcommand\heading[1]{
  \noindent\textbf{\large{#1}}
}

\newcommand\hheading[1]{
  \noindent\textbf{#1}
}

\title{\raggedright{\textbf{Multimodal Whole Slide Foundation Model for Pathology}}}

\author{Tong Ding$^{1,2,3,4,\ast}$, Sophia J. Wagner$^{1,5,6,\ast}$, Andrew H. Song$^{1,2,3,\ast}$, Richard J. Chen$^{1,2,3,\ast}$,  Ming Y. Lu$^{1,2,3,7}$, Andrew Zhang$^{1,2,3,8,+}$, Anurag J. Vaidya$^{1,2,3,8,+}$, Guillaume Jaume$^{1,2,3,+}$, Muhammad Shaban$^{1,2,3}$, Ahrong Kim$^{1,9}$, Drew F.K. Williamson$^{10}$, Bowen Chen$^{1,2,3}$, Cristina Almagro-Perez$^{1,2,3,8}$, Paul Doucet$^{1,2,3}$, Sharifa Sahai$^{1,2,3,12}$, Chengkuan Chen$^{1,2,3}$, Daisuke Komura$^{13}$, Akihiro Kawabe$^{13}$, Shumpei Ishikawa$^{13,14}$, Georg Gerber$^{1}$, Tingying Peng$^{5,6}$, Long Phi Le$^{1,8,\dag}$, Faisal Mahmood$^{1,2,3,11,\dag}$}
\date{}
    
\makeatletter
\let\saved@includegraphics\includegraphics
\AtBeginDocument{\let\includegraphics\saved@includegraphics}

\makeatother

\begin{document}
\maketitle
\begin{affiliations}
 \item Department of Pathology, Mass General Brigham, Harvard Medical School, Boston, MA, USA
 \item Data Science Program, Dana-Farber Cancer Institute, Boston, MA, USA
 \item Cancer Program, Broad Institute of Harvard and MIT, Cambridge, MA, USA
 \item John A. Paulson School of Engineering and Applied Sciences, Harvard University, Cambridge, MA, USA
 \item Helmholtz Munich – German Research Center for Environment and Health, Munich, Germany
 \item School of Computation, Information and Technology, Technical University of Munich, Munich, Germany
  \item Department of Electrical Engineering and Computer Science, Massachusetts Institute of Technology, Cambridge, MA, USA
  \item Harvard-MIT Division of Health Sciences and Technology, Massachusetts Institute of Technology, Cambridge, MA, USA
  \item Department of Pathology, Pusan National University, Busan, South Korea
 \item Department of Pathology and Laboratory Medicine, Emory University School of Medicine, Atlanta, GA, USA
 \item Harvard Data Science Initiative, Harvard University, Cambridge, MA, USA
  \item Department of Systems Biology, Harvard Medical School, Boston, MA, USA
 \item Department of Preventive Medicine, Graduate School of Medicine, The University of Tokyo, Tokyo, Japan
 \item Division of Pathology, National Cancer Center Exploratory Oncology Research \& Clinical Trial Center, Chiba, Japan
 \item[$^{\ast}$] Contributed equally (Co-first)
 \item[$^{+}$] Contributed equally (Co-third)
 \item[$^{\dag}$] Co-senior authors \\
\textbf{Lead Contact}: \\
Faisal Mahmood (FaisalMahmood@bwh.harvard.edu)
 \end{affiliations}

\clearpage
\Heading{Abstract}

\begin{spacing}{1.2}
\noindent
\textbf{The field of computational pathology has been transformed with recent advances in foundation models that encode histopathology region-of-interests (ROIs) into versatile and transferable feature representations via self-supervised learning (SSL). However, translating these advancements to address complex clinical challenges at the patient and slide level remains constrained by limited clinical data in disease-specific cohorts, especially for rare clinical conditions. 
We propose $\ours$, a multimodal whole slide foundation model pretrained using 335,645 WSIs via visual self-supervised learning and vision-language alignment with corresponding pathology reports and 423,122 synthetic captions generated from a multimodal generative AI copilot for pathology. Without any finetuning or requiring clinical labels, $\ours$ can extract general-purpose slide representations and generate pathology reports that generalize to resource-limited clinical scenarios such as rare disease retrieval and cancer prognosis. We evaluate $\ours$ on diverse clinical tasks and find that $\ours$ outperforms both ROI and slide foundation models across machine learning settings such as linear probing, few-shot and zero-shot classification, rare cancer retrieval and cross-modal retrieval, and pathology report generation. The model is publicly accessible at }\url{https://github.com/mahmoodlab/TITAN} 

\end{spacing}


\clearpage
\begin{spacing}{1.35}
\Heading{Introduction}

Foundation models are transforming computational pathology by accelerating the development of AI tools for diagnosis, prognosis, and biomarker prediction from digitized tissue sections\cite{song2023artificial}. Developed using self-supervised learning (SSL) on millions of histology image patches (or regions of interests), these models capture morphological patterns in histology patch embeddings, such as tissue organization and cellular structure\cite{riasatian2021fine, ciga2022self,wang2022transformer, wang2023retccl, kang2023benchmarking, huang2023visual, azizi2023robust, chen2024towards, lu2024visual, vorontsov2024foundation, zimmermann2024virchow, hoptimus0, filiot2023scaling, filiot2024phikonv2largepublicfeature, juyal2024pluto, dippel2024rudolfv}. These representations serve as a ``foundation'' for models that predict clinical endpoints from whole-slide images (WSIs), such as diagnosis or biomarker status\cite{bera2019artificial, echle2021deep, shmatko2022artificial, campanella2024clinical, neidlinger2024benchmarking, yu2016predicting,lipkova2022crane, coudray2018classification, campanella2019clinical, kather2019deep, lu2021ai, fu2020pan, bulten2022artificial, zheng2022graph, skrede2020deep, bejnordi2017diagnostic, wagner2023transformer, foersch2023multistain, courtiol2019deep, lee2022derivation, niehues2023generalizable}. However, translating the capabilities of current patch-based foundation models to address patient- and slide-level clinical challenges still remains complex due to the immense scale of gigapixel WSIs, compounded by the small size of patient cohorts in real-world evidence\cite{yu2018artificial, krishnan2022self}, posing challenges for disease-specific AI model development\cite{van2021deep}.
As an example, in rare diseases with limited training data\cite{gatta2017burden, rarecancers, rarecancers_SEER}, developing effective predictive models is difficult since the slide encoder--which generates slide-level predictions from patch embeddings--still needs to be trained from scratch\cite{lu2024visual,lu2021data}. 
Similarly, given a diagnostically challenging patient slide, retrieving a similar slide via slide search\cite{lew2006content, cruz2011visual, caicedo2009histopathology, sridhar2015content, kalra2020yottixel,chen2022fast,wang2023retccl, zheng2018histopathological, shang2024histopathology} or pathology reports through cross-modal report search\cite{zhang2017mdnet, Lu_2023_CVPR, ikezogwo2024quilt, lu2024visual} typically requires specialized algorithms to bridge the gap between patch and slide embeddings, introducing hurdles towards clinical adoption.


To overcome these limitations, new types of foundation models have recently been proposed for encoding entire WSIs into slide-level general-purpose feature representations\cite{chen2022scaling, xu2024whole,lazard2023giga, hou2024selfsupervisedframeworklearningslide, tran2024histogpt, shaikovski2024prism, xu2024multimodal, jaume2024transcriptomics, song2024morphological, jaume2024multistain}. Instead of training an additional model on top of patch embeddings from scratch\cite{ilse2018attention, lu2021data, li2021dual, chen2021whole, shao2021transmil, wagner2023transformer, xiang2023exploring}, these whole slide representation models can be pretrained to distill pathology-specific knowledge from large WSI collections, simplifying clinical endpoint prediction. The outstanding challenge then becomes developing whole slide foundation models that faithfully encode the tissue microenvironment based on a set of patch embeddings while also handling arbitrarily large WSIs. Although relatively underexplored, slide-level self-supervision can be performed with vision-only pretraining, either through masked image reconstruction\cite{xu2024whole} or intra-slide contrastive learning\cite{lazard2023giga, hou2024selfsupervisedframeworklearningslide, kondepudi2024foundation}, or in a multimodal fashion involving pathology reports, bulk transcriptomics, or immunohistochemistry\cite{tran2024histogpt, shaikovski2024prism, xu2024multimodal, jaume2024transcriptomics, jaume2024multistain, ahmed2024pathalignvisionlanguagemodelslide}. Furthermore, long-range context modeling can either be neglected, essentially treating a WSI as a bag of independent features\cite{lazard2023giga, jaume2024transcriptomics, shaikovski2024prism, xu2024multimodal, wang2024pathology}, or explicitly modeled using Transformers\cite{chen2022scaling,tran2024histogpt, xu2024whole, hou2024selfsupervisedframeworklearningslide}. With efforts to learn general-purpose slide representations intensifying, we believe that adapting successful patch-level recipes to the entire WSI would lead to powerful general-purpose slide representations.

Despite their widespread application potential, previous works on pretraining slide foundation models have several shortcomings. First, these models are predominantly pretrained using vision-only modeling\cite{chen2022scaling,lazard2023giga,hou2024selfsupervisedframeworklearningslide}, which neglects not only rich supervisory signals found in pathology reports, but also precludes multimodal capabilities such as zero-shot visual-language understanding and cross-model retrieval -- which is a fundamental hallmark in foundation models\cite{bommasani2021opportunities, moor2023foundation}. Second, whereas current patch foundation models are trained with millions of histology image patches, slide foundation models are developed with orders of magnitude fewer samples and limited optimization of self-supervised learning recipes, leading to slide representations with restricted generalization capability\cite{shaikovski2024prism,ahmed2024pathalignvisionlanguagemodelslide,xu2024whole,wang2024pathology}. Even with multimodal techniques such as vision-language pretraining that augment the pretraining dataset with pathology reports, current slide foundation models still require end-to-end training or finetuning and lack the capability of learning transferable slide representations for challenging clinical scenarios\cite{ahmed2024pathalignvisionlanguagemodelslide,xu2024whole,wang2024pathology}. Finally, the current models are nascent in transforming pathology AI model development due to their limited evaluations in diagnostically relevant settings such as few-shot learning or slide retrieval.

Here, we introduce \textbf{T}ransformer-based pathology \textbf{I}mage and \textbf{T}ext \textbf{A}lignment \textbf{N}etwork ($\ours$), a multimodal whole-slide vision-language model designed for general-purpose slide representation learning in histopathology. Building on the success of knowledge distillation and masked-image modeling\cite{zhou2021ibot, oquab2023dinov2} for patch encoder pretraining\cite{campanella2024clinical,neidlinger2024benchmarking}, $\ours$ introduces a novel paradigm that leverages millions of high-resolution regions-of-interests (ROIs at $8,192 \times 8,192$ pixels) for large-scale, resolution-agnostic pretraining and scalable WSI encoding. Trained using 336K WSIs across 20 organ types, vision-only $\ours$ produces general-purpose slide representations that can readily be applied to slide-level tasks such as cancer subtyping, biomarker prediction, outcome prognosis, and slide retrieval tasks, outperforming supervised baselines and existing multimodal slide foundation models.
To augment $\ours$ with language capabilities, we further finetune by contrasting with 423K synthetic fine-grained ROI-captions generated with PathChat\cite{lu2024multimodal}, a multimodal generative AI copilot for pathology, and with 183K pathology reports at slide level. By leveraging free-text morphological descriptions, $\ours$ gains the ability to generate pathology reports, perform zero-shot classification, and enable cross-modal retrieval between histology slides and clinical reports.
Pretraining $\ours$ on an extensive repository of multimodal pathology data unlocks new levels of performance compared to existing slide foundation models, particularly in low data regimes, language-guided zero-shot classification, and rare cancer retrieval. 
Additionally, we show the utility of pretraining with synthetic fine-grained morphological descriptions for the first time, hinting at the scaling potential of $\ours$ pretraining with synthetic data\cite{chen2021synthetic,kokosi2022synthetic,carrillo2024generation}.
Through comprehensive evaluation across a large range of clinical tasks, including the first application to rare cancer retrieval across 43 rare cancer types, we demonstrate the efficacy of our vision-language pretraining approach, showcasing the general-purpose capability of our slide foundation model. 

\Heading{Results}

\begin{figure*}
\centering
\includegraphics[width=\textwidth]{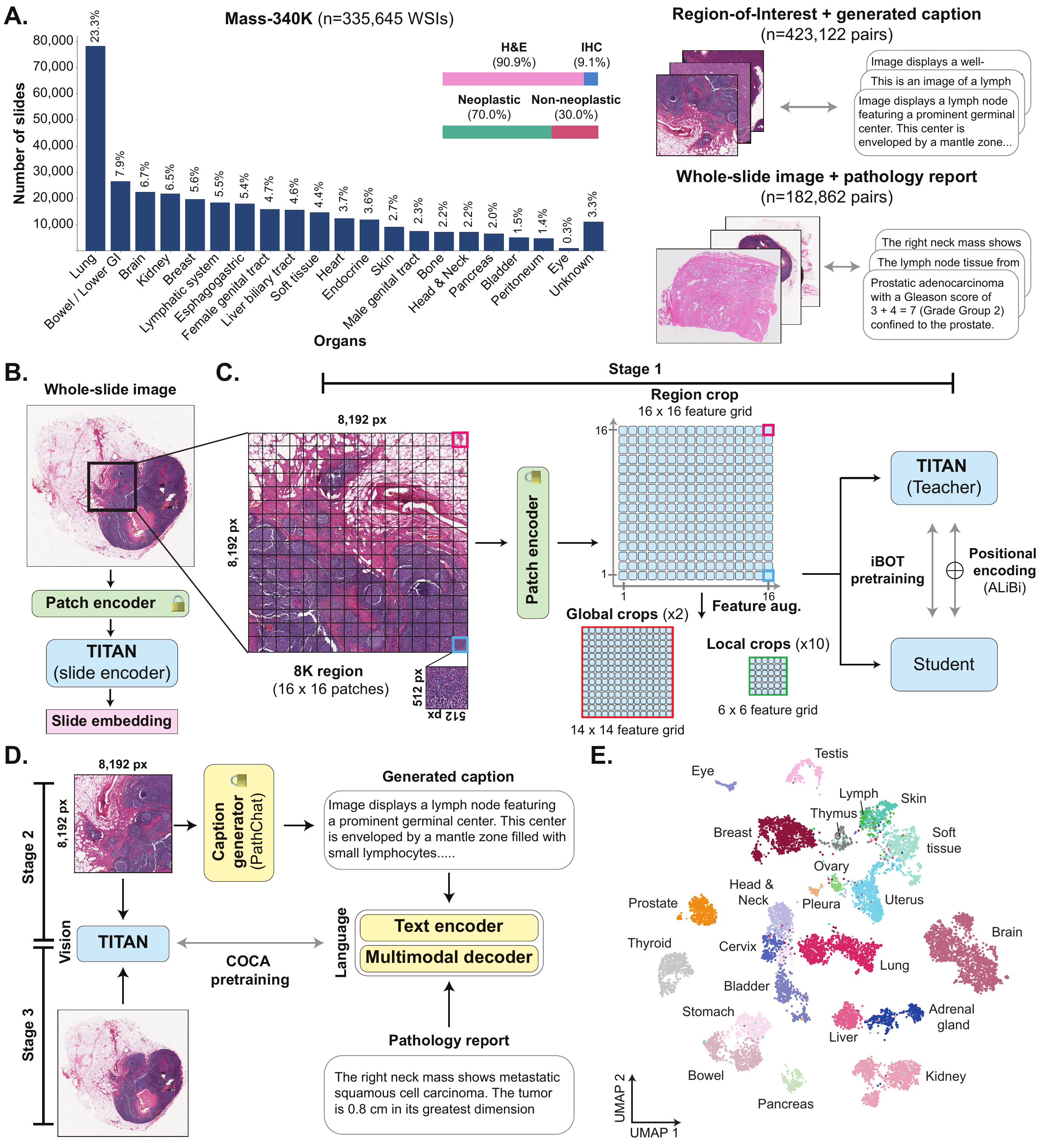}
\caption{\textbf{Overview of $\ours$}.
\textbf{(a)} Tissue site distribution of Mass-340K used for $\oursv$ pretraining (Stage 1). Mass-340K includes 335,645 WSIs across 20 organs with a mix of hematoxylin-and-eosin-stained (90.9\%) and immunohistochemistry-stained tissue sections (9.1\%) or a mix of neoplastic (70.0\%) and non-neoplastic tissue sections (30.0\%). $\oursvcr$ pretraining (Stages 2 and 3) uses a subset of Mass-340K with paired captions and medical reports.
\textbf{(b--d)} Block diagram of $\oursv$ pretraining. 
\textbf{(b)} $\ours$ uses a Vision Transformer to encode a WSI into a slide embedding.
\textbf{(c)} $\oursv$ (Stage 1) is pretrained using self-supervised learning with student--teacher knowledge distillation 
\textbf{(d)} $\oursvcr$ (Stage 2 and 3) is pretrained using vision-language modeling, first by aligning the slide embedding with synthetic captions (Stage 2) and then with medical reports (Stage 3).
\textbf{(e)} UMAP visualization of TCGA slide embeddings obtained with $\oursvcr$, color-coded by organ.
UMAP: uniform manifold approximation and projection.
}
\label{fig:slidessl}
\end{figure*} 

\heading{Scaling self-supervised learning from histology patches to whole slide images} 

\noindent $\ours$ is a Vision Transformer (ViT)\cite{dosovitskiy2020image} that creates a general-purpose slide representation readily deployable in diverse clinical settings. It is pretrained on an internal dataset (termed \textbf{Mass-340K}) consisting of 335,645 WSIs and 182,862 medical reports (\textbf{\Cref{fig:slidessl}A}).
To ensure the diversity of the pretraining dataset, which has proven to be a key factor in successful patch encoders\cite{campanella2024clinical}, Mass-340K is distributed across 20 organs, across different stains (Hematoxylin-and-eosin 90.9\% and immunohistochemistry 9.1\%), and across neoplastic and non-neoplastic tissue (70.0\% and 30.0\%, respectively).
The pretraining strategy consists of three distinct stages to ensure that the resulting slide-level representations capture histomorphological semantics both at the ROI-level (4$\times$4mm$^2$) and at the WSI-level with the help of visual and language supervisory signals: \textbf{Stage 1} vision-only unimodal pretraining with Mass-340K on ROI crops (\textbf{\Cref{fig:slidessl}C}), \textbf{Stage 2} cross-modal alignment of generated morphological descriptions at ROI-level (423K pairs of 8K$\times$8K ROIs and captions), and \textbf{Stage 3} cross-modal alignment at WSI-level (183K pairs of WSIs and clinical reports, \textbf{\Cref{fig:slidessl}D}).
For ease of notation, we refer to the model pretrained with vision-only in Stage 1 as $\oursv$ and to the full model after all three stages of pretraining as $\oursvl$. A detailed description of the pretraining dataset can be found in \textbf{Online Methods} section \textbf{Large-scale pretraining datasets}.

The cornerstone of our approach is emulating the patch encoder designed for input patch images at the slide level.
Instead of using tokens from a partitioned image patch, the slide encoder takes a sequence of patch features encoded by powerful histology patch encoders\cite{wang2022transformer, huang2023visual, chen2024towards, lu2024visual, vorontsov2024foundation, filiot2023scaling, azizi2023robust, hoptimus0, zimmermann2024virchow}. Consequently, all of $\ours$ pretraining stages occur in the embedding space based on pre-extracted patch features, with the patch encoder assuming the role of the ``patch embedding layer" in a conventional ViT  (\textbf{\Cref{fig:slidessl}B}).
To preserve the spatial context of each patch and consequently enable the use of positional encoding in the embedding space, the patch features are spatially arranged in a 2D feature grid replicating the positions of the corresponding patches within the tissue  (\textbf{\Cref{fig:slidessl}C}). 
Following the success of masked image modeling and knowledge distillation in patch encoders\cite{campanella2024clinical}, we apply the iBOT\cite{zhou2021ibot} framework for vision-only pretraining of $\ours$. 
The 2D feature grid setup allows us to directly apply student-teacher knowledge distillation approaches which typically require square crop inputs.

While the conceptual transition to slide-level is simple, this presents a new set of model design and pretraining challenges that precludes clinical translation: 1) Handling long and variable input sequences ($>10^4$ tokens at slide-level vs.\ 196 to 256 tokens at the patch-level), 2) creating multiple views of one sample for self-supervised learning, and 3) ambiguity over positional encoding schemes that capture local and global context in the tissue microenvironment.
First, to tame the computational complexity caused by long input sequence, we construct the input embedding space by dividing each WSI into non-overlapping patches of 512$\times$512 pixels at 20$\times$magnification, followed by extraction of 768-dimensional features for each patch with the extended version of CONCH\cite{lu2024visual}, CONCHv1.5. By increasing the patch size from widely-used 256$\times$256 pixels, we effectively reduce the sequence length by four without impacting the representation quality due to higher resolution patch input, leveraging the robustness of the patch-level foundation models in generalizing to higher resolutions\cite{oquab2023dinov2, chen2024towards, lu2024visual}. 
To address the issue of large and irregular-shaped WSIs, we create views of a WSI by sampling a smaller square crop of features (\textbf{\Cref{fig:slidessl}C}). Specifically, at each epoch for a given WSI, a \textit{region crop} of 16$\times$16 features covering a region of 8,192$\times$8,192 pixels is randomly sampled from the WSI feature grid. From this region crop, two random \textit{global} (14$\times$14) and ten \textit{local} (6$\times$6) crops are sampled for the iBOT training. We augment these feature crops further with vertical and horizontal flipping, followed by posterize feature augmentation\cite{bar2024frozen}.
Finally, to ensure that the limited context pretraining translates to slide-level tasks, we use attention with linear bias (ALiBi) for long context extrapolation of $\ours$ at inference time\cite{alibi}. Originally proposed for long-context inference in large language models, we extended ALiBi to 2D, where the linear bias is based on the relative Euclidean distance between features in the feature grid, which reflects the actual distances between patches in the tissue.
More details of the pretraining dataset and training strategy can be found in \textbf{Online Methods} section \textbf{Vision-only pretraining dataset} and \textbf{Unimodal visual pretraining}, respectively.

To equip our model with language capabilities, we implement two additional multi-resolution pretraining strategies (Stages 2 and 3) using a subset of WSIs in Mass-340K (\textbf{\Cref{fig:slidessl}D}). This is based on the observation that language descriptions exist at multiple morphological scales, from fine-grained descriptions in pathologist annotations or textbooks at the patch- or region-level (Stage 2) to high-level descriptions in pathology reports at the slide-level (Stage 3). For both stages, we use contrastive captioners (CoCa)\cite{yu2022coca} as the pretraining strategy that aligns ROI and slide representations with the corresponding captions and reports, while generating accurate descriptions at ROI-level or reports at slide-level, respectively. The slide encoder (weights initialized with $\oursv$), the text encoder, and the multimodal decoder are all finetuned as part of the pretraining. In Stage 2, we pretrain $\oursv$ with 423,122 pairs of 8K$\times$8K ROIs and synthetic captions generated by the vision-language copilot PathChat\cite{lu2024multimodal}. In Stage 3, we further pretrain the model with 182,862 pairs of WSIs and corresponding pathology reports, resulting in our final model $\oursvcr$.  To diversify the captions and reports with data augmentation, we rewrite the text with a locally deployed large language model (LLM)\cite{yang2024qwen2} and select randomly between several versions for vision-language alignment. A detailed description of the vision-language pretraining dataset and training strategy can be found in \textbf{Online Methods} sections \textbf{Large-scale pretraining datasets} and \textbf{Vision-language continual pretraining}, respectively. 

\begin{figure*}
\centering
\includegraphics[width=\textwidth]{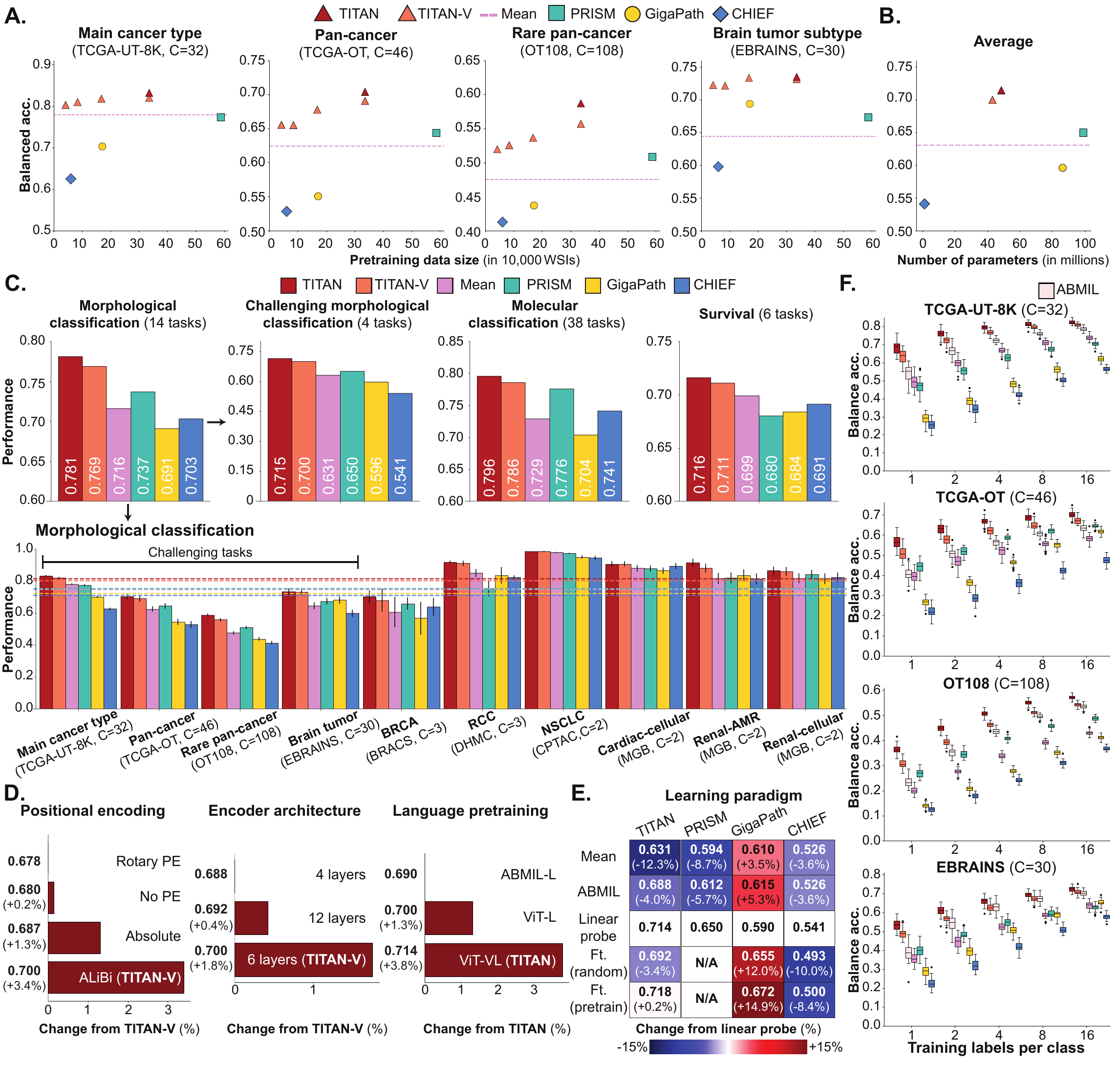}
\caption{\textbf{$\ours$ evaluation.}
\textbf{(a)} Impact of pretraining data size on $\oursv$ and baselines across four challenging subtyping tasks (TCGA-UT-8K, TCGA-OT, OT108 and EBRAINS). $\oursv$ is pretrained with 12.5\%, 25\%, 50\%, and 100\% of Mass-340K.
\textbf{(b)} The average performance of the four tasks against the number of parameters for each baseline.
\textbf{(c)} Linear probe evaluation of $\ours$ and baselines on morphological classification (all and challenging subset), molecular status, and survival prediction tasks. The mean uses the same patch encoder as $\ours$ (CONCHv1.5). Multi-class tasks are evaluated with balanced accuracy, binary tasks with AUROC, and survival tasks with concordance index. For external cohorts (DHMC, CPTAC), the classifier is trained on the corresponding TCGA cohort. All error bars represent standard deviations based on bootstrapping. 
\textbf{(d)} Ablation study comparing the impact of positional encoding, number of Transformer layers, and inclusion of vision-pretraining stage. The performance is averaged across the four subtyping tasks.
\textbf{(e)} Change in performance of $\ours$ and baselines averaged across the four subtyping tasks for different learning paradigms. For mean pooling and ABMIL, the respective patch encoder for each framework is used.
\textbf{(f)} Linear probe few-shot performance $@K$ shots, with $K\in\{1,2,4,8,16\}$, comparing baselines and ABMIL with CONCHv1.5. For each setting, 50 runs were performed. Whiskers extend to data points within 1.5$\times$ the interquartile range.
C: number of classes. Ft.: finetune. ABMIL: attention-based multiple instance learning.
}
\label{fig:diagnostic}
\end{figure*}

\heading{$\ours$ improves region and slide-level diagnostic capabilities}

We begin by evaluating $\oursvcr$, $\oursv$, and existing slide encoders on a large set of diverse slide-level tasks, including morphological subtyping and molecular classification. Following the standard practice in self-supervised learning\cite{oquab2023dinov2,tian2024stablerep}, we employ linear probing by fitting a linear model for classification (logistic regression)
on frozen slide embeddings. 
Specifically, we use the linear weights estimated with the $\ell_2$-regularization parameter tuned on a validation set for evaluating the test performance.
For tasks with multiple cohorts available, we perform cross-validation on one cohort, e.g., from TCGA\cite{weinstein2013tcga, liu2018integrated}, and use the remaining cohorts, e.g., from CPTAC\cite{edwards2015cptac,thangudu2020cptac} or DHMC\cite{wei2019pathologist, zhu2021development}, as an external test cohort.
As baselines, we evaluate the recent vision-language slide foundation models with model weights available, namely PRISM\cite{shaikovski2024prism}, GigaPath\cite{xu2024multimodal}, and CHIEF\cite{wang2024pathology}. Compared to $\ours$, these models employ different pretraining strategies for their slide-level encoders (PRISM: WSI-report contrastive pretraining, GigaPath: masked image reconstruction pretraining, CHIEF: supervised contrastive learning of cancerous vs.\ non-cancerous WSIs), different patch-level encoders pretrained on histology patches trained at different magnifications and patch sizes (PRISM and GigaPath: 256$\times$256 pixels at 20$\times$ magnification, CHIEF: 256$\times$256 pixels at 10$\times$ magnification), 
and utilize a varying number of WSIs for pretraining (PRISM: 1.7$\times$, GigaPath: 0.49$\times$, CHIEF: 0.18$\times$ the WSIs used for $\ours$ pretraining). Additionally, we compare against mean pooling with the same CONCHv1.5 patch encoder as $\ours$, which has shown to be a simple yet powerful unsupervised slide representation framework\cite{song2024morphological, songmultimodal, jaume2024multistain}. 

Furthermore, for a comprehensive evaluation of the baselines, we introduce two tumor classification tasks based on the publicly available repository TCGA with two different context lengths: (i) Main cancer type classification on ROIs (\textit{TCGA-Uniform-Tumor-8K} or \textit{TCGA-UT-8K}), a ROI-level cancer subtyping task with 32 classes, where we manually curated 25,495 tumor-containing regions of 8,192$\times$8,192 pixels at 20$\times$ magnification ($\sim$4$\times$4 mm$^2$) across TCGA, covering the same tissue context as the region crops in $\oursv$ pretraining (\textbf{Extended Data \Cref{fig:edf_tcgaut}}, \textbf{Extended Data~\Cref{tab:dataset_tcga-ut}}) and (ii) pan-cancer classification (\textit{TCGA-OncoTree} or \textit{TCGA-OT}), a slide-level OncoTree code\cite{kundra2021oncotree} classification task with 46 classes, consisting of 11,186 formalin-fixed paraffin-embedded (FFPE) WSIs from TCGA. TCGA-OT is the largest pan-cancer slide-level classification task that is publicly available (\textbf{Extended Data~\Cref{tab:dataset_tcga-ot}}). With the exception of CHIEF, the pretraining datasets of $\ours$ (Mass-340K), PRISM, and GigaPath do not include TCGA and PANDA slides, which allows us to utilize these two datasets as benchmarking tasks without the concern of data leakage\cite{kapoor2023leakage}.
More details on the two tasks can be found in \textbf{Online Methods} section \textbf{Downstream evaluation datasets}. For the benefit of the community, we plan to release TCGA-UT-8K datasets and TCGA-OT labels.

Prior to the expansive evaluations on diverse tasks, we first assess how the pretraining data scale affects the downstream performance of $\oursv$.
We focus on four subtyping tasks -- TCGA-UT-8K, TCGA-OT, OT108, and EBRAINS -- due to the challenging diagnostic complexity with a large number of diagnostic classes.
For the pretraining data scale, we train $\oursv$ with 12.5\%, 25\%, and 50\% of Mass-340K, maintaining the same distribution across the organs as the full dataset.
We observe that the performance increases on all four tasks as more pretraining data is utilized, where $\oursv$ with full Mass-340K exhibits 3.65\%, 3.21\%, and 1.21\% average increase over $\oursv$ pretrained with 12.5\%, 25\%, and 50\% of Mass-340K, respectively (\textbf{\Cref{fig:diagnostic}A}).
Similar to the observation in pretrained patch encoders in pathology\cite{chen2024towards} as well as ViTs for natural images\cite{dosovitskiy2020image, zhai2022scaling}, we observe the \textit{scaling law}\cite{kaplan2020scaling} for $\oursv$ on all four tasks with more pretraining data leads to an increase in performance.
Despite the difference in pretraining recipes, we also observe the same general trend for the three other slide encoders, where PRISM outperforms GigaPath and CHIEF by 9.01\% and 20.1\% on average, having 3.4 times and 9.7 times the number of pretraining WSIs, respectively.
Furthermore, we observe that $\ours$ and $\oursv$, with 48.5 million and 42.1 million parameters respectively, outperform heavier slide encoders PRISM and GigaPath, with 99.0 million and 86.3 million parameters, demonstrating superior parameter efficiency of our model.

We next evaluate $\ours$ on an expansive range of tasks comprised of morphological classification (14 tasks), grading (3 tasks), molecular classification (38 tasks), and survival prediction (6 tasks), where the summary of each task can be found in \textbf{Extended Data \Cref{tab:downstream-links,tab:tasks_morphological,tab:task_grading,tab:task_survivl,tab:task_molecular}}. On average, we observe that $\oursvcr$ and $\oursv$ outperform other slide encoders (\textbf{\Cref{fig:diagnostic}C}). 
$\ours$ especially excels at morphological subtyping tasks across the entire spectrum of diagnostic complexities including fine-grade pan-cancer classification (challenging morphological classification tasks in \textbf{Extended Data \Cref{fig:diagnostic}C}) and non-cancerous tasks such as allograft rejection, with $\oursvcr$ and $\oursv$ achieving an average of +8.4\% and +6.7\%, respectively, in performance on multi-class and binary subtyping tasks, averaged over balanced accuracy for multi-class tasks and AUROC for binary tasks over the next-best performing model, PRISM (\textbf{\Cref{fig:diagnostic}C}).
In particular, $\oursv$ (and $\oursvl$) not only outperforms others on TCGA-UT-8K with 8K$\times$8K context that the model was trained on (+ 6\% and 7.5\% over PRISM), 
but also on WSI-level tasks that involve the entire tissue context, where $\oursv$ benefits from the long-context extrapolation via ALiBi, e.g., TCGA-OT (+ 7\% and 9.5\% over PRISM), 
OT108 (+ 10\% and 16\% over PRISM), 
and EBRAINS (+ 9\% and 9.1\% over PRISM). 
To demonstrate the robustness of $\ours$ across different evaluation schemes, we run further analyses with prototyping evaluation\cite{wang2019simpleshot, snell2017prototypical}, where the label of the query slide is determined by the proximity to the mean of the slide embeddings in each diagnostic class, as well as 20 nearest-neighbors evaluation (\textbf{Extended Data~\Cref{tab:log_reg_TCGAUniformTumor_cancer,tab:log_reg_TCGA_OncoTreeCode,tab:log_reg_OP108_OncoTreeCode,tab:log_reg_ebrains_diagnosis}}). Both $\oursvl$ and $\oursv$ outperform the next best method PRISM by an even larger margin, + 14\% and 9.5\% for SimpleShot and + 15\% and 9.2\% for 20 nearest-neighbor evaluation, averaging balanced accuracy for all morphological subtyping tasks, which manifests that $\ours$ leads to improved representation quality. 
On grading tasks, $\oursvl$ outperforms the next best models CHIEF on average by + 3.2\% and PRISM by + 4\% in quadratic-weighted Cohen's $\kappa$, where the high performance of CHIEF can be attributed to including the dataset PANDA in pretraining.

To evaluate the molecular classification performance, we tested the model on tasks from public datasets (BCNB and MUT-HET) and internal-external paired public datasets (TCGA, CPTAC, and EBRAINS), on immunohistochemistry (IHC) tasks, and MGB internal molecular tasks (\textbf{\Cref{fig:diagnostic}B, Extended Data~\Cref{fig:edf_molecular}, Extended Data~\Cref{tab:log_reg_TCGA-BRCA_subtype,tab:log_reg_TCGA-NSCLC_subtype,tab:log_reg_TCGA-RCC_OncoTreeCode,tab:log_reg_BRACS_label_fine,tab:log_reg_BRACS_label_coarse,tab:log_reg_CRANE_cellular,tab:log_reg_MANTA_HE_label_amr,tab:log_reg_MANTA_HE_label_cell,tab:log_reg_MANTA_HE_label_ifta,tab:log_reg_IMP_grade,tab:log_reg_PANDA_isup_grade,tab:log_reg_mut-het-rcc_BAP1_mutation,tab:log_reg_mut-het-rcc_PBRM1_mutation,tab:log_reg_mut-het-rcc_SETD2_mutation,tab:log_reg_BCNB_er,tab:log_reg_BCNB_pr,tab:log_reg_BCNB_her2,tab:log_reg_TCGA-GBMLGG_idh_status,tab:log_reg_TCGA-BRCA_ER,tab:log_reg_TCGA-BRCA_PR,tab:log_reg_TCGA-BRCA_HER2,tab:log_reg_TCGA-BRCA_PIK3CA,tab:log_reg_TCGA-NSCLC_EGFR,tab:log_reg_TCGA-NSCLC_TP53,tab:log_reg_TCGA-CRC_isMSIH,tab:log_reg_TCGA-CRC_BRAF,tab:log_reg_TCGA-CRC_KRAS,tab:log_reg_MGH_breast_erpr_er_6,tab:log_reg_MGH_breast_erpr_pr_6,tab:log_reg_mgb_brca_er,tab:log_reg_mgb_brca_pr,tab:log_reg_mgb_brca_her2,tab:log_reg_mgb_lung_cdx-2,tab:log_reg_mgb_lung_ck5-6,tab:log_reg_mgb_lung_napsina,tab:log_reg_mgb_lung_p40,tab:log_reg_mgb_lung_p63}}).
We observe that $\oursvl$ consistently performs best with + 0.9\% on BCNB and MUT-HET, + 1.7\% on TCGA, and +3.7\% on internal molecular classification of BRCA and LUAD, in averaged AUROC scores over the next best model PRISM. For IHC quantification, $\oursv$ performs best with + 12\% in quadratic-weighted Cohen's $\kappa$ over the next best model CHIEF since our vision-language alignment does not include IHC reports leading to a slightly lower performance of $\oursvl$ of + 7.3 \% over CHIEF.
In external evaluations, where the linear classifiers trained on TCGA were applied to EBRAINS (IDH) and CPTAC (all other molecular endpoints), $\oursv$ outperformed other slide encoders overall (+ 0.9\% over PRISM), whereas $\oursvl$ performance slightly worse (- 0.3 \%). Since $\oursvl$ still outperforms PRISM by + 4.8\% in SimpleShot and by + 5.4\% in 20-nearest neighbors evaluation (\textbf{Extended Data~\Cref{tab:log_reg_TCGA-GBMLGG_idh_status,tab:log_reg_TCGA-BRCA_ER,tab:log_reg_TCGA-BRCA_PR,tab:log_reg_TCGA-BRCA_HER2,tab:log_reg_TCGA-BRCA_PIK3CA,tab:log_reg_TCGA-NSCLC_EGFR,tab:log_reg_TCGA-NSCLC_TP53,tab:log_reg_TCGA-CRC_isMSIH,tab:log_reg_TCGA-CRC_BRAF}}), this could be attributed to suboptimal generalization of the linear classifier when searching for $\ell_2$-regularization strength over a large range to optimize for linear probing performance. 
We note that the pretraining dataset of CHIEF includes all WSIs from TCGA, which could partially contribute to its performance in respective tasks.

On survival prediction tasks, we utilize disease-specific survival (DSS)\cite{liu2018integrated} as the clinical endpoint and the concordance index (c-index) as the evaluation metric and perform 5-fold cross-validation on six TCGA site-preserving stratified cancer cohorts. Specifically, we fit the linear Cox proportional hazards model on the slide embeddings to predict patient-level survival risk. 
We observe that $\oursvcr$ and $\oursvc$ are generally the best-performing baselines, outperforming the next-best performing model CHIEF by +3.62\% and +2.90\% respectively, even though CHIEF was pretrained on TCGA slides (\textbf{Extended Data Table~\ref{tab:survival_main}}). Interestingly, the mean pooling baseline shows competitive performance. This suggests that the proportion of different morphological phenotypes, which the mean baseline is effectively computing, is an important prognostic factor\cite{song2024morphological, songmultimodal}. 

To further understand how the slide embedding space is organized and consequently affects the downstream performance, we investigate UMAP embeddings of WSIs within our largest and most diverse downstream dataset, TCGA-OT, where we color-code by organs instead of OncoTree codes to reduce visual clutter (\textbf{\Cref{fig:slidessl}E}, \textbf{Extended Data \Cref{fig:edf_umap}}).
We observe that the embeddings form distinct clusters along organs for $\oursvcr$ and $\oursv$, with $\oursvcr$ clusters seemingly better separated than $\oursv$ (\emph{e.g.,} breast further separated from bladder, stomach, and lung), confirming superior subtyping performance against other slide encoders.
The embedding space for both the mean CONCHv1.5 and PRISM are reasonably separated reflecting good subtyping performances, whereas CHIEF and GigaPath are not able to effectively separate different organ embeddings, consequently leading to poor performance.
This demonstrates that the WSIs from diverse organs in our pretraining dataset helps both $\oursvcr$ and $\oursv$ to be able to extract subtle organ-specific morphological cues more effectively.

\heading{Algorithmic design considerations for $\ours$}

To better understand how certain model choices affect the downstream performance, we perform ablation experiments on three design choices of $\ours$: the positional encoding, the number of Transformer layers in $\oursv$, and the inclusion of vision pretraining (\textbf{\Cref{fig:diagnostic}D}).
Similar to previous analyses, we focus on the four subtyping tasks for the ablation experiments.

For the positional encoding, we pretrain $\oursv$ with absolute positional encoding, following the ViT design\cite{dosovitskiy2020image}, rotary positional encoding\cite{su2024roformer} extended to 2D\cite{heo2024rotary} (Rotary PE), and without positional encoding (No PE).
Our results show that ALiBi outperforms the other encoding schemes on all four tasks, with an average improvement of + 2.01\% over absolute positional encoding, the second-best performing method (\textbf{\Cref{fig:diagnostic}D}).
This indicates that diagnostic performance can be enhanced with a suitable choice of positional encoding by contextualizing the patch features, with ALiBi helping $\oursv$ extrapolate effectively to the entire slide, where the context length is over ten times longer.
For the number of layers and consequently the number of parameters for $\oursv$, we observe that 6 layers (43M parameters) on average provide the sweet spot between smaller (4 layers, 29M parameters) and larger models (12 layers, 86M parameters), by outperforming them by 1.72\% and 1.31\%, respectively (\textbf{Extended Data Table~\ref{tab:ablation_model_size_OP108_OncoTreeCode}}).

For the pretraining strategy ablation, we compare the performance between $\oursvcr$ with the full pretraining and $\ourscr$, which only performs vision-language alignment without vision pretraining. We introduce an additional baseline of multiheaded attention-based MIL\cite{ilse2018attention, jaume2024multistain} with the vision-language pretraining  (ABMIL-L), to further understand whether a different slide encoder architecture with the same pretraining recipe can be effective.
We observe that $\oursvcr$ outperforms $\ourscr$ by 2.35\% and ABMIL-L by 3.62\%. This indicates that the vision pretraining (Stage 1) provides better initialization weights than the random weights for vision-language alignment, leading to improved downstream performance, also observed in patch encoder pretraining\cite{lu2024visual}. This also suggests that the better performance of $\oursvcr$ over PRISM, which is pretrained only with the vision-language alignment, could be due to the inclusion of the vision pretraining step. Moreover, the worse performance of ABMIL architecture than ViT with the same pretraining recipe justifies the choice of ViT as the architecture for $\ours$.

\heading{Comparison with different learning paradigms for slide encoding}

To further assess the quality of the slide embeddings produced by $\oursv$, we evaluate different learning paradigms by comparing the linear probe performance of each slide encoder against other MIL models comprised of \textit{mean pooling}, i.e., averaging the patch embeddings, \textit{attention-based MIL} (ABMIL)\cite{ilse2018attention}, and task-specific \textit{finetuning} of the slide encoder from random or the pretrained weights.
For the mean pooling and ABMIL baselines, we use respective patch encoders for each slide encoder framework.
This analysis allows us to gauge whether the pretrained slide encoders have learned meaningful slide representations and consequently outperform the simple yet powerful unsupervised (mean pooling) and supervised (ABMIL) baselines, neither of which involve large-scale pretraining on thousands of WSIs.

We observe several trends with $\ours$ (\textbf{\Cref{fig:diagnostic}E, Extended Data \Cref{fig:edf_learning}, Extended Data~\Cref{tab:finetuning_TCGAUniformTumor_cancer,tab:finetuning_TCGA_OncoTreeCode,tab:finetuning_OP108_OncoTreeCode,tab:finetuning_ebrains_diagnosis}}). 
First, ABMIL outperforms mean pooling, as expected, since ABMIL is supervised and equivalent to weighted averaging of the patch features, which would by default include the simple averaging solution of mean pooling.
Next, the linear probe outperforms ABMIL, which demonstrates that $\oursvcr$ and $\oursv$, having been pretrained on a large repository of multimodal pathology data much larger than what is provided to ABMIL for each downstream task, can encode additional contextual and semantic morphological details of the slide.
This leads to task-agnostic slide embedding of $\oursv$ being better equipped for downstream tasks than tasks-specific supervised slide embeddings of ABMIL.
Finally, we observe that task-specific finetuning of $\oursvcr$ leads mostly to performance improvement over linear probe of $\oursvcr$ and $\oursv$. Furthermore, finetuning the slide encoder from randomly initialized weights yields lower performance (-3.63\% on average) than from the $\oursvl$ pretrained weights. This suggests that the pretrained weights of $\oursv$ can serve as a good initialization set for downstream tasks for typical cohorts with a limited number of patients, in line with previous works\cite{jaume2024transcriptomics, shaikovski2024prism}. One exception is OT108, which could be attributed to the small number of samples for each class (ranging from 4 to 42), which may lead to overfitting.
We observe that similar trends exist for PRISM except for finetuning scenarios, which could not be compared due to PRISM finetuning recipes not being provided.

Interestingly, these trends are not always observed in other slide foundation models. For GigaPath, fine-tuning the slide encoder significantly improves the performance over linear probe (14.9\% on average), but lags behind $\oursvcr$ and $\oursv$ linear probe by 6.25\% and 4.17\% on average, respectively. That finetuning leads to a significant improvement, combined with low linear probe performance, suggests a lack of generalizability off-the-shelf for Gigapath slide embeddings. This is further supported by the fact that the simple mean pooling, which does not leverage pretraining on large WSI repositories, outperforms the linear probe across all four subtyping tasks (\textbf{Extended Data \Cref{fig:edf_learning}}).
For CHIEF, the trends are mixed, with ABMIL performing better than the linear probe on half of the tasks. Interestingly, finetuning the slide encoder always yields worse performance than the linear probe (-8.38\% on average), suggesting that the pretrained CHIEF weights might be a sub-optimal starting point for task-specific fine-tuning.
Nevertheless, finetuning from pretrained weights indeed yields better performance on average than randomly initialized weights for both GigaPath and CHIEF, in line with what is observed for $\ours$.

\heading{Few-shot learning for low data regime}

We also evaluate the data-constrained setting of few-shot learning where only a few examples for each category are provided within the linear probe setting. In the few-shot setting, $\oursv$ remains superior across all tasks and number of shots in balanced accuracy when assessed with linear probe (\textbf{\Cref{fig:diagnostic}F}). To mitigate sampling bias, we aggregate the results over 50 different runs, with random samples used for training, while fixing the test set.
We observe that $\oursvcr$ is the best-performing model across different tasks, demonstrating the strong generalizability of $\oursvcr$. $\oursv$ is the second-best performing model, which supports our results that vision-language alignment benefits the downstream task performance. 
Specifically, $\oursvcr$ and $\oursv$ exhibit especially high performance in one-shot learning, which is on par with other slide encoders trained on more shots 
(\textbf{Extended Data~\Cref{tab:fewshot_log_reg_OP108_OncoTreeCode,tab:fewshot_log_reg_TCGA_OncoTreeCode,tab:fewshot_log_reg_TCGAUniformTumor_cancer,tab:fewshot_log_reg_ebrains_diagnosis}}). Specifically, $\oursvcr$ and $\oursv$ outperform CHIEF based on 16 shots on TCGA tasks by 22.4\% and 13.5\% (TCGA-UT-8K) and 18.7\% and 6.8\% (TCGA-OT) when compared with the median value of 50 runs, respectively, even though CHIEF has been pretrained on TCGA slides.

Interestingly, both $\oursvcr$ and $\oursv$ also outperform ABMIL with the same patch encoder across all settings. While the performance gap with ABMIL shrinks for a higher shot regime as expected, we observe that the gap is indeed wider in the lower shot regime. 
The largest gap for 1-shot is observed in the OT108 task, where $\oursvcr$ outperforms ABMIL by 56.7\%.
These observations underscore the superior data efficiency of a pretrained slide encoder and suggest that $\oursv$ can excel in rare cancer settings with a limited number of samples, such as OT108 in our benchmark, where heavily parameterized supervised approaches are inherently restrained.
This demonstrates the advantage of $\ours$ over other slide encoders such as GigaPath and CHIEF, the intended usages of which are in supervised settings with task-specfic finetuning, rather than off-the-shelf usage with the frozen slide embeddings.
The same trend is observed even when evaluated with prototyping, where $K$ samples (shots) from each class are averaged to construct the prototype (\textbf{Extended Data~\Cref{tab:fewshot_prototypes_OP108_OncoTreeCode,tab:fewshot_prototypes_TCGA_OncoTreeCode,tab:fewshot_prototypes_TCGAUniformTumor_cancer,tab:fewshot_prototypes_ebrains_diagnosis}}). More details on the experiments can be found in \textbf{Online Methods} section \textbf{Few-shot classification}.

\heading{Language-aligned $\ours$ enables cross-modal capabilities}

\begin{figure*}
\centering
\includegraphics[width=\textwidth]{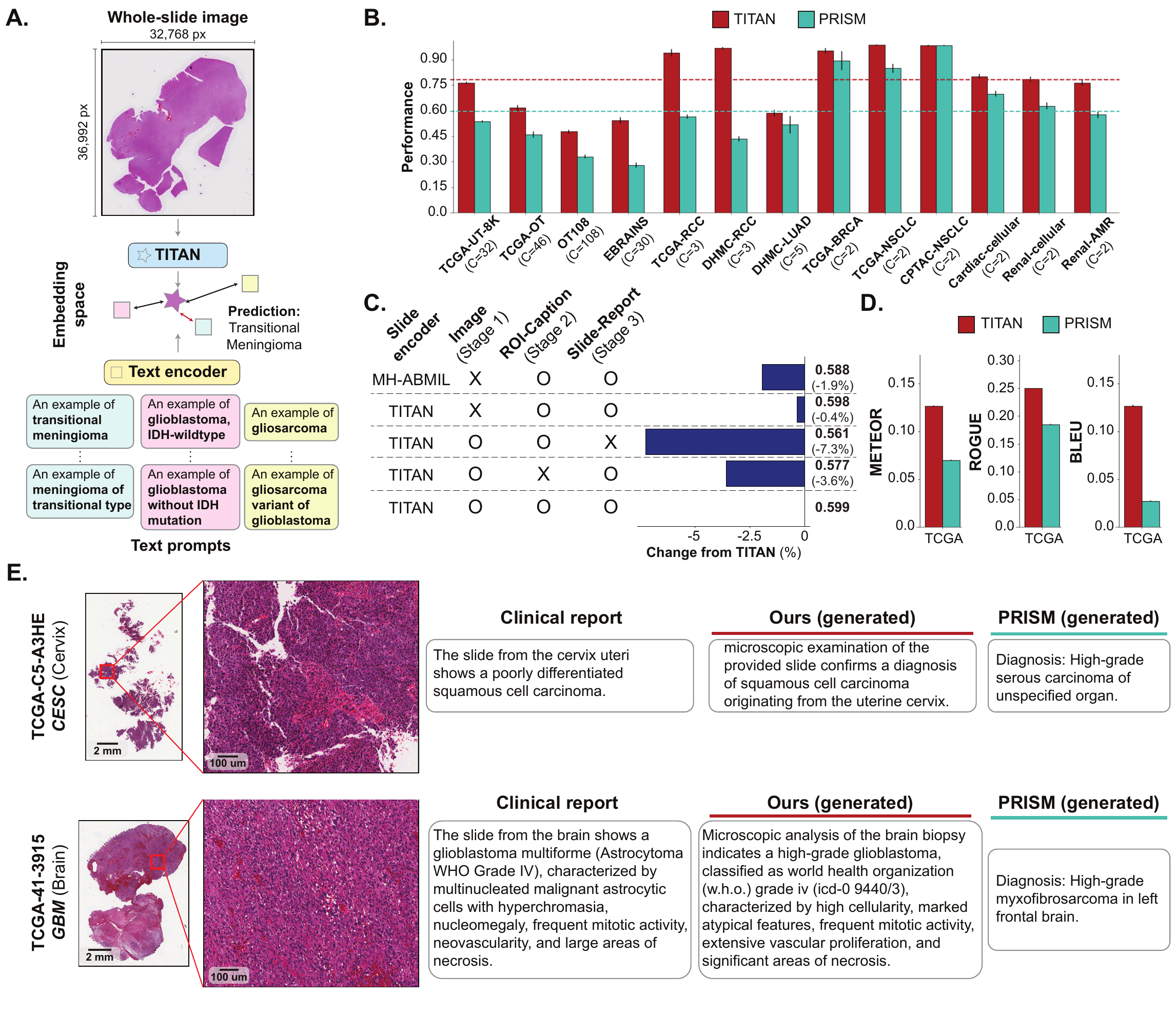}
\caption{\textbf{Visual-language evaluation of $\ours$.} \textbf{(a)} A schematic for zero-shot evaluation. The query slide is classified by identifying the closest text prompt embedding in the slide embedding space.  \textbf{(b)} Zero-shot performance of $\oursvcr$ and PRISM. All multi-class tasks are evaluated with balanced accuracy and binary tasks are evaluated with AUROC. All error bars represent standard deviations based on bootstrapping. \textbf{(c)} Ablation study comparing different pretraining strategies, and assessed with zero-shot performance averaged across TCGA-UT-8K, TCGA-OT, OT108, and EBRAINS. Evaluations are based on the percentage changes of balanced accuracy from the reference zero-shot performance of $\oursvcr$. \textbf{(d)} Report generation evaluation on TCGA-Slide-Reports, and evaluated  using METEOR, ROGUE, and BLEU. \textbf{(e)} TCGA examples of generated reports of $\oursvcr$ and PRISM, with the corresponding clinical reports. Additional examples of generated reports are available in \textbf{Extended Data \Cref{fig:edf_reports}}. C: number of classes.}
\label{fig:vislang}
\end{figure*} 

\noindent 
We further assess the language capabilities of $\oursvcr$ by aligning the slide representations of $\oursv$ to language-based morphological descriptions.
Specifically, we assess the cross-modal zero-shot classification\cite{radford2021learning, Lu_2023_CVPR, ikezogwo2024quilt} and report-generation capabilities of $\oursvcr$ and study the effect of Stage 2 pretraining for caption alignment with fine-grained morphological descriptions and Stage 3 pretraining with coarse clinical reports describing the relevant microscopic findings.

To evaluate the quality of vision-language alignment, we first perform cross-modal zero-shot experimentation on 13 subtyping tasks of varying difficulties (\textbf{\Cref{fig:vislang}A}). In cross-modal zero-shot evaluation, the diagnostic labels expressed as text prompts are encoded with the text encoder. The diagnostic prediction of the query slide is then decided by the closest label embedding to the slide embedding encoded with $\oursvcr$. 
The cross-modal zero-shot experiment evaluates how the embedding space with the visual pretraining can be further aligned with the language modality. 
We compare the zero-shot performance against PRISM, also equipped with cross-modal capabilities.
Gigapath is not included as the language-aligned extension has not been publicly released.  
We observe that $\oursvcr$ performs best across these tasks, outperforming PRISM by a large margin on multi-class classification tasks (balanced accuracy +56.52\%) and binary subtyping tasks (AUROC +13.8\%), for both cancer subtyping tasks and non-cancerous tasks (\textbf{\Cref{fig:vislang}B}, \textbf{Extended Data~\Cref{tab:zeroshot_TCGAUniformTumor_cancer,tab:zeroshot_TCGA_OncoTreeCode,tab:zeroshot_OP108_OncoTreeCode,tab:zeroshot_ebrains_diagnosis,tab:zeroshot_TCGA-RCC_OncoTreeCode,tab:zeroshot_cptac_dhmc_rcc_OncoTreeCode,tab:zeroshot_dhmc_luad_label,tab:zeroshot_TCGA-BRCA_subtype,tab:zeroshot_TCGA-NSCLC_subtype,tab:zeroshot_cptac_nsclc_subtype,tab:zeroshot_MANTA_HE_label_amr,tab:zeroshot_MANTA_HE_label_cell,tab:zeroshot_CRANE_cellular}}). The performance gap between $\oursvcr$ and PRISM is the widest on the 30-class EBRAINS subtyping task, where the balanced accuracy of $\oursvcr$ is more than double that of PRISM (balanced accuracy of +121.9\%).
The text prompts used for zero-shot experiments can be found in \textbf{Extended Data~\Cref{tab:tcga_nsclc_brca_prompts,tab:ot108_tcga_prompts_1,tab:ot108_tcga_prompts_2,tab:ot108_tcga_prompts_3,tab:ot108_tcga_prompts_4,tab:ot108_tcga_prompts_5,tab:ot108_tcga_prompts_6,tab:ot108_tcga_prompts_7,tab:ot108_tcga_prompts_8,tab:ot108_tcga_prompts_9,tab:ot108_tcga_prompts_10,tab:ot108_tcga_prompts_11,tab:ot108_tcga_prompts_12,tab:tcga_ut_prompt_1,tab:tcga_ut_prompt_2,tab:tcga_ut_prompt_3,tab:tcga_ut_prompt_4,tab:ebrains_prompt_1,tab:ebrains_prompt_2,tab:ebrains_prompt_3,tab:renal_prompts,tab:heart_prompts}}.

To further understand how different design considerations affect the zero-shot performance of $\oursvcr$, we ablate over pretraining stages and the slide encoder architecture (\textbf{\Cref{fig:vislang}C}). In total, we experiment with four different variations of $\oursvcr$ and present the average performance over four challenging subtyping tasks at slide level, TCGA-UT-8K, TCGA-OT, OT108, and EBRAINS (results for each dataset can be found in \textbf{Extended Data~\Cref{tab:ablation_vl_TCGAUniformTumor_cancer,tab:ablation_vl_TCGA_OncoTreeCode,tab:ablation_vl_OP108_OncoTreeCode,tab:ablation_vl_ebrains_diagnosis}}).
We observe that $\oursvcr$ maintains the best overall zero-shot performance. Of the three pretraining stages, Stage 1 vision-pretraining contributes the least (balanced accuracy of -0.4\% against $\oursvcr$), followed by Stage 2 ROI-caption alignment (-3.6\% against $\oursvcr$) and Stage 3 slide-report alignment (-7.3\% against $\oursvcr$). This demonstrates that aligning vision and language at both fine-grained and global levels, thereby combining the insights independently derived at patch-level\cite{huang2023visual, lu2024visual} and slide-level\cite{xu2024whole, shaikovski2024prism}, is necessary, which is lacking in report-only aligned baselines such as PRISM and GigaPath. 
Finally, the variant using a multi-headed-ABMIL (MH-ABMIL) network as vision backbone with vision-language alignment pretraining lags behind $\oursvcr$ with and without vision-pretraining by 1.94\% and 1.54\%, indicating that our ViT architecture using self-attention with ALiBi provides better downstream performance than attention-based networks.

Finally, we assess $\oursvcr$'s capabilities of generating pathological reports, utilizing the text decoder trained during CoCa pretraining. To this end, we introduce a report generation task on TCGA, consisting of 10,108 FFPE WSIs with paired slide-level reports parsed from 9,523 patient-level TCGA reports released by a previous study\cite{kefeli2024tcga}. We evaluate the models using three metrics METEOR\cite{banerjee2005meteor}, ROGUE\cite{lin2004rouge}, and BLEU\cite{papineni2002bleu}. We observe that $\oursvcr$ outperforms PRISM by a large margin, on average by 161\% across the three metrics (\textbf{\Cref{fig:vislang}D}). In addition, $\oursvcr$ outperforms $\oursvr$ without Stage 2 pretraining, agreeing with the previous experiments on the importance of the ROI-level vision-language alignment. 
Examples of the generated reports for $\oursvcr$ considered high-quality by the pathologists are shown in \textbf{\Cref{fig:vislang}E}, often capable of correctly capturing key attributes such as tissue site, diagnosis and tumor grade as well as key representative morphology. Additional examples are illustrated in \textbf{Extended Data \Cref{fig:edf_reports}}. 
More details on the dataset can be found in \textbf{Online Methods} section \textbf{Downstream evaluation datasets}.

\begin{figure*}[]
\centering
\includegraphics[width=\textwidth]{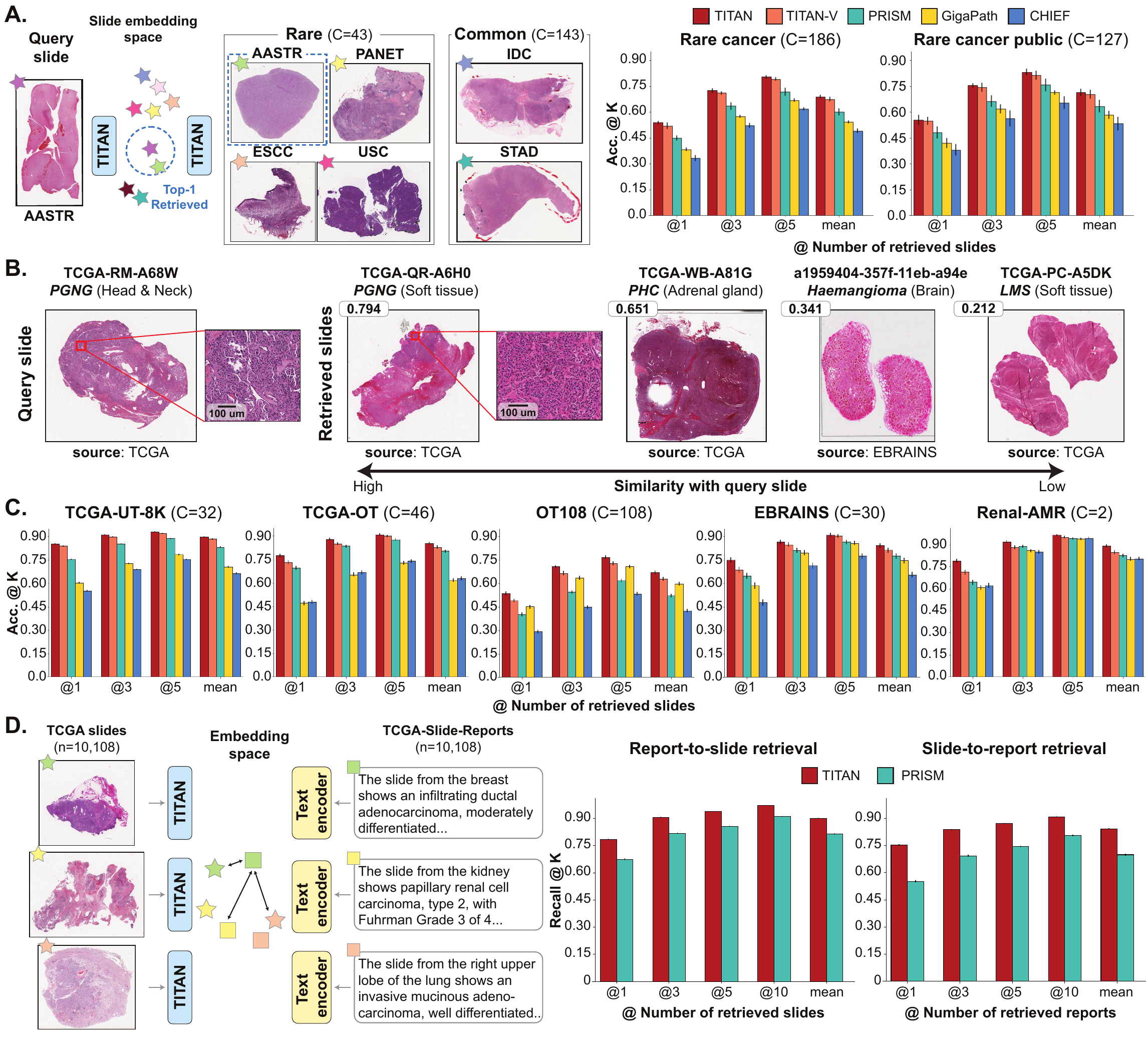}
\caption{\textbf{Retrieval capabilities of $\ours$.}
\textbf{(a)} Slide retrieval results on rare cancer retrieval tasks assessed with accuracy$@K$, with $K=\{1,3,5\}$. Rare-Cancer (internal rare cancer cohort) consists of TCGA, EBRAINS, and the MGB internal cohort, with 43 rare and 143 common cancer types for a total of 186 classes. Rare-Cancer-Public (public rare cancer cohort) consists of TCGA and EBRAINS only, with 29 rare and 98 common cancer types for a total of 127 classes.
\textbf{(b)} Example of rare cancer retrieval on Rare-Cancer with the query slide and four representative retrieved slides. The number indicates the cosine similarity between the query and the retrieved slide. Additional examples of rare cancer retrieval are available in \textbf{Extended Data \Cref{fig:edf_rare}}.
\textbf{(c)} Slide retrieval results on five subtyping tasks. Mean represents the average performance across three shots.
\textbf{(d)} Report-to-slide and slide-to-report cross-modal retrieval performance assessed with recall @ K, with $K=\{1, 3, 5, 10\}$ on TCGA cohort of 10,108 pairs of WSIs and reports for $\oursvcr$ and PRISM. Mean represents the average performance across four shots.
All error bars represent standard deviations based on bootstrapping.
C: number of classes. 
}
\label{fig:retrieval}
\end{figure*}

\heading{$\ours$ enables rare cancer retrieval and cross-modal retrieval} 

Considering cases with similar morphological features and diagnoses is essential for pathologists to make informed decisions, in particular when dealing with complex or rare cases\cite{cruz2011visual,caicedo2009histopathology, komura2018machine, kalra2020yottixel, chen2022fast, hegde2019similar, wang2023retccl, dippel2024rudolfv, shang2024histopathology}. Retrieving similar histology slides or pathology reports facilitates identifying relevant cases from large archival databases, and has become an essential clinical decision support function in digital pathology workflows. This is especially beneficial for rare cancers that affect fewer than 15 individuals per 100,000 annually\cite{rarecancers,rarecancers_SEER, gatta2017burden}, for which pathologists can identify non-specific malignancies based on WSIs with similar morphologies and their corresponding pathology reports. Slide foundation models readily provide WSI representations for vector database indexing, significantly simplifying the task of histology slide retrieval compared to patch foundation models, which provide more than 10$^4$ representations per WSI and consequently renders slide-level retrieval non-trivial. 

Given a query slide and labeled set of support slides (indexed into a vector database by a slide foundation model), histology slide search is evaluated by assessing the accuracy performance in retrieving similarly labeled slides from the support set. This setting is non-parametric and solely relies on how the slide representations are clustered along different diagnostic labels. Specifically, we test whether the $K$-closest neighbors of a query slide in the embedding space––determined using cosine similarity with $K=\{1,3,5\}$––include slides sharing the same diagnostic label as the query slide. Performance is assessed using Accuracy@$K$, which measures whether at least one of the $K$ neighboring slides has the same diagnostic label as the query. 
We also provide MVAcc@$K$ which requires the majority vote of top-$K$ neighboring slides is of the same diagnostic label as the query and is therefore more stringent criteria than Accuracy@$K$. 
For the rare cancer retrieval task, we create a large database of 186 cancer types with 19,626 WSIs, Rare-Cancer, by combining the \textit{rare cancer set} of 43 cancer types (3,039 WSIs) with the \textit{common cancer set} of 143 more common cancer types (16,587 WSIs) from TCGA, EBRAINS, and MGB internal data (\textbf{\Cref{fig:retrieval}A}, \textbf{Extended Data~\Cref{tab:dataset_rare-cancers}}).
To assess the performance, we create a query set as the subset of the \textit{rare cancer set}, ensuring all 43 rare cancer types are represented. The support set, from which similar slides are retrieved, is constructed by incorporating the remaining WSIs of the \textit{rare cancer set} into the \textit{common cancer set}, ensuring all 186 cancer types are represented. This design emulates the real-world setting of clinicians interacting with an extensive cancer database encompassing a diverse mix of rare and common cancer types. This procedure is repeated five times, with a different query set each time.
We additionally create a public version with 127 cancer types and 14,062 WSIs, Rare-Cancer-Public, using the data from TCGA and EBRAINS resulting in 29 rare cancer types (1,982 WSIs) and a lower diversity in the set of common cancers with 98 types (12,080 WSIs). The same evaluation procedure as for Rare-Cancer is repeated (\textbf{Extended Data~\Cref{tab:dataset_rare-cancer-public}}).

We observe that $\oursvcr$ and $\oursv$ outperform other slide encoders with +14.8\% and +12.3\% in Accuracy@$K$ and +18.1\% and +13.4\% in MVAcc@$K$ to the next best model PRISM (\textbf{Extended Data~\Cref{tab:slide_retrieval_rare}}) on average.
The trends in performance are preserved on the public version of the rare cancer task with slightly higher performance levels as the task is easier with a support set containing fewer cancer types (\textbf{Extended Data~\Cref{tab:slide_retrieval_rare_public}}).
An example of rare cancer retrieval is demonstrated in \textbf{Figure~\ref{fig:retrieval}B}, where the closest slide to the paraganglioma (PGNG) query is also of PGNG with a high similarity of 0.794 and less similar slides are of different cancer type (Haemangioma from brain, similarity of 0.341). One of the retrieved slides is Pheochromocytoma (PHC) with a high similarity of 0.651, agreeing with the clinical understanding that both are morphologically tightly connected as rare neuroendocrine tumors\cite{neumann2019pheochromocytoma}. Additional examples of rare cancer retrieval can be found in \textbf{Extended Data \Cref{fig:edf_rare}}, where the retrieved slides of high similarity are indeed from the same diagnostic label or organ.
Even when further assessed with multi-class cancer subtyping tasks, from relatively simple AMR for renal allograft ($C=2$) to challenging OT108 ($C=108$), we observe that both $\oursvcr$ and $\oursv$ outperform other slide encoders (\textbf{\Cref{fig:retrieval}C}, \textbf{Extended Data~\Cref{tab:slide_retrieval_TCGAUniformTumor_cancer,tab:slide_retrieval_TCGA_OncoTreeCode,tab:slide_retrieval_OP108_OncoTreeCode,tab:slide_retrieval_ebrains_diagnosis,tab:slide_retrieval_TCGA-BRCA_subtype,tab:slide_retrieval_TCGA-NSCLC_subtype,tab:slide_retrieval_MANTA_HE_label_amr}}).

Encouraged by the unimodal retrieval performance, we further investigate the cross-modal retrieval performance of $\oursvcr$, as the slide and report embedding spaces are aligned from the pretraining steps.
We perform the cross-modal experiments on TCGA-Slide-Reports, our proposed dataset for report generation with 10,108 slide-report pairs (\textbf{Extended Data~\Cref{tab:dataset_tcga-slide-reports}}).
For the report-to-slide (slide-to-report) retrieval task, we test whether any of the $K$-closest slides (reports) for the query report (slide) in the embedding space have the same diagnostic label as the query, the metric referred to as Recall@$K$ with $K=\{1,3,5,10\}$. observe that $\oursvcr$ outperforms PRISM on both retrieval tasks across all $K$ retrievals with +10.5\% and +20.5\% on average for report-to-slide and slide-to-report retrieval tasks, respectively (\textbf{\Cref{fig:retrieval}D}, \textbf{Extended Data~\Cref{tab:t2i_retrieval,tab:i2t_retrieval}}). The largest gap to PRISM is observed for slide-to-report retrieval when only a single report was retrieved, where $\oursvcr$ outperforms by 36.4\%. 
The strong performance of $\oursvcr$ even with a single report (0.75) hints at the clinical potential, where for a diagnostically challenging slide clinicians can benefit from sifting through retrieved past medical reports that describe identical diagnoses, and vice versa.
More details on the experiments can be found in \textbf{Online Methods} section \textbf{Slide retrieval} and \textbf{Cross-modal retrieval}.

\Heading{Discussion}

We introduce a multimodal whole-slide foundation model for pathology, $\ours$, that combines and elevates successful recipes of self-supervised learning (SSL) from the patch level to the slide level. Methodologically, $\ours$ employs histology knowledge distillation in the feature space (vision-only) and contrastive learning by aligning regions of interest (ROIs) with synthetic captions and whole slide images (WSIs) with reports (vision-language). 
Pretrained on 336K WSIs, $\ours$, a Vision Transformer (ViT) architecture equipped with ALiBi positional encoding for long-context extrapolation, produces powerful general-purpose slide representations for a large variety of downstream tasks even without task-specific finetuning.
From cancer subtyping to molecular classification, $\ours$ consistently outperforms other state-of-the-art slide encoders, such as PRISM\cite{shaikovski2024prism}, GigaPath\cite{xu2024whole}, and CHIEF\cite{wang2024pathology}. This superiority is maintained in data-constrained settings such as rare disease classification and histology slide retrieval, which underscores the representation quality of $\ours$. Further aligning the vision-pretrained $\ours$ with 423K ROI-level captions generated by PathChat and 183K pathology reports equips the model with multimodal capabilites such as zero-shot diagnosis, slide-report retrieval, and report generation. 
We observe that aligning the slide embedding with both the fine-grained (ROI captions) and coarse-level (pathology reports) morphological descriptions is crucial for handling the multiscale information inherent in tissue slides--an insight made possible for the first time through the use of generated ROI captions. 
Similar to the unimodal setting, $\ours$ outperforms PRISM, another language-equipped model, on all cross-modal tasks.
To advance the field of slide-representation learning\cite{wagner2022make}, we curated and plan to release two challenging multi-class morphology classification tasks beyond patch-level from the publicly available repository TCGA: TCGA-UniformTumor-8K (TCGA-UT-8K) for 32-class tumor-ROI subtyping and TCGA-OncoTree (TCGA-OT) for 46-class WSI-level OncoTree code classification.

Detailed ablation analyses reveal further insights into the properties of $\ours$.
We observe that Stage 1 unimodal pretraining of $\oursv$ captures morphological concepts already with much less data than existing slide encoders, as demonstrated in the few-shot data efficiency experiments. In particular, $\oursv$ consistently outperforms its mean pooling and task-specific attention-based pooling baselines that utilize the same patch encoder as $\oursv$, proving that unimodal pretraining effectively captures the context of patch features in contrast to existing unimodal slide encoders.
Next, in addition to unlocking language-related capabilities, we observe that the vision-language alignment further enhances the representation quality of our vision-only model. In particular, $\oursvl$ improves over $\oursv$ on average for slide-level tasks with the strongest improvements in evaluation settings that solely rely on the structure of the slide embedding space without any parameter tuning, such as prototyping or $k$-nearest neighbor settings. A further sign of the improved representations is that $\oursvl$ outperforms all other baselines, including $\oursv$, on slide retrieval and few-shot tasks.
While slide embeddings from pretrained $\ours$ are already promising, especially in the low-data regime, task-specific fine-tuning of the pretrained model can further enhance the downstream performance for tasks with a large enough patient cohort, pointing to the flexibility of $\ours$ when applied to diverse clinical and data settings.

Providing multimodal slide embedding off-the-shelf presents immediate clinical potential to assist clinicians in their routine diagnostic workflows\cite{moor2023foundation}. Presented with challenging patient tissue slides to diagnose, pathologists and oncologists can hugely benefit from being able to retrieve and analyze diagnostically similar slides or clinical reports, likely leading to a reduction in patient misdiagnosis and interobserver variability. As shown in the extensive set of experiments, $\ours$ can accurately retrieve similar diagnostic slides and reports for challenging scenarios from a large number of cancer types ($>$ 100), as well as rare cancer types\cite{rarecancers_SEER} where the corresponding slides have scarce representation in the database. That all of these could be performed off-the-shelf with pretrained $\ours$ without a dedicated algorithm for each task underscores both the generalizability of $\ours$ slide embeddings, as well as how slide-level tasks can become simpler with the advent of pretrained slide encoders. 

Despite the encouraging performance of $\ours$, our framework has a few shortcomings. First, Mass-340K contains fewer slides compared to other pretraining datasets used for patch encoders\cite{hoptimus0,nechaev2024hibou,zimmermann2024virchow} and slide encoders such as PRISM\cite{shaikovski2024prism}. We believe that the already strong performance of $\ours$, merged with concurrently ongoing effort to expand Mass-340K, will further allow improved slide-level and cross-modal performance. Next, pretraining on the region crops of 8K$\times$8K and extrapolating with ALiBi to the entire WSI may still not capture the full contextual information. Larger pretraining contexts such as 16K$\times$16K, and other positional encodings for extrapolation could address this limitation. Finally, preprocessing of clinical reports presents a further challenge for vision-language alignment; Striking a balance between relevant information for contrastive learning while only including morphology-related information 
is non-trivial and involves a lot of manual tuning despite the automated processing pipelines.
Restructuring the reports into distinctive morphology and molecular characteristics could facilitate contrasting with only relevant information.

Promisingly, $\ours$ can be scaled up in terms of data and architecture to improve performance. WSIs and corresponding medical reports are routinely available and stored in the clinical workflow. The synthetic region-level captions can be generated with the generative AI model in an unlimited manner, providing the model with a wealth of text guidance. Using this additional data, a heavier ViT slide encoder architecture than what is currently being utilized for $\ours$ can potentially improve the performance, as was already demonstrated at the level of patch encoders\cite{zimmermann2024virchow, hoptimus0, nechaev2024hibou}. In addition, the improved patch representation quality from more powerful patch encoders will likely improve the quality of the downstream slide encoder.
We envision $\ours$ and its future iterations being incorporated into practitioners' everyday toolkits for routine application and comparison with other task-specific supervised frameworks, together reaching new levels of performance in clinically important tasks.

\Heading{Online Methods}

\heading{Pretraining dataset}

For large-scale visual pretraining, we curated Mass-340K, a diverse dataset consisting of 335,645 WSIs across 20 organs, with $90\%$ hematoxylin and eosin (H\&E) stained slides and $10\%$ immunohistochemistry (IHC) slides, sourced from the combination of in-house histology slides and the GTEx consortium\cite{gtex2015genotype}. To explore the effects of data scale at the pretraining stage, we formed three additional partitions of Mass-340K, containing 12.5\%, 25\%, and 50\% of the original dataset. These partitions were sampled to maintain the ratio of different data sources and preserve organ distribution. 

\hheading{Synthetic caption generation using PathChat}

For the initial stage of vision-language alignment (Stage 2 of $\ours$), we used synthetic captions generated by PathChat, a state-of-the-art multimodal LLM designed for pathology\cite{lu2024multimodal}. To go beyond the typically brief clinical reports focused on the final diagnosis, we prompted PathChat to generate detailed morphological descriptions of ROIs, providing important training data for models to capture complex pathological features. Using PathChat, we generated synthetic captions for 423,122 diverse 8,192$\times$8,192 ROIs sampled from Mass-340K. 
Since PathChat cannot process inputs of size 8,192$\times$8,192 pixels directly, we divide each ROI into 64 1,024$\times$1,024 patches. To retain the most representative morphological features, we applied K-means clustering with $K=16$ to the 64 patches and then randomly sampled one patch from each cluster. The resulting 16 morphologically-representative 1,024$\times$1,024 patches were subsequently fed to PathChat. To further enhance the diversity of these captions, we utilized Qwen2-7B-Instruct\cite{yang2024qwen2} to rewrite the generated captions, ensuring varied language structures and expressions. Detailed prompts for both PathChat and Qwen2, along with examples of generated and diversified captions, are provided in\textbf{ Extended Table~\ref{tab:pathchat_prompt}-\ref{tab:caption_rewrite}}.

\hheading{Curation of slide-report dataset}

For the second stage of vision-language alignment (Stage 3 of $\ours$), we curated a dataset of 182,862 slide-report pairs from a combination of in-house clinical reports and pathology notes from the GTEx consortium\cite{gtex2015genotype}. However, clinical reports are often noisy and are typically organized at the patient level, hence contain information on multiple slides from the same patient, complicating the slide-report alignment. To address this, we utilized a locally served Qwen2-7B-Instruct\cite{yang2024qwen2} model to extract slide-specific descriptions and remove sensitive information unrelated to pathological diagnosis, such as gross descriptions, hospital and doctor names, and patient clinical history. Additionally, we applied the same rewriting strategy used for synthetic captions to diversify the report text. Example prompts used for report cleaning and rewriting can be found in \textbf{Extended Data Table~\ref{tab:report_parsing}-\ref{tab:report_rewrite}}.

\heading{Unimodal visual pretraining}

\hheading{Preprocessing}

Similar to the previous studies\cite{lu2021data, lu2024visual, chen2024towards}, WSIs were preprocessed by tissue segmentation, tiling, and feature extraction using a pretrained patch encoder. We used the CLAM toolbox\cite{lu2021data} for tissue segmentation and tiling. Tissues were segmented by binary thresholding of the saturation channel in HSV color space at a low resolution. Following this, we applied median blurring, morphological closing, and filtering of contours below a minimum area to smooth tissue contours and eliminate artifacts. Non-overlapping 512$\times$512 pixel patches were then extracted from the segmented tissue regions of each WSI at 20$\times$ magnification. For feature extraction, we used CONCHv1.5, an extended version of CONCH\cite{lu2024visual}, which was trained with 1.26 million image-caption pairs using the CoCa training objective for 20 epochs. 
The choice of CONCHv1.5 for feature extraction was due to the fact the model was pretrained on histology regions with diverse stains and tissue types, including FFPE, frozen tissue, and immunohistochemistry, thereby yielding region features that are robust against diverse tissue processing protocols.
Refer to \textbf{Extended Data Table \ref{tab:hparams_coca_base}} for detailed hyperparameters of the patch encoder.

To enhance the effectiveness of the ROI sampling strategy during Stage 1 training of $\oursv$, an additional preprocessing step was performed to group the segmented tissue contours based on their spatial proximity within the slide. This addresses the challenging cases where multiple tissue regions are interspersed with background areas, particularly for biopsy samples where tissue fragments are often widely dispersed and for samples with multiple slices placed on the same slide. Specifically, we grouped tissue contours into clusters based on their coordinates, resulting in tissue groups that contain densely packed tissue regions with minimal background regions between them. Furthermore, tissue groups that contained fewer than 16 patches were filtered out. This grouping operation produced a total of 345,782 tissue groups from Mass-340K.

\hheading{Pretraining protocol}

For training $\oursv$ on Mass-340K, we use iBOT, a state-of-the-art self-supervised learning method based on the combination of student-teacher knowledge distillation and masked image modeling\cite{zhou2021ibot}. As iBOT is applied in the patch embedding space, instead of the typical use case of the raw image space, we adapt the pretraining recipes as follows.

\noindent\textbf{View generation.} During training, we create region crops randomly sampled from the tissue groups, each of which corresponds to a feature grid of size 16$\times$16, corresponding to a field of view of 8,192$\times$8,192 at 20$\times$ magnification (\textbf{Figure~\ref{fig:slidessl}B}). 
The random sampling of region crops, instead of precomputing fixed regions, increases the diversity of the training set and effectively acts as an additional data augmentation, as the model encounters different parts of the same WSI at each training epoch.
A region crop contains 256 features, which is equivalent in length to training on images of 256$\times$256 pixels with a token size of 16$\times$16 in the typical natural image setting. From this region crop, two global views (14$\times$14 crops) and ten local views (6$\times$6 crops) are generated by cropping within the region crop without scaling or interpolation and fed to iBOT training.

To achieve realistic augmentations in the embedding space, previous methods have employed offline image augmentations in the pixel space\cite{lazard2023giga, wagner2023transformer} by extracting multiple patch features from different views of a given patch. While effective, this approach limits the number of additional views and becomes computationally infeasible for large training datasets. Additionally, choosing color space augmentations adapted to the use of histopathology that go beyond standard color transformations introduces additional computational overhead. A few recent approaches addressed the difficulty with training generative networks on the feature space to transform the features\cite{zaffar2023embeddingaugmentation, shao2023augdiff}, but also introduced additional computational cost for training. Instead, we apply frozen feature augmentations, which have been shown to work well for a few-shot classification task in the feature space of pretrained Vision Transformers\cite{bar2024frozen}. 

\noindent\textbf{Positional encoding.} Traditional multiple instance learning methods consider the patches to be permutation-invariant within the slide. Despite the promising results, this approach ignores the tissue context which can be essential for capturing the interaction in the tumor micro-environments and can thus affect the model's performance\cite{jaume2022integrating}. In this context, for $\ours$, we employ positional encodings in the patch embedding space to break permutation invariance and encode tissue context. Furthermore, $\ours$ adopts the strategy of \textit{Train short, test long} to ease the computational burden, which also requires positional information via positional encodings. Trained at the region crops (ROIs) of 8,192$\times$8,192 pixels (\textit{Train short}), we directly apply $\ours$ on the whole slide during inference (\textit{Test long}). We used Attention with Linear Biases (ALiBi), a method originally proposed for 1D sequence in large language models (LLMs)\cite{alibi}. Absolute positional encoding, another popular alternative that works well for images at training sizes, was shown to have weak extrapolation abilities \cite{alibi}. Unlike other positional encodings applied to the input features, ALiBi adds a bias to the query-key dot product during the computation of attention scores. ALiBi effectively penalizes the attention score for tokens which are further apart from each other. Formally, let $q_i \in \mathbb{R}^{d}$ and $k_j \in \mathbb{R}^{d}$ represent the $i$-th query and $j$-th key, respectively. The attention score, which is typically computed as $\operatorname{softmax}\left( q_i k_j^{\text{T}}\right)$, is modified with 1D ALiBi as $\operatorname{softmax}\left( q_i k_j^T - m|i-j|\right)$, where $m$ is a predefined slope specific to each attention head.
Since the feature grids and the resulting views are of 2D grid structure, we extend ALiBi to 2D by incorporating the Euclidean distance between patches $i$ and $j$. The 2D ALiBi can be written as
\begin{equation} \operatorname{softmax} \left( q_i k_j^{\text{T}} - m \sqrt{(i_x - j_x)^2 + (i_y - j_y)^2} \right), \end{equation}
where $i_x,\ i_y$ and $j_x,\ j_y$ are the 2D grid coordinates of patches $i$ and $j$. The $x$ and $y$ coordinates are defined as the 2D patch coordinates (at magnification 20$\times$) divided by the patch size of 512.

\noindent\textbf{Network architecture and training details.} For the slide encoder, we use a Vision Transformer (ViT)\cite{dosovitskiy2020image} with 6 Transformer layers, 12 attention heads of dimension 64 resulting in an embedding dimension of 768, and a hidden dimension of 3,072. This smaller architecture, compared to typical ViTs used in patch encoders, ic chosen base on previous studies\cite{chen2022scaling}, which suggest that a compact network suffices for slide representation learning on the embedding space, especially given the limited data scale of WSIs compared to histology patch datasets at the scale of billions. 
The patch embedding layer is replaced by an MLP to process the feature inputs.
We train the model for 270 epochs (equivalent to 91,260 iterations), distributed across four NVIDIA A100 80GB graphics processing units (GPUs) with a local batch size of 256 per GPU. For all training hyperparameters, refer to \textbf{Extended Data Table \ref{tab:hparams_ibot}.}

\heading{Vision-language continual pretraining}

To enhance the unimodal capabilities of $\oursv$, we further explored the multimodal vision-language alignment of $\oursv$ with clinical text. Training a multimodal foundation model, however, faces several limitations related to data and compute. First, paired slide-report data are scarce compared to the scale of millions of image-caption pairs for patches. Additionally, real-world clinical reports typically only contain brief diagnostic information, unlike the detailed morphological descriptions in educational captions for histology regions of interest (ROI) images. Finally, contrastive learning-based cross-modal training typically requires a large batch size, which is computationally infeasible for WSIs. 

To address these issues, we propose a two-stage continual pretraining approach (referred to \textbf{Stage 2} and \textbf{Stage 3} for $\ours$) that progressively aligns the model with increasing context. We first align synthetic captions for 8,192$\times$8,192 ROIs, followed by real clinical reports for WSIs. With emphasis on detailed morphological descriptions, the first vision-language alignment stage allows the model to learn fine-grained pathological concepts using a large batch size. In the next stage, we further augment the model's understanding of diagnostic terminology and reasoning, targeted to enhance its zero-shot understanding in downstream tasks. The second stage also serves as a “high-resolution fine-tuning” phase, adapting the model from the local contexts of ROIs  to the full-scale global context of WSIs. Altogether, these two stages are designed to gradually build the model’s ability to comprehend and generate meaningful vision-language representations for WSIs. 

\hheading{Network architecture and training details}

Following the success of previous studies\cite{lu2024visual}, we use CoCa\cite{yu2022coca}, a state-of-the-art visual-language foundation model pretraining method, for both stages of vision-language alignment. The model consists of an image encoder, a text encoder, and a multimodal text decoder. Using our unimodal $\oursv$ as the image backbone, we add two attentional pooler components on top. The first attentional pooler uses a single query (contrastive query) to pool a single global representation of the feature grids and enable cross-modal contrastive learning with text embeddings. This global WSI representation can then be used for zero-shot or unsupervised evaluation of $\oursvcr$ on downstream tasks. The second attentional pooler uses $n=128$ queries (reconstruction queries) to generate a set of 128 image tokens designed for interacting with the multimodal text decoder for caption generation. We use the pretrained text encoders and multimodal decoders of CONCHv1.5\cite{lu2024visual}, each consisting of 12 Transformer layers with an embedding dimension of 768 and a hidden dimension of 3,072.

For both stages, we used 8 NVIDIA A100 80GB GPUs. During Stage 2 vision-caption pretraining, we used a local batch size of 196 per GPU, with gradient accumulation of 2 resulting in an effective batch size of 3,136. For Stage 3 vision-report pretraining, we randomly crop the WSIs to 64$\times$64 feature grids to allow for larger batch sizes while maintaining a large field-of-views, corresponding to 32,768$\times$32,768 pixels, wich already covers most slides in our pretraining dataset. We used a local batch size of 16 per GPU, with a gradient accumulation of 2 to achieve an effective batch size of 256. To avoid deteriorating the quality of the pretrained vision encoder, we used a smaller learning rate and weight decay, as well as a slow warm-up strategy for the vision backbone, following previous work\cite{beyer2024paligemma}. For all hyperparameters, refer to \textbf{Extended Data Table \ref{tab:hparams_coca_caption}-\ref{tab:hparams_coca_report}.}

\heading{Evaluation setting}

\hheading{Baselines}

We compare $\oursv$ against 1) \textit{unsupervised} baselines with four other slide encoders, Prov-GigaPath (referred to as GigaPath in the manuscript)\cite{xu2024whole}, PRISM\cite{shaikovski2024prism}, CHIEF\cite{wang2024pathology}, and the mean pooling baselines with features from the respective patch encoders,  2) \textit{supervised} baselines, and 3) our vision-language model $\oursvcr$ against zero-shot baseline PRISM.

\noindent \textbf{Unsupervised baselines.} GigaPath uses LongNet architecture as the slide encoder, a ViT\cite{dosovitskiy2020image} in the ``base configuration", replacing the vanilla dense attention with dilated attention. It was trained on 171,189 in-house WSIs from Providence via masked autoencoder\cite{he2022masked}. As patch encoder, GigaPath uses ViT-G/14 pretrained with DINOv2\cite{oquab2023dinov2} on the same in-house dataset. While GigaPath further performed continual vision-language pretraining, we only assess the unimodal model, as the multimodal model is not publicly available.
For performance analysis, we use the output of the Transformer layer 11 as slide representation, which yields the best results on downstream tasks and also agrees with the provided finetuning recipe. 
PRISM\cite{shaikovski2024prism} employs the Perceiver architecture\cite{jaegle2021perceiver} as the slide encoder with CoCa-based vision and language alignment\cite{yu2022coca} on 195,344 specimen-report pairs, where each specimen contains one or more WSIs with a total of 587,196 WSIs.
As for the patch encoder, PRISM uses Virchow~\cite{vorontsov2024foundation}, a ViT-H/14 pretrained with DINOv2\cite{oquab2023dinov2} on an in-house dataset. 
CHIEF\cite{wang2024pathology} applies attention-based feature aggregation, trained via slide-level contrastive learning and anatomic site information. The patch encoder is based on CTransPath\cite{wang2022transformer}, a self-supervised SwinTransformer\cite{liu2021swin} trained on 15 million patches. 
In addition to the pretrained slide encoders, we evaluate mean pooling as baseline, where the patch features are averaged within each slide, as it serves as a strong unsupervised baseline despite its simplicity\cite{jaume2024transcriptomics, jaume2024multistain, song2024morphological}. While we mainly compare with mean pooling based on CONCHv1.5 patch features, we also provide results for mean pooling with the corresponding patch encoders of each slide encoder for a subset of analyses.

\noindent \textbf{Supervised baselines.} We compare $\ours$ against attention-based MIL (ABMIL)\cite{ilse2018attention, lu2021data} and finetuning of the pretrained slide encoders. For ABMIL, the model was trained with a batch size of 1 using the AdamW optimizer with weight decay $10^{-5}$ and a Cosine annealing learning rate scheduler with peak learning rate $10^{-4}$ over 20 epochs. The patch encoders were selected accordingly for each analysis. For GigaPath finetuning, we used the publicly available code, which uses a batch size of 1, AdamW optimizer with weight decay 0.05, and Cosine annealing learning rate scheduler with warm-up and base learning rate $2\cdot10^{-3}$ over 5 epochs. For CHIEF finetuning, we also used the publicly available finetuning code. For tasks with a validation set, the best model is chosen based on the validation loss.

\noindent \textbf{Cross-modal baselines.}
For cross-modal zero-shot retrieval and clinical report generation, we compare $\oursvcr$ against PRISM~\cite{shaikovski2024prism}.

\hheading{Linear and K-nearest neighbors probe evaluation}

To evaluate the transfer capabilities and representation quality of slide encoders, we adopt recent work in representation learning with self-supervised frameworks and perform linear (logistic regression) and \textit{k}-nearest neighbor (\textit{k}-NN) probing. For linear probing, we minimize cross-entropy loss using the scikit-learn L-BFGS solver with $\ell_2$-regularization, selecting $\ell_2$ from 45 logarithmically spaced values between $10^{-6}$ and $10^5$ based on the validation loss. The maximum number of L-BFGS iterations is set to 500. For datasets without validation set, e.g., in small datasets or in few-shot experiments, we choose default values of $\ell_2=1$ with 1,000 iterations. 
We additionally evaluated with \textit{k}-NN probing, a non-parametrized measure to quantify the representation quality of fixed embeddings. We apply it in two settings: First, we follow SimpleShot to create a prototypical class representation by averaging all slide embeddings per diagnostic class\cite{wang2019simpleshot}. Second, we use the scikit-learn implementation of \textit{k}-NN with $\textit{k}=20$ following stability observations from self-supervised learning literature\cite{dino, oquab2023dinov2}. In both settings, Euclidean distance is used as the distance metric based on the centered and normalized slide embeddings. 

\hheading{Slide retrieval}

To further evaluate the representation quality of different slide encoders, we perform content-based slide retrieval using slide-level classification datasets, where we retrieve slides with the same class label as a given query slide. Specifically, we extract slide features for all WSIs. The training and validation sets are combined to serve as the database of candidate slides (keys), and we treat each slide in the test set as a query slide. Prior to retrieval, we preprocess both keys and queries to center the slide embeddings by subtracting their Euclidean centroid, followed by $\ell_2$ normalization. The similarity between the query and each candidate in the database is computed using the $\ell_2$ distance metric, where a smaller distance indicates a higher similarity. The retrieved slides are then sorted based on their similarities to the query. The class labels are used to evaluate the retrieval performance using Acc@$K$ for $K\in \{1,3,5\}$, which measures whether at least one of the top K retrieved slides shared the same class label as the query, and MVAcc@5, which considers the majority class label among the top 5 retrieved slides. Detailed descriptions of these metrics are provided in the section \textbf{Evaluation Metrics}.

\hheading{Cross-modal retrieval}

Leveraging the vision-language aligned embedding space, we also evaluate cross-modal retrieval performance on TCGA-Slide-Reports. Specifically, we assess both slide-to-report and report-to-slide retrieval tasks. All slides and reports are embedded into a shared space using the vision and text encoders, respectively, followed by $\ell_2$ normalization. Retrieval is performed by calculating pairwise cosine similarity between the slide and report embeddings. 
Our class-based approach mirrors the uni-modal slide retrieval, where retrieval is successful if the retrieved slide or report belongs to the same diagnostic class as the query. 
Performance is quantified using Recall@K for $K \in \{1, 3, 5, 10\}$ for the class-based approach, which measures the proportion of queries for which the correct result appears among the top-K retrieved items.
Additionally, we report the mean recall, computed as the average of the Recall@K values across the four $K$ levels. Further details on these metrics are provided in Section \textbf{Evaluation Metrics}.

\hheading{Few-shot slide classification}

We evaluate few-shot classification by varying the number of shots $k$ in $\{1, 2, 4, 8, 16,32\}$. For each $k$, we select $k$ shots per class or all samples per class if the class has less than $k$ samples. We follow  previous studies that used the SimpleShot~\cite{wang2019simpleshot} framework for evaluation of the few-shot learning performance of self-supervised models\cite{chen2024towards}. SimpleShot computes a prototypical representation per class by averaging all samples within that class. The distances to the class prototypes are then computed on the test set. All embeddings are centered and normalized based on the few-shot samples. To make the evaluation better comparable to supervised baselines such as ABMIL, we additionally assess few-shot classification with linear probing. As no validation set is available in few-shot experiments,  we use the default scikit-learn recipe with regularization strength $\ell_2=1$ and up to 1,000 iterations of the L-BFGS solver. 

\hheading{Survival analysis}

For survival analysis, we employed the linear Cox proportional hazards model on the disease-specific survival (DSS) clinical endpoint. 
We note that this is different from typical MIL survival prediction with negative log likelihood\cite{zadeh2020bias, song2024morphological}, as we deal with a single embedding for the slide (as opposed to a bag of patch embeddings) and patients can be batched (as opposed to the single patient per batch due to memory usage).
To reduce the impact of batch effects, we performed a five-fold site-preserved stratification\cite{howard2021impact}. Due to the small cohort size for reliable survival prediction modeling, we used four folds for training and the remaining fold for evaluation, without employing the validation fold. A hyperparameter $\alpha$ was searched over 25 logarithmically spaced values between $10^{1}$ and $10^5$, with the $\ell_2$ coefficient defined as $C=\alpha$. For each combination of encoder and cancer type, we chose $C$ that yielded the best average test metric across the five folds. For fitting and testing the Cox model, we used the scikit-surv package.

\hheading{Zero-shot slide classification}

For zero-shot slide classification, we adopted the method described in CLIP\cite{radford2021learning} to use the similarities between a given slide and the text prompts of each class as its prediction logits. Specifically, for a class $c\in\{1,2,\dots,C\}$, we first created the text prompts for each class, followed by extracting their $\ell_2$-normalized text embeddings $\mathbf{v}_c$ using the text encoder. Since the model could be sensitive to the specific choice of text prompts, we created an ensemble of prompts for each class. The complete set of prompt ensembles are provided in \textbf{Extended Data Table \ref{tab:zero_shot_prompts}}. For each WSI, we similarly computed a $\ell_2$-normalized embedding $\mathbf{u}_i$ using the slide encoder. We then calculated the cosine similarity between the slide embedding and each class text embedding. The predicted class for a slide was the one with the highest cosine similarity score:
\begin{equation}
    {\hat y}_i = \underset{c}{\text{argmax}}\;{\mathbf{u}_i}^T\mathbf{v}_c
\end{equation}

\hheading{Report generation}

Slide captioning provides concise and interpretable summaries of visual findings in pathology, potentially enhancing clinical workflows. The generative objective of CoCa enabled the model's capabilities of generating pathological reports, which we explored on 10,108 slide-report pairs from TCGA. We performed zero-shot captioning using $\oursvcr$ and compared the quality of the generated report against PRISM \cite{shaikovski2024prism}. Specifically, we use a beam search decoding strategy with 5 beams and 1 beam group, where the model explores five potential sequences at each step and retains only the most likely sequence within a single group to maximize quality while minimizing redundancy. 

\hheading{Evaluation metrics}

We report balanced accuracy and weighted F1-score for all classification tasks with more than two classes. For ordinal multiclass classification tasks, we report balanced accuracy and quadratic weighted Cohen's $\kappa$. For binary classification tasks, we report balanced accuracy and area-under-the-receiver-operator-curve (AUROC). For survival tasks, we report the concordance index (c-index), which measures the agreement between the model's predicted risks and the actual survival times. For slide retrieval tasks, we report Acc@$K$ for $K\in{1,3,5}$, which measures if at least one slide among the top $K$ retrieved slides has the same class label as the query. We also report MVAcc@5, which is a more strict metric that considers whether the majority vote of the top 5 retrieved slides is in the same class as the query. For cross-modal retrieval tasks, we report Recall@$K$ for $K\in{1,3,5,10}$, which measures the proportion of queries for which the correct result appears in the top-$K$ retrieved items. We also report mean recall, which is calculated as the average of the four Recall@$K$ values. For report generation, we compare the generated reports with the ground truth pathological reports using METEOR, ROUGE, and BLEU. METEOR\cite{banerjee2005meteor} is a metric that evaluates text quality through unigram matching by considering both precision and recall while also accounting for synonyms, stemming, and word order between the candidate and reference texts. ROUGE\cite{lin2004rouge} compares the overlap of n-grams, word sequences, and word pairs between the generated and reference texts, focusing on recall. We use, ROUGE-1, which specifically measures the overlap of unigrams. BLEU\cite{papineni2002bleu} measures the quality of generated text based on unigram overlap, focusing on precision. We use BLEU-1, which evaluates the extent of word-level matches between the generated and reference texts.

\hheading{Statistical analysis}

For the datasets with five-fold splits, where we employ 5-fold cross-validation, we report the mean performance and the standard deviations across all folds. For the datasets with a single split, we use non-parametric bootstrapping with 1,000 samples to calculate the mean and standard deviation. 

\heading{Downstream evaluation datasets}

For the evaluation of $\ours$ on a diverse set of downstream tasks (\textbf{Extended Data~\Cref{tab:tasks_morphological,tab:task_grading,tab:task_molecular}}, we re-arrange the pre-extracted CONCHv1.5 features from  512$\times$512 patches to feature grids cropped around the tissue regions of the WSIs. Additionally, background masks are created to mask out features corresponding to background patches. Each WSI is then one single input image to $\ours$. For downstream tasks with patient-level annotations, we create the patient embeddings by averaging all slide embeddings of $\ours$ corresponding to one patient. In the following, we detail all datasets used in our downstream evaluations including splits and targets. We first describe the five datasets that we introduce in our study, TCGA-UniformTumor-8K, TCGA-OncoTree, TCGA-Slide-Reports, Rare-Cancer, and Rare-Cancer-Public, followed by existing datasets in alphabetical order.

\noindent\textbf{TCGA-UniformTumor-8K (TCGA-UT-8K)} is a pan-cancer subtyping dataset at region-level consisting of 25,495 8,192$\times$8,192 pixel cancer regions of 9,662 H\&E FFPE diagnostic histopathology WSIs from TCGA. The tumor regions were manually annotated by two expert pathologist, with slide exclusion due to poor staining, poor focus, lacking cancerous regions and incorrect cancer types. Approximately three representative tumor regions per WSI were annotated with pixel-level contours. For each contour, we center cropped a 8,192$\times$8,192 image region in order to contain both the dense tumor and surrounding tissue context. We split the regions into train-val-test split (13,853:3,434:8,208 slides) preserving the source site. Refer to \textbf{Extended Data Table \ref{tab:dataset_tcga-ut}} for a detailed overview of all classes contained in this dataset.

\noindent\textbf{TCGA-OncoTree (TCGA-OT)} is a pan-cancer subtyping dataset of 11,186 H\&E FFPE diagnostic histopathology WSIs from TCGA\cite{weinstein2013tcga}. All WSIs are classified into 46 classes according to the OncoTree classification system such that every class is represented by at least 50 samples. We select all diagnostic H\&E FFPE WSIs from TCGA with primary tumors. Concretely, we exclude frozen tissue slides, slides without magnification information, metastatic or recurrent tumor slides, slides without tumor tissue, and IHC slides. For training and evaluation, we split the dataset into training-validation-test folds of 8,226:1,612:1,348 samples while preserving the source site, i.e., all slides from one source site are in one split. Refer to \textbf{Extended Data~\Cref{tab:dataset_tcga-ot}} for a detailed overview of all classes contained in this dataset.

\noindent\textbf{TCGA-Slide-Reports} is a pan-cancer slide-report dataset of H\&E FFPE diagnostic histopathology WSIs from TCGA~\cite{weinstein2013tcga}. The dataset consists of 10,108 WSIs with paired pathological reports at slide-level. The dataset is built on the TCGA-Reports dataset, which consists of 9,523 patient-level reports released by a previous study~\cite{kefeli2024tcga}. The dataset TCGA-Reports was created using 11,108 pathology report PDFs, corresponding to 11,010 patients, available on the TCGA data portal. The raw reports were preprocessed by removing 82 patients with multiple reports, 399 patients with non-primary tumors, 72 patients with no survival data, 381 “Missing Pathology” reports, and 212 “TCGA Pathologic Diagnosis Discrepancy Form” reports, resulting in 9,850 reports. Optical character recognition (OCR) was then performed for text extraction from the PDFs, followed by the removal of “Consolidated Diagnostic Pathology Form” reports, “Synoptic Translated” forms, within-report TCGA metadata insertions, and clinically irrelevant reports, resulting in 9,523 patient-level reports. While these reports are clean and clinically relevant, they often contain descriptions of multiple tissue blocks per patient. This lack of one-to-one mapping between slides and reports poses a challenge for slide-level report generation and cross-modal retrieval, which require distinct slide-to-report alignment. Since block IDs are unavailable in TCGA metadata, we used the slide-level diagnoses to map diagnoses in each tissue block description. Specifically, if a block’s diagnosis matched the slide-level diagnosis, we designated it as corresponding to the slide. This process was automated with GPT4o-mini, producing a final set of 10,108 slide-report pairs. These paired slides are all H\&E FFPE WSIs from primary tumors adhering to the same exclusion criteria as mentioned for TCGA-OT. We excluded all frozen tissue slides, slides without magnification information, metastatic or recurrent tumor slides, slides without tumor tissue, and IHC slides. 
Refer to \textbf{Extended Data~\Cref{tab:dataset_tcga-slide-reports}} for a detailed overview of the diagnosis distribution of this dataset.

\noindent\textbf{Rare-Cancer-Public} is a pan-cancer dataset of H\&E FFPE diagnostic WSIs from TCGA\cite{weinstein2013tcga}. The dataset consists of 1,982 WSIs, with 1,548 WSIs from TCGA and 434 WSIs from EBRAINS, representing 28 rare cancer types. According to the National Institute of Health, rare cancers are defined as those occurring in fewer than 15 individuals per 100,000 annually\cite{rarecancers}. The OncoTree codes of WSIs from TCGA and EBRAINS were manually curated for this criterion by two expert pathologists (A.K., D.F.K.W.). EBRAINS provides more granular diagnostic classifications than the OncoTree codes, enabling the dataset to include finer distinctions for rare brain tumors. 
We split the dataset into five folds on the patient level. The dataset was divided into five patient-level folds. To assess the retrieval performance for rare cancers within a clinically representative dataset, we use one fold of the rare cancer dataset as the query set and the remaining folds combined with the common cancer types as a support set. In total, the support and query datasets contain 14,062 slides, including 11,646 WSIs from TCGA and 2,416 from EBRAINS.

\noindent\textbf{Rare-Cancer} is an in-house extension of the public dataset Rare-Cancer-Public with MGB internal cases. This dataset comprises 43 rare cancer types and 3,039 H\&E FFPE diagnostic histopathology WSIs, where 1,056 additional cases were added from Brigham and Women's Hospital (BWH). The entire dataset including common cancer types contains 19,626 WSIs with 5,564 WSIs from BWH from 186 OncoTree codes. 

\noindent\textbf{BCNB} consist of 1,058 H\&E FFPE WSIs of early breast cancer core-needle biopsies\cite{bcnb_dataset}. All cases are annotated with ER 
 (WT: 227, MUT: 831), PR (WT: 268, MUT: 790), HER2 (WT: 781, MUT: 277) expressions. We split the dataset label-stratified by a ratio of  60:20:20 (676:170:212 slides).

\noindent\textbf{BRACS} consists of 547 H\&E FFPE WSIs of benign (including normal), atypical, and malignant breast tumors from 189 patients\cite{brancati2022bracs}. The cases are annotated in coarse and fine-grained subtypes of three classes (benign tumors: 265, atypical tumors: 89, malignant tumors:	193) and six classes (atypical ductal hyperplasia: 48, ductal carcinoma in situ:	61, flat epithelial atypia:	41, invasive carcinoma: 132, normal: 44, pathological benign: 147, usual ductal hyperplasia: 74). We split the dataset label-stratified on patient level into five splits by a ratio of  60:20:20 (approx. 302:94:151 slides).

\noindent\textbf{Cardiac allograft rejection} consists of 5,021 H\&E FFPE WSIs of 1,688 patient biopsies collected from BWH\cite{lipkova2022crane}. Each biopsy is labeled for the presence of cardiac rejection characterized by acute cellular rejection (no rejection: 866 patients, rejection: 822 patients) . We split the dataset  label-stratified on patient level into train, val, and test splits by ratio 70:10:20 (3547:484:990 slides).

\noindent\textbf{DHMC LUAD} consists of 143 H\&E FFPE WSIs of lung adenocarcinoma from the Department of Pathology and Laboratory Medicine at Dartmouth-Hitchcock Medical Center (DHMC)\cite{dhmc_luad}. All WSIs are labeled into five classes of the predominant patterns of lung adenocarcinoma (acinar: 59, lepidic: 19, micropapillary: 9, papillary: 5, solid: 51). Given the limited size of the dataset, we use it exclusively for evaluation in a zero-shot setting, where we use the entire dataset as test set.

\noindent\textbf{DHMC RCC} consists of 563 H\&E FFPE WSIs of renal cell carcinoma (RCC) from DHMC\cite{dhmc_rcc}. All slides are labeled into the four predominant patterns of RCC including one benign class (renal oncocytoma, chromophobe RCC, clear cell RCC, papillary RCC). We use the three RCC subtypes as external test set for the three class subtyping task TCGA RCC. 

\noindent\textbf{EBRAINS} consists of 2,319 H\&E FFPE diagnostic histopathology WSIs from the EBRAINS Digital Tumor Atlas sourced from the University of Vienna \cite{ebrains}. Due to small sample size we exclude two classes and predict a fine-grained 30 class brain tumor subtyping task. All brain tumors in these tasks are designated as rare cancers by the RARECARE project and the NCI-SEER program. For training and evaluation, we approximately label-stratified the dataset into a train–validation–test fold with a 50:25:25 ratio (1,151:595:573 slides). Additionally, we use 873 samples with annotations for isocitrate dehydrogenase (IDH) mutation as external test set for IDH mutation prediction on the TCGA-GBMLGG cohort. 

\noindent\textbf{IMP-CRC} consists of 5,333 H\&E FFPE colorectal biopsy and polypectomy WSIs retrieved from the data archive of IMP Diagnostics laboratory, Portugal\cite{imp_biopsies1,imp_biopsies2,imp_biopsies3}. All cases are classified within one of three categories: Non-neoplastic (847 slides), low-grade lesions (2847 slides), i.e., conventional adenomas with low-grade dysplasia, and high-grade lesions (1639 slides), i.e., conventional adenomas with high-grade dysplasia, intra-mucosal carcinomas, and invasive adenocarcinomas. We split the dataset label-stratified by a ratio of 60:20:20 into train:val:test set (3546:887:900 slides).

\noindent\textbf{MGB-BRCA} consists of 1,264 H\&E FFPE WSIs of biopsies and resections invasive breast cancers (BRCA) from BWH\cite{vaidya2024demographic,jaume2024multistain}. Each case is annotated with three IHC status prediction tasks: estrogen receptor (ER) status prediction (0: 261, 1: 613), progesterone receptor (PR) status prediction (0: 37, 1: 504), and human epidermal growth factor receptor 2 (HER2) status prediction (0: 665, 1: 151), where ER, PR, and HER2 status were manually extracted from pathology reports.

\noindent\textbf{MGB-LUAD} consists of 1,939 H\&E FFPE WSIs of lung adenocarcinoma from BWH\cite{vaidya2024demographic,jaume2024multistain}. The WSIs are annotated by five molecular tasks with ground truth from IHC: protein 40 (P40) status prediction (0: 113, 1: 72), protein 63 (P63) status prediction (0: 72, 1: 81), Napsin A status prediction (0: 60, 1: 66), caudal type homeobox 2 (CDX2) status prediction (0: 55, 1: 24), and cytokeratin 5 and 6 (CK-5\&6) status prediction (0: 29, 1: 29).

\noindent\textbf{MGH-BRCA} consists of 1,071 IHC FFPE WSIs of invasive breast  breast carcinoma from Mass General Hospital (MGH)\cite{jaume2024multistain}. The cases contain annotations for IHC quantification in six expression levels of ER abundance (1: 168, 2: 169, 3: 219, 4:	170, 5: 175, 6:	169) and PR abundance (1: 2603, 2: 2397, 3: 1209, 4: 1118, 5: 1124, 6: 1101).

\noindent\textbf{MUT-HET} consists of 1,291 H\&E FFPE WSIs of clear-cell RCC (ccRCC), each representing a single patient treated at the Mayo Clinic\cite{joseph2016clear,acosta2022muthet}. All cases are labeled with the following mutations, determined from matched IHC slides: BAP1 mutation (WT: 1130, MUT: 162), PBRM1 mutation (WT: 622, WT: 670), and SETD2 mutation (WT: 943, MUT: 349). We split the dataset into five splits with train:val:test ratio of 60:20:20 (774:258:259 slides) in each split.

\noindent\textbf{OT-108} is an in-house pan-cancer subtyping dataset consisting of 5,564 H\&E FFPE diagnostic WSIs from BWH classified into 108 classes according to the OncoTree classification~\cite{kundra2021oncotree}. We split the dataset into train-val-test (3,164:780:1,620 slides). The test set is balanced across the classes and contains 15 slides per class.

\noindent\textbf{PANDA} consists of 10,616 H\&E FFPE diagnostic histopathology WSIs of core needle biopsies of prostate cancer sourced from the Radboud University Medical Center and the Karolinska Institute. Each slide is assigned a score recommended by the International Society of Urological Pathology (ISUP) that defines prostate cancer grade (6-class grading task). For quality control, we follow prior work\cite{pati2023weakly} in excluding slides which were erroneously annotated or had noisy labels resulting in overall 9,555 slides (grade 0: 2,603, grade 1: 2,399, grade 2: 1,209, grade 3: 1,118, grade 4: 1,124, grade 5: 1,102). For training and evaluation, we label-stratified PANDA into 80:10:10 train-validation-test folds (7,645:954:953 slides).

\noindent\textbf{Renal allograft rejection} consists of 4,847 H\&E FFPE WSIs of renal allograft biopsies from 1,118 patients collected at BWH between 2013 and 2022. Each case has associated labels for antibody-mediated rejection (AMR) status (AMR: 286 patients, no AMR: 832 patients), cellular-mediated rejection (cellular rejection: 341, no cellular rejection: 777), and interstitial fibrosis and tubular atrophy (IFTA) status (advanced IFTA: 162 patients, mild IFTA: 706 patients, moderate IFTA: 250 patients). We split the dataset label-stratified into train:val:test set (3002:376:824 slide).

\noindent\textbf{TCGA BRCA} consists of 1,049 invasive breast carcinoma (BRCA) H\&E FFPE diagnostic histopathology WSIs from TCGA. The WSIs are classified into two classes: invasive ductal carcinoma (IDC) 
and invasive lobular carcinoma (ILC). 

\noindent\textbf{TCGA NSCLC} consists of 1,043 H\&E FFPE diagnostic histopathology WSIs from TCGA of 946 patients with non-small cell lung cancer (NSCLC). The WSIs are classified into two classes: lung adenocarcinoma (LUAD, 531 slides) and lung squamous cell carcinoma (LUSC, 512 slides). We split the dataset into 5-fold cross validation, stratified by labels with ratio 60:20:20 (e.g., 659:191:193 for fold 0). \textbf{CPTAC-NSCLC} serves as external dataset with 1,091 H\&E FFPE diagnostic histopathology WSIs from CPTAC of 422 patients with NSCLC.

\noindent\textbf{TCGA LUAD} consists of 524 H\&E FFPE diagnostic histopathology WSIs from TCGA of 462 patients with lung adenocarcinoma (LUAD). We predict the mutations in the genes \textit{EGFR} (wildtype (WT): 404 patients, mutated (MUT): 58 patients), \textit{KRAS} (WT: 317, MUT: 145), \textit{STK11} (WT: 391, MUT: 71), and \textit{TP53} (WT: 222, MUT: 240). We split the dataset into 5-fold cross validation, stratified by labels with ratio 60:20:20 (e.g., 659:191:193 for fold 0). \textbf{CPTAC-LUAD} serves as external dataset with 324 H\&E FFPE diagnostic histopathology WSIs from CPTAC of 108 patients with LUAD.

\noindent\textbf{TCGA CRC} consists of 549 H\&E FFPE diagnostic histopathology WSIs from TCGA of 543 patients with colorectal cancer (CRC). We predict the presence of microsatellite instability (MSI: 61 patients, microsatellite stable (MSS): 353 patients), mutations in the genes \textit{BRAF} (WT: 429 patients, MUT: 58 patients) and \textit{KRAS} (WT: 286 patients, MUT: 201 patients), and tumor staging (T1: 16 slides, T2: 97 slides, T3: 372 slides, T4: 64 slides). \textbf{CPTAC-COAD} with 107 H\&E FFPE diagnostic histopathology WSIs from CPTAC of 103 patients with colon adenocarcinoma serves as external validation dataset for all tasks (MSI: 24 patients, MSS: 79 patients, \textit{BRAF} WT: 16 patients, \textit{BRAF} MUT: 87 patients, \textit{KRAS} WT: 36 patients, \textit{KRAS} MUT: 58 patients, T2: 17 slides, T2: 77 slides, T4: 13 slides).

\noindent\textbf{TCGA GBMLGG} consists of 1,123 H\&E FFPE diagnostic histopathology WSIs from TCGA of 558 patients with gliomas, more specifically glioblastomas multiforme and lower-grad gliomas (GBMLGG). The WSIs are classified into two classes: Isocitrate Dehydrogenase (IDH) mutation (425 slides) and no IDH mutation (698 slides). \textbf{EBRAINS} serves as an external cohort for this task (IDH MUT: 333 slides, IDH WT 540 slides).

\heading{Computing Software and Hardware}

We used Python (version 3.9.16) for all experiments and analyses in the study, which can be replicated using open-source libraries as outlined below. We used PyTorch (version 2.0.1, CUDA 11.8) for deep learning model training and inference. To train $\oursv$ and $\ours$, we modified the public implementation of iBOT (github.com/bytedance/ibot) and CoCa (github.com/mlfoundations/open\_clip). We used four and eight $\times$ 80GB NVIDIA A100 GPUs configured for multi-GPU training using distributed data-parallel (DDP) for $\oursv$ and $\ours$ training, respectively. All downstream experiments were conducted on single 24GB NVIDIA 3090 GPUs. All WSI processing was supported by OpenSlide (version 4.3.1), openslide-python (version 1.2.0), and CLAM (github.com/mahmoodlab/CLAM). 
We used Scikit-learn (version 1.2.2) for its implementation of K-Nearest Neighbors, and the logistic regression implementation and SimpleShot implementation provided by the LGSSL codebase (github.com/mbanani/lgssl). For survival tasks, we used scikit-survival (Version 0.23.1). Implementations of other slide encoders benchmarked in the study are found at the following links: GigaPath (github.com/prov-gigapath/prov-gigapath), PRISM (huggingface.coco/paige-ai/Prism), and CHIEF (github.com/hms-dbmi/CHIEF). For training weakly-supervised ABMIL models, we adapted the training scaffold code from the CLAM codebase (github.com/mahmoodlab/CLAM). Matplotlib (version 3.8.4) and Seaborn (version 0.13.2) were used to create plots and figures. Usage of other miscellaneous Python libraries is listed in the \textbf{Reporting Summary}.

\heading{Data availability}

GTEx data used in pretraining can be accessed through the GTEx portal (\href{https://www.gtexportal.org/home/}{https://www.gtexportal.org/home/}). For benchmarks, TCGA and CPTAC data can be accessed through the NIH genomic data commons\\ (\href{https://portal.gdc.cancer.gov}{https://portal.gdc.cancer.gov}) and proteomics data commons (\href{https://proteomic.datacommons.cancer.gov/pdc}{https://proteomic.datacommons.cancer.gov}) respectively. Coordinates and labels of  TCGA-UniformTumor-8K dataset is made publicly available in the $\ours$ GitHub repository. All other publicly-available datasets benchmarked in this work can be can accessed in their respective data portals: EBRAINS (\href{https://doi.org/10.25493/WQ48-ZGX }{https://doi.org/10.25493/WQ48-ZGX}),\\ DHMC-RCC  (\href{https://bmirds.github.io/KidneyCancer/}{https://bmirds.github.io/KidneyCancer}), DHMC-LUAD (\href{https://bmirds.github.io/LungCancer/}{https://bmirds.github.io/LungCancer/}), BRACS (\href{https://www.bracs.icar.cnr.it/}{https://bracs.icar.cnr.it}), PANDA (\href{https://panda.grand-challenge.org/data/}{https://panda.grand-challenge.org}),\\ IMP (\href{https://rdm.inesctec.pt/dataset/nis-2023-008}{https://rdm.inesctec.pt/dataset/nis-2023-008}), BCNB (\href{https://bupt-ai-cz.github.io/BCNB/}{https://bupt-ai-cz.github.io/BCNB/}), MUT-HET-RCC (\href{https://aacrjournals.org/cancerres/article/82/15/2792/707325/Intratumoral-Resolution-of-Driver-Gene-Mutation}{https://aacrjournals.org/cancerres/article/82/15/2792/707325/Intratumoral-Resolution-of-Driver-Gene-Mutation}). Links for all public datasets are also presented in \textbf{Extended Data Table \ref{tab:downstream-links}}. 

\heading{Code availability}

Code and model weights for loading both $\ours$ and $\oursv$ can be accessed for academic research purposes at 
\href{https://github.com/mahmoodlab/TITAN}{https://github.com/mahmoodlab/\ours}.

\paragraph{Ethics Statement}
The retrospective analysis of internal pathology images and associated reports used in this study received approval from Mass General Brigham institutional review board. Prior to the computational analysis and model development, all internal digital data, including whole slide images (WSIs), pathology reports, and electronic medical records, were anonymized. Since the study did not involve direct patient participation or recruitment, informed consent was waived for the analysis of archival pathology slides. 


\section*{Author Contributions}
T.D., S.J.W., A.H.S, R.J.C., F.M. conceived the study and designed the experiments. L.P.L., T.D., R.J.C., B.C. curated the Mass-340K whole-slide images and corresponding pathology reports. R.J.C., S.J.W., A.H.S., A.J.V., G.J., C.A.P., P.D. scanned the whole-slide images. T.D., S.J.W., R.J.C., A.H.S developed the stage 1 vision-only $\ours$ model. T.D., R.J.C, M.Y.L., S.J.W., A.H.S. developed the stage 2 and stage 3 vision-language $\ours$ models. T.D., S.J.W., and M.Y.L. developed the codebase for zero-shot vision-language slide understanding. T.D., S.J.W., A.H.S., A.Z., A.J.V., G.J. implemented the benchmarking codebase for pretrained slide models. A.K., D.F.K.W. evaluated the synthetic captions and generated reports and helped with the study design for slide retrieval. A.K., D.F.K.W., C.C. curated the rare disease retrieval dataset. R.J.C., D.K., A.K., S.I. curated and annotated the TCGA-Uniform-8K dataset. S.S. curated the renal allograft rejection dataset. T.D., S.J.W., A.H.S, R.J.C., F.M. prepared the manuscript. All authors contributed to the writing. L.P.L., F.M. supervised the research.

\section*{Acknowledgements}
This work was funded in part by the Brigham and Women’s Hospital (BWH) President’s Fund, Mass General Hospital (MGH) Pathology and by the National Institute of Health (NIH) National Institute of General Medical Sciences (NIGMS) through R35GM138216. 
S.J.W. was supported by the Helmholtz Association under the joint research school “Munich School for Data Science - MUDS” and the Add-on Fellowship of the Joachim Herz Foundation. This work was supported by a fellowship of the German Academic Exchange Service (DAAD).
M.Y.L. was supported by the Tau Beta Pi Fellowship and the Siebel Foundation.
This work was additionally supported by the AMED Practical Research for Innovative Cancer Control under grant number JP 24ck0106873 to SI. The content is solely the responsibility of the authors and does not reflect the official views of the NIH, NIGMS, NCI, DoD.


\end{spacing}
\newpage
\Heading{References}
\bibliographystyle{nature}
\bibliography{main}

\newpage
\setcounter{figure}{0}
\renewcommand{\figurename}{\textbf{Extended Data Figure}}

\setcounter{table}{0}
\renewcommand{\tablename}{\textbf{Extended Data Table}}

\clearpage
\begin{figure*}
\centering
\includegraphics[width=\textwidth]{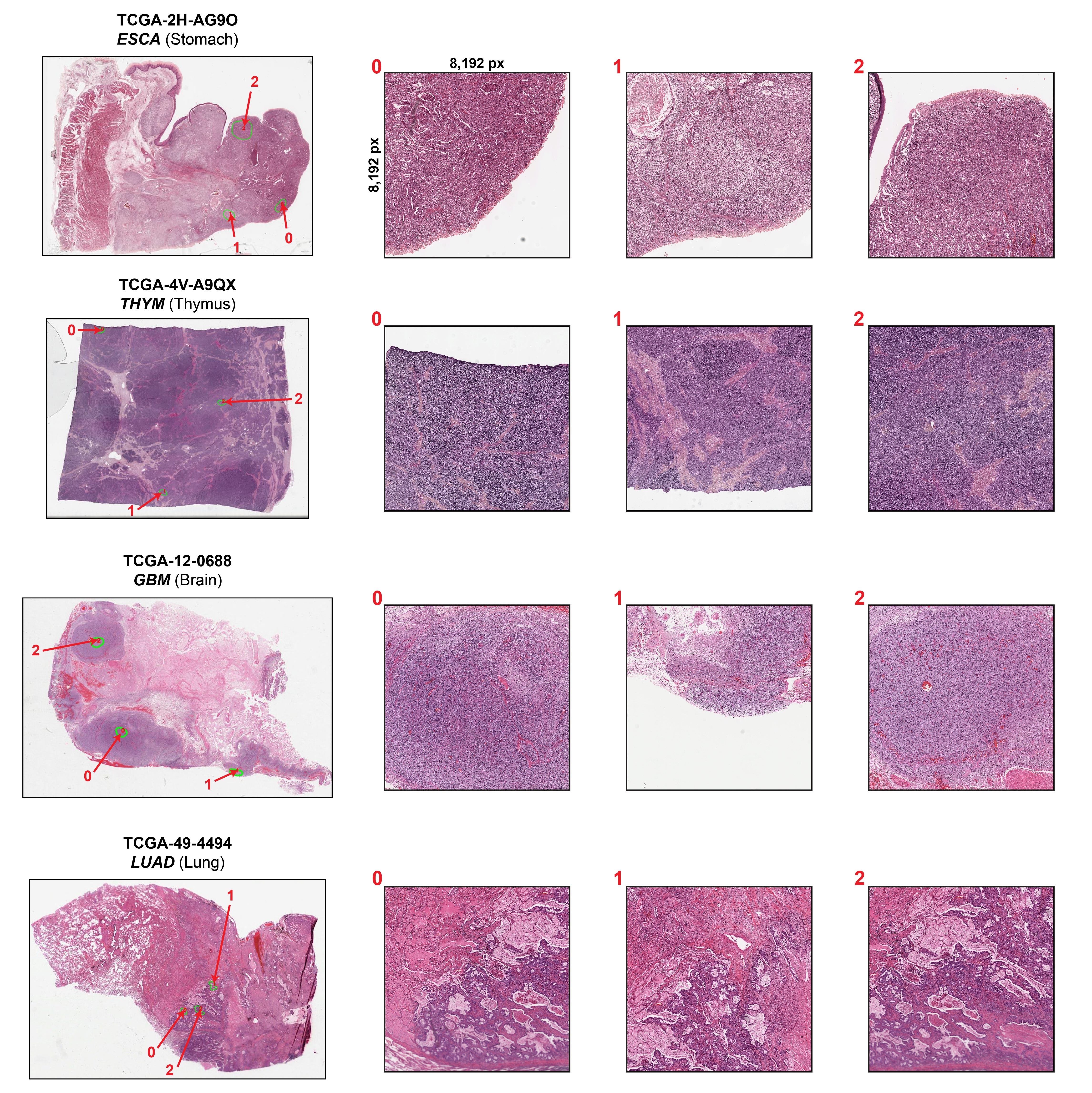}
\caption{\textbf{Examples of TCGA-UT-8K dataset}. Examples of TCGA-UT-8K, which are ROIs of $8,192\times 8,192$ pixel selected by the pathologists. The green contours illustrate the cancer region annotations, with the red number indicating the ROI index within a given TCGA slide.}
\label{fig:edf_tcgaut}
\end{figure*}

\begin{figure*}
\centering
\includegraphics[width=\textwidth]{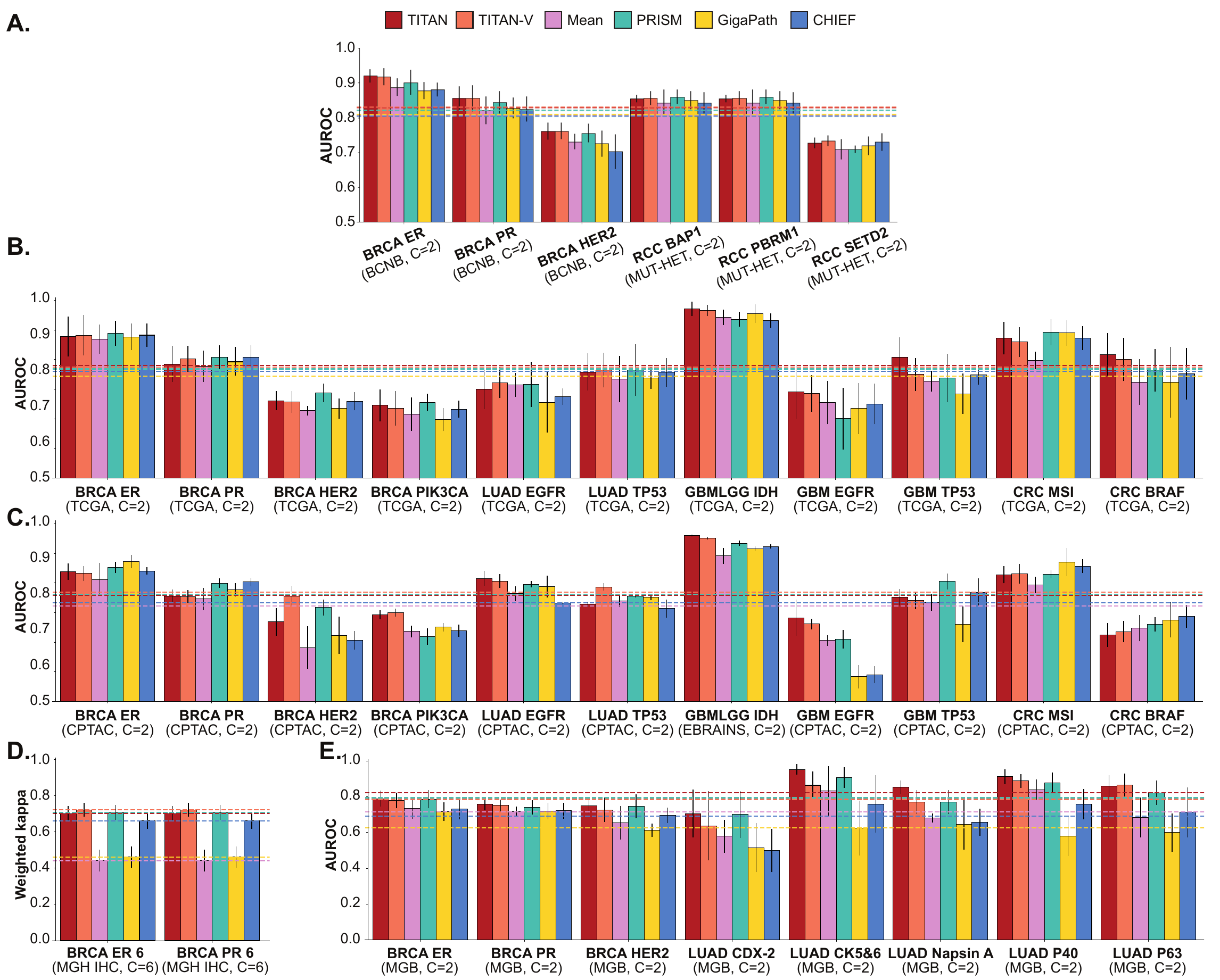}
\caption{\textbf{Linear probe results for molecular classification tasks}. \textbf{(a)} Linear models are fitted and evaluated on binary molecular status predictions for BCNB and MUT-HET. \textbf{(b)} Linear models are fitted and evaluated on five fold-splits on TCGA, \textbf{(c)} the same models are evaluated on the corresponding external datasets from CPTAC and EBRAINS. \textbf{(d)} 6-level ER and PR prediction from immunohistochemistry (IHC) slides from Mass General Hospital (MGH). \textbf{(e)} molecular classification tasks for BRCA and LUAD from Mass General Brigham (MGB).}
\label{fig:edf_molecular}
\end{figure*}

\begin{figure*}
\centering
\includegraphics[width=\textwidth]{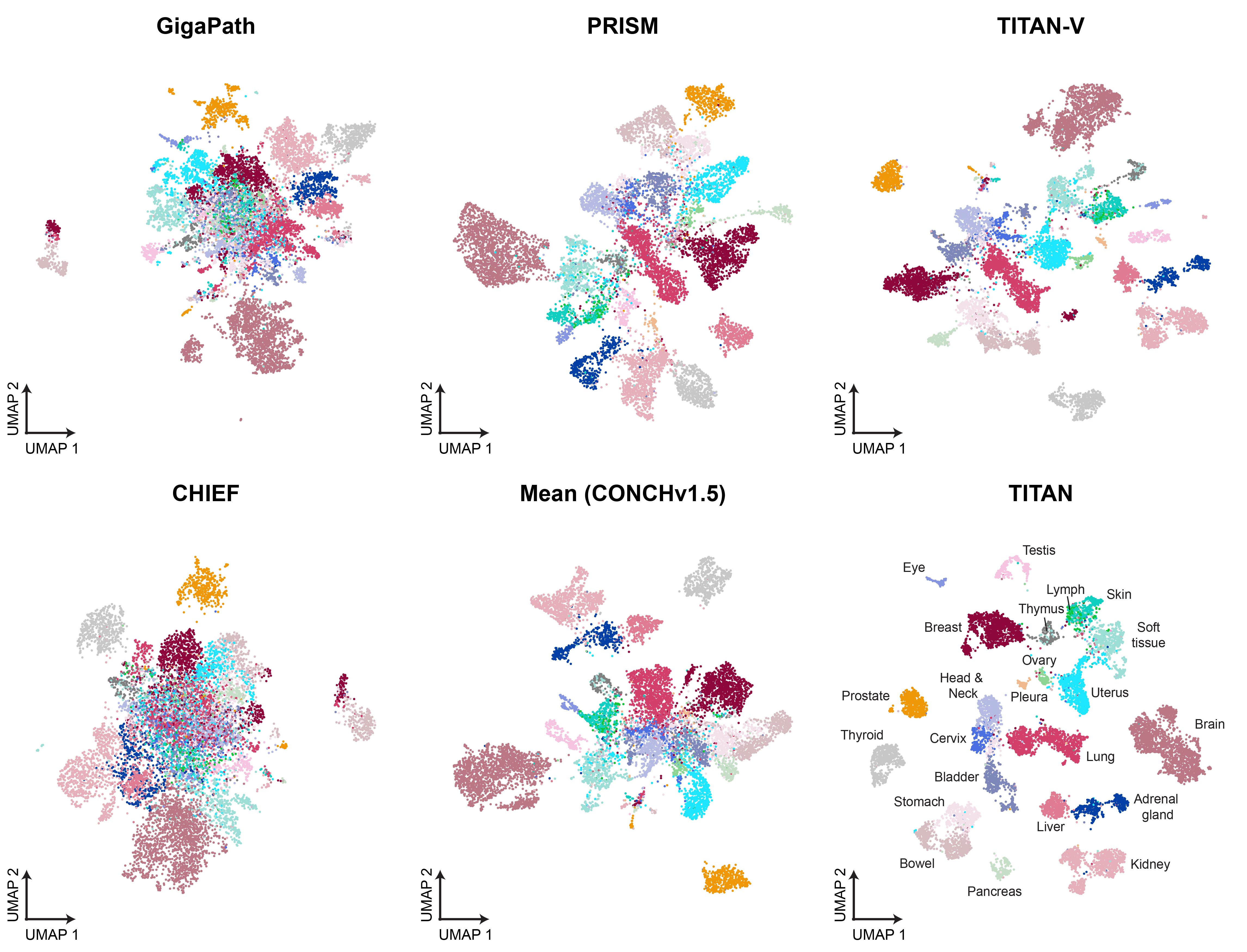}
\caption{\textbf{UMAP of slide embedding space for TCGA-OT}. UMAP visualization of slide embeddings in TCGA-OT cohort for all slide encoder baselines including $\oursvcr$ and $\oursv$, color-coded by different organs for visual decluttering.}
\label{fig:edf_umap}
\end{figure*} 

\begin{figure*}
\centering
\includegraphics[width=\textwidth]{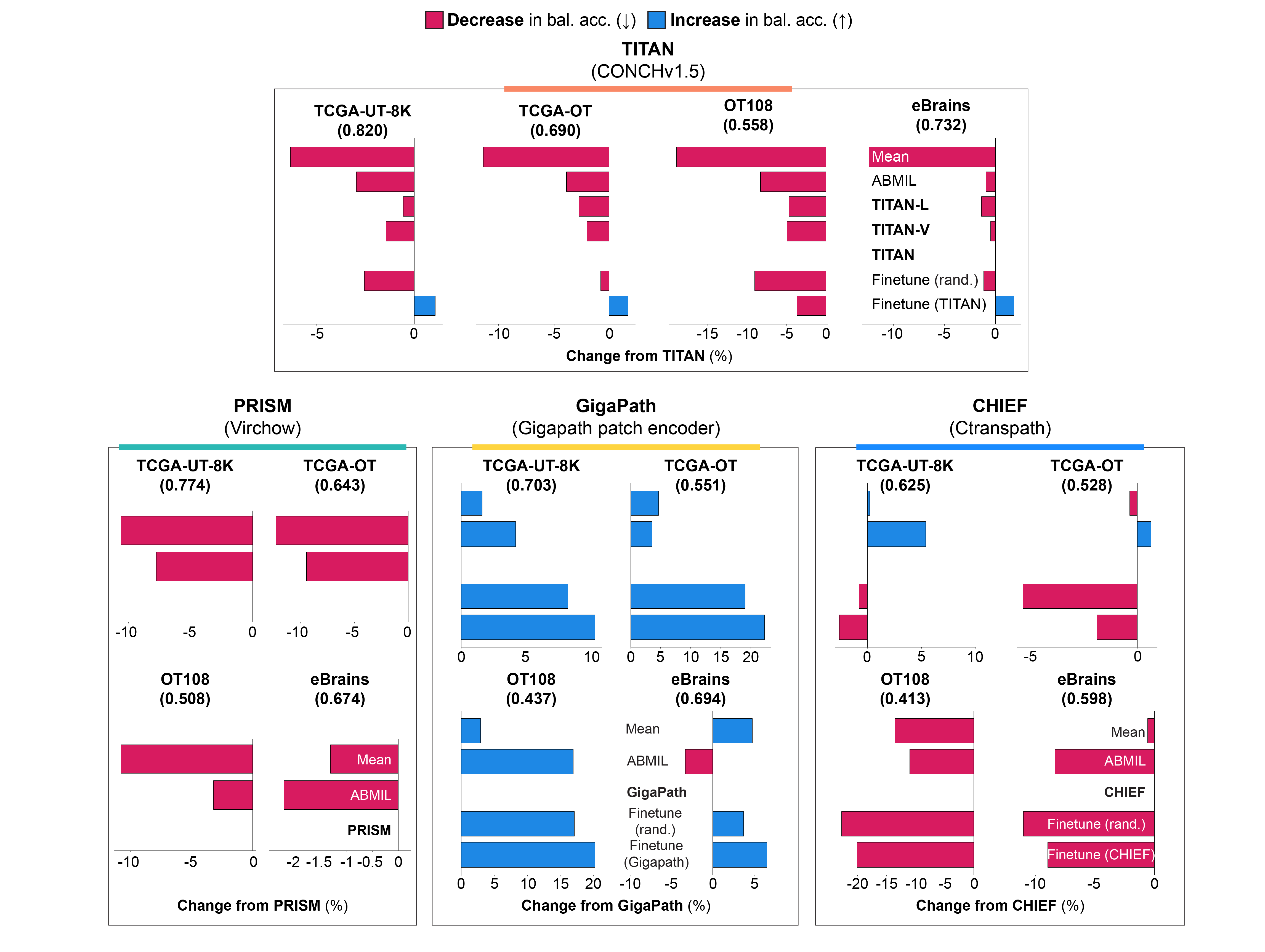}
\caption{\textbf{Ablation experiments on different learning paradigms}. Change in balanced accuracy performance for several learning paradigms on four subtyping tasks with respect to the linear probe. The baselines include mean pooling, ABMIL, linear probe, and finetuned from pretrained or randomly initialized weights. The number under each task name indicates the linear probe performance. $\ourscr$ represents the variation of $\ours$ without vision-pretraining. For mean pooling and ABMIL, we use the respective patch encoder for each framework, as specified under each slide encoder name. Finetuning results are not provided for PRISM, as the finetuning recipes were not available.}
\label{fig:edf_learning}
\end{figure*} 

\begin{figure*}
\centering
\includegraphics[width=\textwidth]{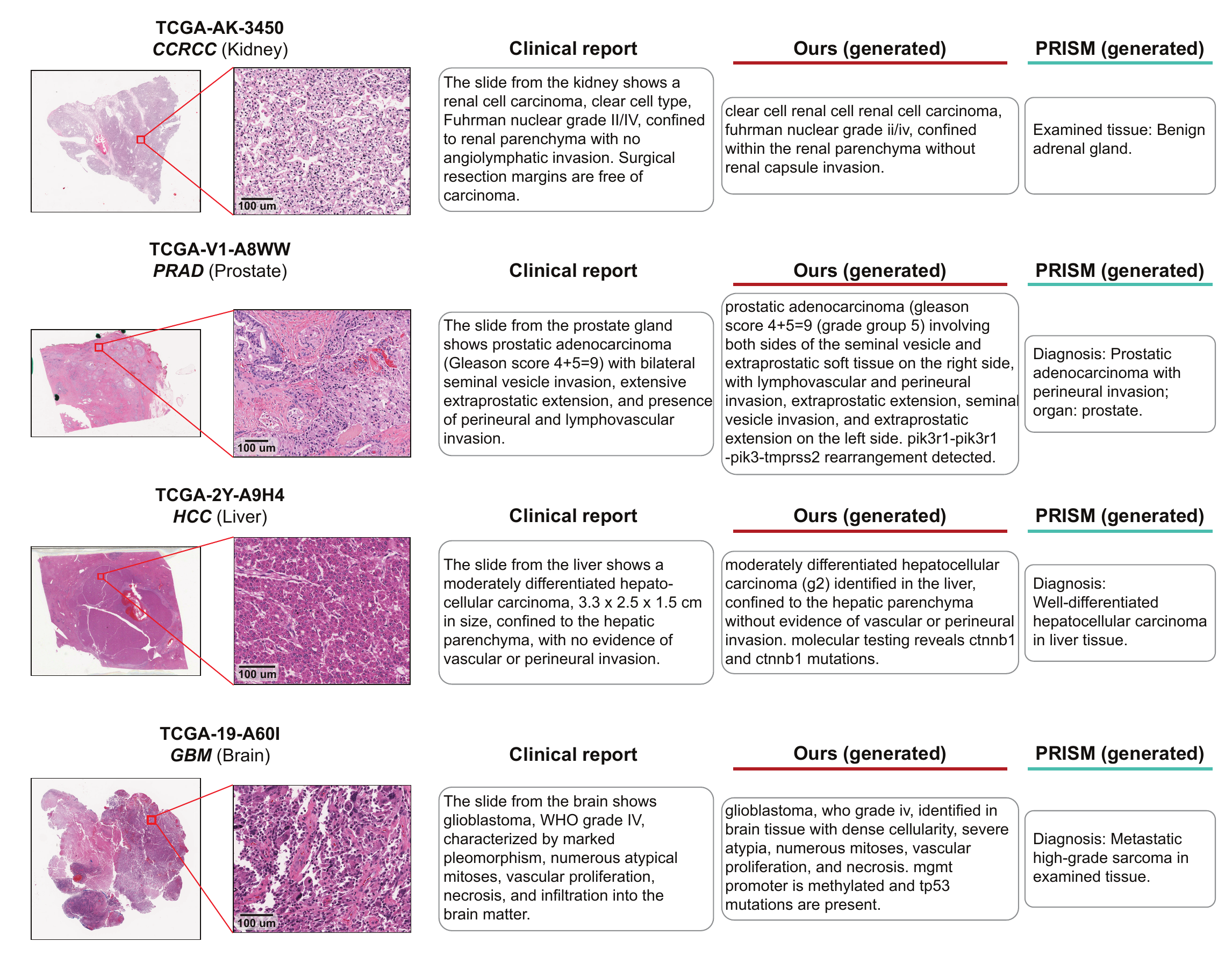}
\caption{\textbf{Examples of generated reports}. TCGA examples of generated reports of $\oursvcr$ and PRISM, with the corresponding clinical reports.}
\label{fig:edf_reports}
\end{figure*} 

\begin{figure*}
\centering
\includegraphics[width=\textwidth]{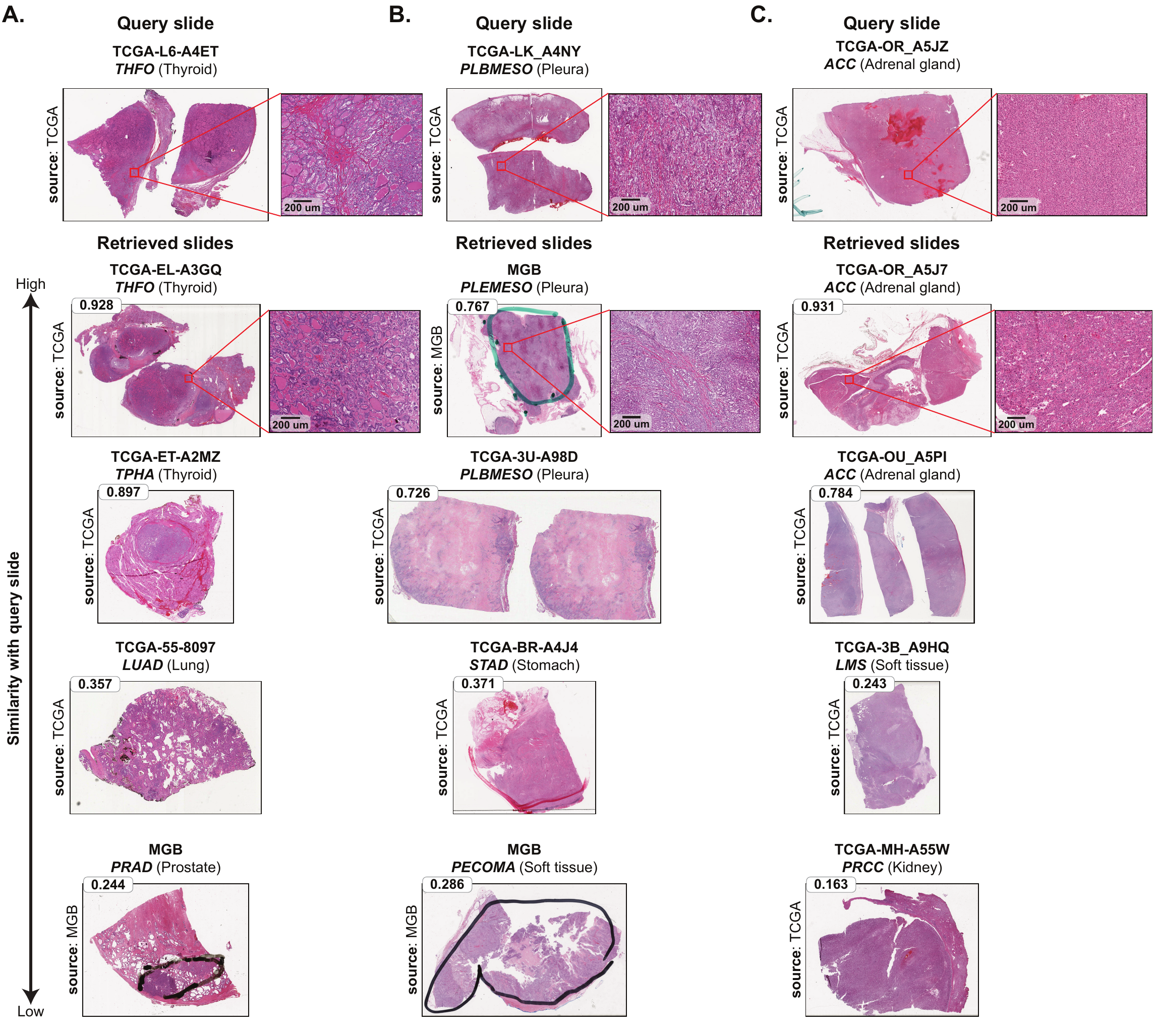}
\caption{\textbf{Rare cancer retrieval with $\ours$}. \textbf{(a-c)} Examples of slide retrieval on Rare-Cancer. The number for each retrieved slide represents the cosine similarity between the query and the retrieved slide. The retrieved slides with high similarity are either of the same diagnostic label or from the same organ as the query slide.}
\label{fig:edf_rare}
\end{figure*} 


\clearpage

\clearpage
\begin{table}[]
\centering
\footnotesize
\begin{tabular}{l|l|l|l}
\toprule
\textbf{Tissue site}             & \textbf{Class}                                                   & \textbf{OncoTree code} & \textbf{\#samples (train:val:test)} \\  \midrule
Adrenal gland           & Adrenocortical carcinoma                                & ACC           & 493 (371:74:48)            \\ \midrule
Biliary tract           & Cholangiocarcinoma                                      & CHOL          & 90 (57:3:30)               \\ \midrule
Bladder                 & Bladder urothelial carcinoma                            & BLCA          & 943 (535:103:305)          \\ \midrule
Bowel                   & Colon adenocarcinoma                                    & COAD          & 798 (623:120:55)           \\ \midrule
\multirow{2}{*}{Brain}  & Glioblastoma multiforme                                 & GBM           & 2283 (1223:342:718)        \\ 
                       &  Lower grade glioma                                      & --              & 2098 (1113:278:707)        \\ \midrule
Breast                  & Invasive carcinoma                                      & BRCA          & 2196 (1086:261:849)        \\ \midrule
Cervix                  & Squamous cell carcinoma and endocervical adenocarcinoma & CESC, ECAD    & 591 (340:78:173)           \\ \midrule
Esophagus               & Esophageal carcinoma                                    & --              & 294 (117:27:150)           \\ \midrule
Eye                     & Uveal melanoma                                          & UM    & 147 (65:16:66)             \\ \midrule
Head and neck           & Head and neck squamous cell carcinoma                   & HNSC          & 1159 (555:158:446)         \\ \midrule
\multirow{3}{*}{Kidney} & Renal clear cell carcinoma                              & CCRCC         & 798 (406:100:292)          \\
                        & Papillary renal cell carcinoma                          & PRCC          & 443 (249:43:151)           \\
                        & Chromophobe Renal Cell Carcinoma                        & CHRCC         & 180 (103:33:44)            \\ \midrule
Liver                   & Hepatocellular carcinoma                                & HCC           & 841 (565:128:148)          \\ \midrule
\multirow{2}{*}{Lung}   & Lung adenocarcinoma                                     & LUAD          & 1217 (676:144:397)         \\
                        & Lung squamous cell carcinoma                            & LUSC          & 1213 (606:195:412)         \\ \midrule
Lymph                   & Diffuse large B-cell lymphoma                           &  --             & 72 (37:12:23)              \\ \midrule
--           & Mesothelioma                                                   &     --          & 163 (75:31:57)             \\ \midrule
Ovary                   & Serous Cystadenocarcinoma                               &  --             & 220 (95:24:101)            \\ \midrule
Pancreas                & Pancreatic adenocarcinoma                               & PAAD          & 341 (190:42:109)           \\ \midrule
--                       & Pheochromocytoma and paraganglioma                      & PHC, PGNG     & 128 (81:17:30)             \\ \midrule
Prostate                & Prostate adenocarcinoma                                 & PRAD          & 815 (411:117:287)          \\ \midrule
Rectum                  & Rectum adenocarcinoma                                   & READ          & 150 (93:29:28)             \\ \midrule
Soft tissue                       & Sarcoma                                                 &    --           & 1270 (850:186:234)         \\ \midrule
Skin                    & Cutaneous Melanoma                                      & SKCM          & 931 (313:94:524)           \\ \midrule
Stomach                 & Stomach adenocarcinoma                                  & STAD          & 2306 (1482:335:489)        \\ \midrule
Testis                  & Testicular germ cell tumor                              & --              & 551 (375:79:97)            \\ \midrule
Thymus                  & Thymoma                                                 & THYM          & 328 (138:34:156)           \\ \midrule
Thyroid                 & Thyroid carcinoma                                       & --              & 1063 (528:166:369)         \\ \midrule
\multirow{2}{*}{Uterus} & Uterine corpus endometrial carcinoma                    & UCEC    & 1172 (424:144:604)         \\
                        & Uterine carcinosarcoma                                  & UCS           & 201 (71:21:109)      \\
\bottomrule
\end{tabular}
\caption{\textbf{Overview of the dataset TCGA-UniformTumor-8K} with 32 classes grouped by tissue site and sorted by largest class. This dataset is exclusively curated for ROIs (8,192$\times$8,192 pixels) subtyping task. Not every class has a one-to-one mapping to a tissue site or a single OncoTree code.}
\label{tab:dataset_tcga-ut}
\end{table}

\begin{table}[h]
\centering
\scriptsize
\begin{tabular}{l|l|l}
\toprule
\textbf{Tissue site}                    & \textbf{OncoTree code} & \textbf{\#samples (train:val:test)} \\ \midrule
\multirow{2}{*}{Adrenal gland} & ACC            & 227 (158:50:19)            \\
                               & PHC            & 163 (118:27:18)            \\ \midrule
Bladder                        & BLCA           & 457 (376:49:32)            \\ \midrule
\multirow{3}{*}{Bowel}         & COAD           & 375 (278:49:48)            \\
                               & READ           & 156 (113:25:18)            \\
                               & MACR           & 63 (48:11:4)               \\ \midrule
\multirow{6}{*}{Brain}         & GBM            & 858 (798:50:10)            \\
                               & OAST           & 217 (165:45:7)             \\
                               & ODG            & 203 (152:46:5)             \\
                               & AASTR          & 164 (113:39:12)            \\
                               & AOAST          & 155 (112:41:2)             \\
                               & ASTR           & 104 (64:33:7)              \\ \midrule
\multirow{2}{*}{Breast}        & IDC            & 838 (743:49:46)            \\
                               & ILC            & 211 (154:36:21)            \\ \midrule
Cervix                         & CESC           & 229 (134:40:55)            \\ \midrule
Eye                            & UM             & 79 (34:10:35)              \\ \midrule
Head and neck                  & HNSC           & 472 (338:47:87)            \\ \midrule
\multirow{3}{*}{Kidney}        & CCRCC          & 519 (455:49:15)            \\
                               & PRCC           & 297 (197:46:54)            \\
                               & CHRCC          & 109 (65:19:25)             \\ \midrule
Liver                          & HCC            & 362 (303:46:13)            \\ \midrule
\multirow{2}{*}{Lung}          & LUAD           & 531 (305:48:178)           \\
                               & LUSC           & 512 (364:48:100)           \\ \midrule
Melanoma                       & MEL            & 393 (347:31:15)            \\ \midrule
Ovary                          & HGSOC          & 107 (75:23:9)              \\ \midrule
Pancreas                       & PAAD           & 194 (135:32:27)            \\ \midrule
Pleura                         & PLEMESO        & 62 (43:17:2)               \\ \midrule
Prostate                       & PRAD           & 449 (382:44:23)            \\ \midrule
Skin                           & SKCM           & 75 (36:29:10)              \\ \midrule
\multirow{5}{*}{Soft tissue}   & MFH            & 165 (119:40:6)             \\
                               & LMS            & 155 (108:37:10)            \\
                               & MFS            & 141 (91:41:9)              \\
                               & DDLS           & 87 (56:29:2)               \\
                               & THYM           & 34 (18:6:10)               \\
\midrule
\multirow{5}{*}{Stomach}       & STAD           & 193 (114:46:33)            \\
                               & ESCC           & 92 (60:12:20)              \\
                               & TSTAD          & 87 (45:14:28)              \\
                               & DSTAD          & 74 (57:14:3)               \\
                               & ESCA           & 66 (44:10:12)              \\ \midrule
\multirow{2}{*}{Testis}        & NSGCT          & 128 (78:43:7)              \\
                               & SEM            & 91 (55:32:4)               \\ \midrule
Thymus                         & THYM           & 146 (68:35:43)             \\ \midrule
\multirow{2}{*}{Thyroid}       & THPA           & 408 (253:45:110)           \\
                               & THFO           & 109 (41:11:57)             \\ \midrule
\multirow{3}{*}{Uterus}        & UEC            & 422 (308:44:70)            \\
                               & USC            & 120 (64:40:16)             \\
                               & UCS            & 87 (42:34:11)             \\
    \bottomrule
    \end{tabular}
    \caption{\textbf{Overview of the dataset TCGA-OT} with 46 OncoTree codes grouped by tissue site and sorted by largest classes. This dataset is used for slide-level evaluations. Some cancer types can occur at multiple tissue sites and are listed in the tissue sites with the most samples, e.g., Leiomyosarcoma (LMS) contains samples from the uterus, stomach, bone, ovary, and head and neck. Additionally, all melanomas (MEL) are listed as a separate site with samples from skin (241), lymph (110), soft tissue (21), bowel (8), spleen (2), adrengal gland, brain, head and neck, thorax, and vulva (each 1). For every OncoTree code, we list the total number of samples and the number of samples contained in train, val, and test folds. }
    \label{tab:dataset_tcga-ot}
\end{table}

\begin{table}[]
\centering
\scriptsize
\begin{tabular}{l|l|r}
\toprule
\textbf{Tissue Site}                        & \textbf{Oncotree code} & \multicolumn{1}{l}{\textbf{\#samples}} \\
\midrule \multirow{3}{*}{Adrenal Gland}              & ACC                    & 221                                    \\
                                            & PHC                    & 155                                    \\
                                            & MNET                   & 1                                      \\
\midrule \multirow{4}{*}{Biliary tract}              & IHCH                   & 30                                     \\
                                            & PHCH                   & 4                                      \\
                                            & EHCH                   & 2                                      \\
                                            & CHOL                   & 1                                      \\
\bottomrule
\end{tabular}
    \caption{\textbf{Overview of the dataset TCGA-Slide-Reports} with 89 OncoTree codes grouped by 30 tissue sites.}
    \label{tab:dataset_tcga-slide-reports}
\end{table}

\begin{table}[]
\centering
\scriptsize
\begin{tabular}{l|l|r}
\toprule
\midrule \multirow{2}{*}{Nervous system}             & DLBCLNOS               & 2                                      \\
                                            & MPNST                  & 1                                      \\
\midrule \multirow{2}{*}{Ovary}                      & HGSOC                  & 63                                     \\
                                            & LMS                    & 1                                      \\
\midrule \multirow{3}{*}{Pancreas}                   & PAAD                   & 182                                    \\
                                            & PANET                  & 8                                      \\
                                            & UCP                    & 1                                      \\
\midrule \multirow{2}{*}{Peritoneum}                 & DDLS                   & 1                                      \\
\bottomrule
\end{tabular}
    \caption{\textbf{Overview of the dataset TCGA-Slide-Reports.} Continued.}
    \label{tab:dataset_tcga-slide-reports}
\end{table}

\begin{table}[]
\centering
\footnotesize
\begin{tabular}{l|l|r}
\toprule
\textbf{Tissue site}                                     & \textbf{OncoTree code} & \multicolumn{1}{l}{\textbf{\# samples}} \\
\midrule \multirow{2}{*}{Adrenal gland}                           & ACC                    & 248                                      \\
                                                         & PAAD                   & 1                                        \\
\midrule \multirow{3}{*}{Biliary tract}                           & IHCH                   & 32                                       \\
                                                         & GBC                    & 27                                       \\
                                                         & CHOL                   & 5                                        \\
\midrule \multirow{2}{*}{Bladder}                                 & UTUC                   & 28                                       \\
                                                         & LUCA                   & 1                                        \\
\bottomrule
\end{tabular}
    \caption{\textbf{Overview of the dataset Rare-Cancers} with 43 OncoTree codes grouped by 28 tissue sites.}
    \label{tab:dataset_rare-cancers}
\end{table}

\begin{table}[]
\centering
\footnotesize
\begin{tabular}{l|l|r}
\toprule
\multirow{14}{*}{Liver}                                  & PANET                  & 23                                       \\
                                                         & CHOL                   & 14                                       \\
                                                         & GBC                    & 11                                       \\
                                                         & PAAD                   & 10                                       \\
                                                         & ANSC                   & 6                                        \\
                                                         & LUNE                   & 3                                        \\
                                                         & ESCC                   & 3                                        \\
                                                         & LUCA                   & 2                                        \\
\bottomrule
\end{tabular}
    \caption{\textbf{Overview of the dataset Rare-Cancers.} Continued.}
    \label{tab:dataset_rare-cancers}
\end{table}

\begin{table}[]
\centering
\footnotesize
\begin{tabular}{l|l|r}
\toprule
\multirow{7}{*}{Peritoneum}                              & USC                    & 3                                        \\
                                                         & GBC                    & 3                                        \\
                                                         & CHOL                   & 3                                        \\
                                                         & EOV                    & 2                                        \\
                                                         & CCOV                   & 2                                        \\
                                                         & PANET                  & 1                                        \\
                                                         & PAAD                   & 1                                        \\
\midrule \multirow{3}{*}{Pleura}                                  & PLBMESO                & 19                                       \\
\end{tabular}
    \caption{\textbf{Overview of the dataset Rare-Cancers.} Continued.}
    \label{tab:dataset_tcga-slide-reports}
\end{table}

\begin{table}[]
\centering
\footnotesize
\begin{tabular}{l|l|r}
\toprule
\textbf{Tissue Site}                                     & \textbf{Oncotree code} & \multicolumn{1}{l}{\textbf{\#samples}} \\
\midrule
Adrenal gland                                            & ACC                    & 227                                    \\ \midrule
Biliary tract                                            & IHCH                   & 32                                     \\ \midrule
\multirow{3}{*}{Brain}                                   & ODG                    & 203                                    \\
                                                         & AASTR                  & 164                                    \\
                                                         & ASTR                   & 104                                    \\ \midrule
Breast                                                   & MBC                    & 13                                     \\ \midrule
\multirow{2}{*}{Cervix}                                  & ECAD                   & 24                                     \\
\bottomrule
\end{tabular}
    \caption{\textbf{Overview of the dataset Rare-Cancers-Public} with 29 OncoTree codes grouped by 19 tissue sites.}
    \label{tab:dataset_rare-cancer-public}
\end{table}

\clearpage
\begin{table}[ht]
\centering
\scriptsize
\begin{tabular}{l|ccccccccc}
\toprule
 \textbf{Dataset}         & \textbf{Tissue site}         & \textbf{Target}                      & \textbf{Label-level}   &   \textbf{\# Classes} & \textbf{\# Patients}   & \textbf{\# WSIs}   & \textbf{Metric}     & \textbf{External test}     \\
\midrule
 OT108                    & Pan-cancer & OncoTree               & WSI     &          108 & 5,510         & 5,564        & Bal. acc. &    \\
 TCGA-OT                  & Pan-cancer & OncoTree               & WSI     &           46 & 9,149         & 11,186       & Bal. acc. &  \\
 TCGA-UT-8K               & Pan-cancer & Subtype               & WSI     &           32 & 7,784         & 24,392        & Bal. acc. & $\checkmark$     \\
 EBRAINS                  & Brain      & Diagnosis                  & WSI     &           30 & 2,147         & 2,319    & Bal. acc. &      \\
 BRACS                    & Breast     & Subtype                     & WSI     &            7 & 189           & 547     & Bal. acc. &         \\
 BRACS                    & Breast     & Subtype (coarse)              & WSI     &            3 & 189           & 547   & Bal. acc. &        \\
 TCGA-BRCA                & Breast     & Subtype                    & WSI     &            2 & 984           & 1,049    & AUROC     &        \\
\bottomrule
\end{tabular}
\caption{\textbf{Overview of the morphological tasks} sorted by organ.}
\label{tab:tasks_morphological}
\end{table}

\begin{table}[th]
    \centering
    \scriptsize
\begin{tabular}{l|ccccccccc}
\toprule
 \textbf{Dataset}         & \textbf{Tissue site}         & \textbf{Target}                      & \textbf{Label-level}   &   \textbf{\# Classes} & \textbf{\# Patients}   & \textbf{\# WSIs}   & \textbf{Metric}        \\
\midrule
 IMP & Colorectal & Dysplasia grading                                       & WSI     &            3 & 5,333         & 5,333    & Cohen's $\kappa$ \\
 Renal allograft rejection            & Kidney     & IFTA status & Patient &            3 & 1,118         & 4,847               & Cohen's $\kappa$ \\
 PANDA            & Prostate   & Gleason grading                           & WSI     &            6 & 9,555         & 9,555     & Cohen's $\kappa$ \\
\bottomrule
\end{tabular}
\caption{\textbf{Overview of the grading tasks} sorted by organ. The metric Cohen's $\kappa$ is quadratic weighted.}
    \label{tab:task_grading}
\end{table}

\begin{table}[th]
    \centering
    \scriptsize
\begin{tabular}{l|ccccccccc}
\toprule
 \textbf{Dataset}         & \textbf{Tissue Site}         & \textbf{Target}                      & \textbf{Label-level}   & \textbf{\# Patients}   & \textbf{\# WSIs}  \\
\midrule
 TCGA-BRCA                       & Breast       & Survival & Patient        & 1,022       & 1,092   \\
 TCGA-BLCA                       & Bladder      & Survival & Patient        & 360         & 424     \\
 TCGA-CRC                        & Colorectal   & Survival & Patient        & 545         & 554     \\
 TCGA-KIRC                       & Kidney       & Survival & Patient        & 502         & 508     \\
 TCGA-UCEC                       & Uterus       & Survival & Patient        & 504         & 564     \\
 TCGA-NSCLC                      & Lung         & Survival & Patient        & 844         & 920     \\
\bottomrule
\end{tabular}
\caption{\textbf{Overview of the survival prediction tasks} for measuring disease-specific survival (DSS) with c-index sorted by organ. All tasks are evaluated in five-fold site-preserving cross-validation splits.}
    \label{tab:task_survivl}
\end{table}

\begin{table}[ht]
    \centering
    \scriptsize
\begin{tabular}{l|ccccccccc}
\toprule
 \textbf{Dataset}     & \textbf{Tissue Site}      & \textbf{Target}                                  & \textbf{Label-level}   &   \textbf{\# Classes} & \textbf{\# Patients}   & \textbf{\# WSIs}   & \textbf{Metric}         & \textbf{External test}   \\
\midrule
 TCGA-GBM    & Brain      & \textit{{EGFR}}                            & Patient &   2 & 228           & 560       & AUROC            &        \\
 CPTAC-GBM   & Brain      & \textit{{EGFR}}                            & Patient &   2 & 99            & 243       & AUROC            & $\checkmark$    \\
 TCGA-GBM    & Brain      & \textit{{TP53}}                            & Patient &   2 & 228           & 560       & AUROC            &       \\
 CPTAC-GBM   & Brain      & \textit{{TP53}}                            & Patient &   2 & 99            & 243       & AUROC            &         \\
 TCGA-GBMLGG & Brain      & IDH                                        & WSI     &   2 & 558      & 1,123     & AUROC                 &       \\
 EBRAINS     & Brain      & IDH                                        & WSI     &   2 & 795      & 873       & AUROC                 & $\checkmark$    \\
  BCNB        & Breast     & ER                                        & Patient &   2 & 1,058         & 1,058     & AUROC            &       \\
\bottomrule
\end{tabular}
\caption{\textbf{Overview of the molecular tasks} sorted by organ. The metric Cohen's $\kappa$ is quadratic weighted.}
\label{tab:task_molecular}
\end{table}
\clearpage
\begin{table}[h]
    \centering
    \footnotesize
        \begin{tabular}{ll}
            \toprule
            \textbf{Dataset} & \textbf{Link} \\
            \midrule
            TCGA & \href{https://portal.gdc.cancer.gov/}{https://portal.gdc.cancer.gov/} \\
            CPTAC & \href{https://proteomic.datacommons.cancer.gov/pdc}{https://proteomic.datacommons.cancer.gov/pdc} \\
            EBRAINS & \href{https://doi.org/10.25493/WQ48-ZGX}{https://doi.org/10.25493/WQ48-ZGX} \\
            DHMC-RCC & \href{https://bmirds.github.io/KidneyCancer}{https://bmirds.github.io/KidneyCancer} \\
            DHMC-LUAD & \href{https://bmirds.github.io/LungCancer/}{https://bmirds.github.io/LungCancer/} \\
            BRACS & \href{https://www.bracs.icar.cnr.it/}{https://www.bracs.icar.cnr.it/} \\
            PANDA & \href{https://panda.grand-challenge.org}{https://panda.grand-challenge.org} \\
            IMP & \href{https://rdm.inesctec.pt/dataset/nis-2023-008}{https://rdm.inesctec.pt/dataset/nis-2023-008} \\
            BCNB & \href{https://bupt-ai-cz.github.io/BCNB/}{https://bupt-ai-cz.github.io/BCNB/} \\
            MUT-HET-RCC & \href{https://aacrjournals.org/cancerres/article/82/15/2792/707325/Intratumoral-Resolution-of-Driver-Gene-Mutation}{https://aacrjournals.org/cancerres/article/82/15/2792/707325/Intratumoral-Resolution-of-Driver-Gene-Mutation} \\
            \bottomrule
            \end{tabular}
        \caption{\textbf{Summary of publicly available datasets.}}
\label{tab:downstream-links}
\end{table}

\clearpage
\input{tables/linear_probe}
\clearpage

\begin{table*}[!ht]
\centering
\small
\caption{\textbf{Survival prediction} Results for $\ours$ and other slide encoders for measuring disease-specific survival (DSS) with c-index on six TCGA cancer cohorts. Except for CHIEF, all slide encoders have not used TCGA as the pretraining dataset. The best and second-best performances are denoted by \textbf{bold} and \underline{underlined}, respectively.}
\begin{tabular}{l|cccccc}
\toprule
 &  CHIEF & GigaPath &                             PRISM &   Mean &                          $\oursv$ &     $\oursvcr$ \\
\midrule
BLCA   &  0.599 $\pm$ \scriptsize{0.019} &  0.589 $\pm$ \scriptsize{0.054} &    \underline{0.656} $\pm$ \scriptsize{0.023} & 0.630 $\pm$ \scriptsize{0.058} & \textbf{0.657} $\pm$ \scriptsize{0.055} &  \underline{0.656} $\pm$ \scriptsize{0.040} \\
BRCA   &  \underline{0.737} $\pm$ \scriptsize{0.039} &  0.687 $\pm$ \scriptsize{0.083} &  0.685 $\pm$ \scriptsize{0.076} & 0.711 $\pm$ \scriptsize{0.048} &  0.713 $\pm$ \scriptsize{0.044} &  \textbf{0.757} $\pm$ \scriptsize{0.015} \\
CRC    &  0.680 $\pm$ \scriptsize{0.079} &  0.628 $\pm$ \scriptsize{0.085}  &  0.622 $\pm$ \scriptsize{0.125} & \textbf{0.720} $\pm$ \scriptsize{0.089} & \underline{0.710} $\pm$ \scriptsize{0.110} &  0.705 $\pm$ \scriptsize{0.097} \\
KIRC   &  0.736 $\pm$ \scriptsize{0.061} &  0.751 $\pm$ \scriptsize{0.072} &  \underline{0.769} $\pm$ \scriptsize{0.061} & \textbf{0.774} $\pm$ \scriptsize{0.070} &  \textbf{0.774} $\pm$ \scriptsize{0.060} &  0.768 $\pm$ \scriptsize{0.064} \\
NSCLC  &  \underline{0.633} $\pm$ \scriptsize{0.037} &  \textbf{0.670} $\pm$ \scriptsize{0.031} &   0.600 $\pm$ \scriptsize{0.074} & 0.596 $\pm$ \scriptsize{0.049} &  0.624 $\pm$ \scriptsize{0.082} &  0.623 $\pm$ \scriptsize{0.041} \\
UCEC   &  0.758 $\pm$ \scriptsize{0.109} &  0.779 $\pm$ \scriptsize{0.105} &   0.747 $\pm$ \scriptsize{0.026} & 0.761 $\pm$ \scriptsize{0.133} &  \textbf{0.789} $\pm$ \scriptsize{0.091} &  \underline{0.787} $\pm$ \scriptsize{0.067} \\
\midrule
Avg. ($\uparrow$) & 0.691 & 0.684 & 0.680 & 0.699 & \underline{0.711} & \textbf{0.716}\\
\bottomrule
\end{tabular}
\label{tab:survival_main}
\end{table*}
\clearpage
\begin{table}[h!]
\centering
\footnotesize
\begin{tabular}{l|c|ccc}
\toprule
 Encoder            & Bal. acc.                                 & Weighted F1                               & AUROC                                     \\
\midrule
 ABMIL (CONCHv1.5) (finetuned) & 0.807\scriptsize{$\pm$0.0058}             & 0.854\scriptsize{$\pm$0.0039}             & \underline{0.991}\scriptsize{$\pm$0.0004} \\
 GigaPath                      & 0.700\scriptsize{$\pm$0.0069}             & 0.782\scriptsize{$\pm$0.0046}             & 0.979\scriptsize{$\pm$0.0010}             \\
 GigaPath (finetuned)          & 0.772\scriptsize{$\pm$0.0061}             & 0.841\scriptsize{$\pm$0.0039}             & 0.983\scriptsize{$\pm$0.0008}             \\
 PRISM                         & 0.774\scriptsize{$\pm$0.0062}             & 0.824\scriptsize{$\pm$0.0042}             & 0.989\scriptsize{$\pm$0.0005}             \\
 PRISM (finetuned)             & 0.746\scriptsize{$\pm$0.0065}             & 0.811\scriptsize{$\pm$0.0045}             & 0.987\scriptsize{$\pm$0.0007}             \\
 CHIEF                         & 0.625\scriptsize{$\pm$0.0079}             & 0.723\scriptsize{$\pm$0.0051}             & 0.973\scriptsize{$\pm$0.0011}             \\
 CHIEF (finetuned)             & 0.609\scriptsize{$\pm$0.0078}             & 0.708\scriptsize{$\pm$0.0051}             & 0.960\scriptsize{$\pm$0.0017}             \\
\bottomrule
\end{tabular}\caption{\textbf{Finetuning results} for tumor subtype ($C=2$) prediction on TCGA-UT-8K for all baselines. Finetuned results are trained on the specific task, while non-finetuned results are logistic regression fits of the frozen slide embeddings. The best result is marked in bold and the second best is underlined.}
\label{tab:finetuning_TCGAUniformTumor_cancer}
\end{table}

\begin{table}[h!]
\centering
\footnotesize
\begin{tabular}{l|c|ccc}
\toprule
 Encoder            & Bal. acc.                                 & Weighted F1                               & AUROC                                     \\
\midrule
 ABMIL (CONCHv1.5) (finetuned) & 0.677\scriptsize{$\pm$0.0204}             & 0.735\scriptsize{$\pm$0.0126}             & 0.982\scriptsize{$\pm$0.0026}             \\
 GigaPath                      & 0.543\scriptsize{$\pm$0.0178}             & 0.659\scriptsize{$\pm$0.0134}             & 0.976\scriptsize{$\pm$0.0021}             \\
 GigaPath (finetuned)          & 0.664\scriptsize{$\pm$0.0175}             & 0.736\scriptsize{$\pm$0.0126}             & 0.979\scriptsize{$\pm$0.0022}             \\
 PRISM                         & 0.643\scriptsize{$\pm$0.0181}             & 0.732\scriptsize{$\pm$0.0129}             & 0.986\scriptsize{$\pm$0.0014}             \\
 PRISM (finetuned)             & 0.548\scriptsize{$\pm$0.0172}             & 0.644\scriptsize{$\pm$0.0134}             & 0.980\scriptsize{$\pm$0.0015}             \\
 CHIEF                         & 0.528\scriptsize{$\pm$0.0205}             & 0.640\scriptsize{$\pm$0.0142}             & 0.968\scriptsize{$\pm$0.0028}             \\
 CHIEF (finetuned)             & 0.518\scriptsize{$\pm$0.0200}             & 0.630\scriptsize{$\pm$0.0143}             & 0.944\scriptsize{$\pm$0.0043}             \\
\bottomrule
\end{tabular}\caption{\textbf{Finetuning results} for OncoTree code ($C=2$) prediction on TCGA-OT for all baselines. Finetuned results are trained on the specific task, while non-finetuned results are logistic regression fits of the frozen slide embeddings. The best result is marked in bold and the second best is underlined.}
\label{tab:finetuning_TCGA_OncoTreeCode}
\end{table}

\begin{table}[h!]
\centering
\footnotesize
\begin{tabular}{l|c|ccc}
\toprule
 Encoder            & Bal. acc.                                 & Weighted F1                               & AUROC                                     \\
\midrule
 ABMIL (CONCHv1.5) (finetuned) & 0.538\scriptsize{$\pm$0.0109}             & 0.521\scriptsize{$\pm$0.0134}             & 0.977\scriptsize{$\pm$0.0016}             \\
 GigaPath                      & 0.437\scriptsize{$\pm$0.0110}             & 0.423\scriptsize{$\pm$0.0124}             & 0.949\scriptsize{$\pm$0.0030}             \\
 GigaPath (finetuned)          & 0.525\scriptsize{$\pm$0.0109}             & 0.508\scriptsize{$\pm$0.0134}             & 0.972\scriptsize{$\pm$0.0019}             \\
 PRISM                         & 0.508\scriptsize{$\pm$0.0110}             & 0.481\scriptsize{$\pm$0.0133}             & 0.976\scriptsize{$\pm$0.0018}             \\
 PRISM (finetuned)             & 0.492\scriptsize{$\pm$0.0110}             & 0.478\scriptsize{$\pm$0.0128}             & 0.967\scriptsize{$\pm$0.0022}             \\
 CHIEF                         & 0.413\scriptsize{$\pm$0.0109}             & 0.394\scriptsize{$\pm$0.0130}             & 0.944\scriptsize{$\pm$0.0029}             \\
 CHIEF (finetuned)             & 0.330\scriptsize{$\pm$0.0108}             & 0.319\scriptsize{$\pm$0.0121}             & 0.916\scriptsize{$\pm$0.0038}             \\
\bottomrule
\end{tabular}\caption{\textbf{Finetuning results} for OncoTree code ($C=2$) prediction on OT108 for all baselines. Finetuned results are trained on the specific task, while non-finetuned results are logistic regression fits of the frozen slide embeddings. The best result is marked in bold and the second best is underlined.}
\label{tab:finetuning_OP108_OncoTreeCode}
\end{table}

\begin{table}[h!]
\centering
\footnotesize
\begin{tabular}{l|c|ccc}
\toprule
 Encoder            & Bal. acc.                                 & Weighted F1                               & AUROC                                     \\
\midrule
 ABMIL (CONCHv1.5) (finetuned) & 0.728\scriptsize{$\pm$0.0203}             & 0.778\scriptsize{$\pm$0.0184}             & 0.982\scriptsize{$\pm$0.0024}             \\
 GigaPath                      & 0.680\scriptsize{$\pm$0.0217}             & 0.746\scriptsize{$\pm$0.0197}             & 0.978\scriptsize{$\pm$0.0027}             \\
 GigaPath (finetuned)          & 0.725\scriptsize{$\pm$0.0209}             & 0.780\scriptsize{$\pm$0.0185}             & 0.983\scriptsize{$\pm$0.0028}             \\
 PRISM                         & 0.674\scriptsize{$\pm$0.0200}             & 0.732\scriptsize{$\pm$0.0191}             & 0.978\scriptsize{$\pm$0.0026}             \\
 PRISM (finetuned)             & 0.652\scriptsize{$\pm$0.0224}             & 0.707\scriptsize{$\pm$0.0196}             & 0.967\scriptsize{$\pm$0.0039}             \\
 CHIEF                         & 0.598\scriptsize{$\pm$0.0237}             & 0.670\scriptsize{$\pm$0.0206}             & 0.969\scriptsize{$\pm$0.0032}             \\
 CHIEF (finetuned)             & 0.544\scriptsize{$\pm$0.0247}             & 0.622\scriptsize{$\pm$0.0208}             & 0.941\scriptsize{$\pm$0.0076}             \\
\bottomrule
\end{tabular}\caption{\textbf{Finetuning results} for tumor type ($C=2$) prediction on EBRAINS for all baselines. Finetuned results are trained on the specific task, while non-finetuned results are logistic regression fits of the frozen slide embeddings. The best result is marked in bold and the second best is underlined.}
\label{tab:finetuning_ebrains_diagnosis}
\end{table}
\clearpage
\clearpage

\begin{table}[h!]
\centering
\footnotesize
\begin{tabular}{l|cc|cc|cc}
\toprule
  & \multicolumn{2}{c|}{Logistic regression} & \multicolumn{2}{c|}{SimpleShot} & \multicolumn{2}{c}{20-nearest neighbors} \\ Encoder   & Bal. acc.                                 & Weighted F1                               & Bal. acc.                                 & Weighted F1                               & Bal. acc.                                 & Weighted F1                               \\
\midrule
 $\oursv$ (abs)                                                                                                                        & 0.816\scriptsize{$\pm$0.0059}             & \underline{0.870}\scriptsize{$\pm$0.0037} & 0.813\scriptsize{$\pm$0.0053}             & 0.858\scriptsize{$\pm$0.0038}             & 0.811\scriptsize{$\pm$0.0050}             & 0.839\scriptsize{$\pm$0.0038}             \\
 $\oursv$ (none)                                                                                                                       & \underline{0.819}\scriptsize{$\pm$0.0054} & 0.867\scriptsize{$\pm$0.0038}             & \underline{0.815}\scriptsize{$\pm$0.0054} & \underline{0.859}\scriptsize{$\pm$0.0039} & \textbf{0.814}\scriptsize{$\pm$0.0048}    & \textbf{0.851}\scriptsize{$\pm$0.0037}    \\
 $\oursv$ (rope)                                                                                                                       & 0.818\scriptsize{$\pm$0.0059}             & \textbf{0.873}\scriptsize{$\pm$0.0037}    & 0.812\scriptsize{$\pm$0.0053}             & \underline{0.859}\scriptsize{$\pm$0.0040} & 0.804\scriptsize{$\pm$0.0056}             & 0.836\scriptsize{$\pm$0.0040}             \\
\bottomrule
\end{tabular}\caption{\textbf{Ablation study of positional encodings.} Linear probing results for tumor subtype ($C=32$) prediction on TCGA-UT-8K. The best result is marked in bold and the second best is underlined.}
\label{tab:ablation_pos_enc_TCGAUniformTumor_cancer}
\end{table}

\begin{table}[h!]
\centering
\footnotesize
\begin{tabular}{l|cc|cc|cc}
\toprule
  & \multicolumn{2}{c|}{Logistic regression} & \multicolumn{2}{c|}{SimpleShot} & \multicolumn{2}{c}{20-nearest neighbors} \\ Encoder   & Bal. acc.                                 & Weighted F1                               & Bal. acc.                                 & Weighted F1                               & Bal. acc.                                 & Weighted F1                               \\
\midrule
 $\oursv$ (abs)                                                                                                                        & \underline{0.673}\scriptsize{$\pm$0.0157} & \underline{0.753}\scriptsize{$\pm$0.0126} & 0.631\scriptsize{$\pm$0.0184}             & 0.721\scriptsize{$\pm$0.0130}             & \underline{0.706}\scriptsize{$\pm$0.0190} & \textbf{0.777}\scriptsize{$\pm$0.0116}    \\
 $\oursv$ (none)                                                                                                                       & 0.664\scriptsize{$\pm$0.0188}             & 0.749\scriptsize{$\pm$0.0127}             & \underline{0.633}\scriptsize{$\pm$0.0175} & \underline{0.726}\scriptsize{$\pm$0.0129} & \textbf{0.708}\scriptsize{$\pm$0.0187}    & \textbf{0.777}\scriptsize{$\pm$0.0112}    \\
 $\oursv$ (rope)                                                                                                                       & 0.672\scriptsize{$\pm$0.0170}             & 0.735\scriptsize{$\pm$0.0127}             & 0.627\scriptsize{$\pm$0.0164}             & 0.719\scriptsize{$\pm$0.0129}             & 0.679\scriptsize{$\pm$0.0207}             & 0.765\scriptsize{$\pm$0.0115}             \\
\bottomrule
\end{tabular}\caption{\textbf{Ablation study of positional encodings.} Linear probing results for OncoTree code ($C=46$) prediction on TCGA-OT. The best result is marked in bold and the second best is underlined.}
\label{tab:ablation_pos_enc_TCGA_OncoTreeCode}
\end{table}

\begin{table}[h!]
\centering
\footnotesize
\begin{tabular}{l|cc|cc|cc}
\toprule
  & \multicolumn{2}{c|}{Logistic regression} & \multicolumn{2}{c|}{SimpleShot} & \multicolumn{2}{c}{20-nearest neighbors} \\ Encoder   & Bal. acc.                                 & Weighted F1                               & Bal. acc.                                 & Weighted F1                               & Bal. acc.                                 & Weighted F1                               \\
\midrule
 $\oursv$ (abs)                                                                                                                        & 0.532\scriptsize{$\pm$0.0106}             & 0.508\scriptsize{$\pm$0.0133}             & 0.446\scriptsize{$\pm$0.0102}             & 0.411\scriptsize{$\pm$0.0130}             & \underline{0.506}\scriptsize{$\pm$0.0102} & \underline{0.497}\scriptsize{$\pm$0.0128} \\
 $\oursv$ (none)                                                                                                                       & \underline{0.536}\scriptsize{$\pm$0.0103} & \underline{0.512}\scriptsize{$\pm$0.0130} & \underline{0.450}\scriptsize{$\pm$0.0104} & \underline{0.414}\scriptsize{$\pm$0.0130} & 0.488\scriptsize{$\pm$0.0108}             & 0.483\scriptsize{$\pm$0.0130}             \\
 $\oursv$ (rope)                                                                                                                       & 0.526\scriptsize{$\pm$0.0103}             & 0.505\scriptsize{$\pm$0.0132}             & 0.446\scriptsize{$\pm$0.0102}             & \underline{0.414}\scriptsize{$\pm$0.0129} & 0.495\scriptsize{$\pm$0.0109}             & \underline{0.497}\scriptsize{$\pm$0.0129} \\
\bottomrule
\end{tabular}\caption{\textbf{Ablation study of positional encodings.} Linear probing results for OncoTree code ($C=108$) prediction on OT108. The best result is marked in bold and the second best is underlined.}
\label{tab:ablation_pos_enc_OP108_OncoTreeCode}
\end{table}

\begin{table}[h!]
\centering
\footnotesize
\begin{tabular}{l|cc|cc|cc}
\toprule
  & \multicolumn{2}{c|}{Logistic regression} & \multicolumn{2}{c|}{SimpleShot} & \multicolumn{2}{c}{20-nearest neighbors} \\ Encoder   & Bal. acc.                                 & Weighted F1                               & Bal. acc.                                 & Weighted F1                               & Bal. acc.                                 & Weighted F1                               \\
\midrule
 $\oursv$ (abs)                                                                                                                        & \underline{0.728}\scriptsize{$\pm$0.0203} & \underline{0.781}\scriptsize{$\pm$0.0178} & \textbf{0.658}\scriptsize{$\pm$0.0190}    & \textbf{0.713}\scriptsize{$\pm$0.0209}    & \underline{0.722}\scriptsize{$\pm$0.0212} & \textbf{0.735}\scriptsize{$\pm$0.0185}    \\
 $\oursv$ (none)                                                                                                                       & 0.693\scriptsize{$\pm$0.0197}             & 0.743\scriptsize{$\pm$0.0195}             & 0.648\scriptsize{$\pm$0.0195}             & 0.702\scriptsize{$\pm$0.0209}             & 0.714\scriptsize{$\pm$0.0204}             & \underline{0.728}\scriptsize{$\pm$0.0189} \\
 $\oursv$ (rope)                                                                                                                       & 0.702\scriptsize{$\pm$0.0198}             & 0.763\scriptsize{$\pm$0.0187}             & 0.591\scriptsize{$\pm$0.0212}             & 0.666\scriptsize{$\pm$0.0212}             & 0.675\scriptsize{$\pm$0.0205}             & 0.669\scriptsize{$\pm$0.0198}             \\
\bottomrule
\end{tabular}\caption{\textbf{Ablation study of positional encodings.} Linear probing results for tumor type ($C=30$) prediction on EBRAINS. The best result is marked in bold and the second best is underlined.}
\label{tab:ablation_pos_enc_ebrains_diagnosis}
\end{table}

\begin{table}[h!]
\centering
\footnotesize
\begin{tabular}{l|cc|cc|cc}
\toprule
  & \multicolumn{2}{c|}{Logistic regression} & \multicolumn{2}{c|}{SimpleShot} & \multicolumn{2}{c}{20-nearest neighbors} \\ Encoder   & Bal. acc.                                 & Weighted F1                               & Bal. acc.                                 & Weighted F1                               & Bal. acc.                                 & Weighted F1                               \\
\midrule
 $\oursv$ (12.5)                                                                                                                       & 0.803\scriptsize{$\pm$0.0061}             & 0.862\scriptsize{$\pm$0.0039}             & 0.784\scriptsize{$\pm$0.0061}             & 0.829\scriptsize{$\pm$0.0043}             & 0.791\scriptsize{$\pm$0.0058}             & 0.826\scriptsize{$\pm$0.0041}             \\
 $\oursv$ (25)                                                                                                                         & 0.810\scriptsize{$\pm$0.0059}             & 0.860\scriptsize{$\pm$0.0040}             & 0.791\scriptsize{$\pm$0.0063}             & 0.841\scriptsize{$\pm$0.0042}             & 0.809\scriptsize{$\pm$0.0057}             & 0.837\scriptsize{$\pm$0.0041}             \\
 $\oursv$ (50)                                                                                                                         & \underline{0.818}\scriptsize{$\pm$0.0057} & \underline{0.865}\scriptsize{$\pm$0.0039} & \underline{0.802}\scriptsize{$\pm$0.0058} & \underline{0.850}\scriptsize{$\pm$0.0040} & \textbf{0.816}\scriptsize{$\pm$0.0050}    & \underline{0.839}\scriptsize{$\pm$0.0041} \\
\bottomrule
\end{tabular}\caption{\textbf{Ablation study of dataset size.} Linear probing results for tumor subtype ($C=32$) prediction on TCGA-UT-8K. The best result is marked in bold and the second best is underlined.}
\label{tab:ablation_datasize_TCGAUniformTumor_cancer}
\end{table}

\begin{table}[h!]
\centering
\footnotesize
\begin{tabular}{l|cc|cc|cc}
\toprule
  & \multicolumn{2}{c|}{Logistic regression} & \multicolumn{2}{c|}{SimpleShot} & \multicolumn{2}{c}{20-nearest neighbors} \\ Encoder   & Bal. acc.                                 & Weighted F1                               & Bal. acc.                                 & Weighted F1                               & Bal. acc.                                 & Weighted F1                               \\
\midrule
 $\oursv$ (12.5)                                                                                                                       & 0.656\scriptsize{$\pm$0.0184}             & 0.741\scriptsize{$\pm$0.0123}             & 0.610\scriptsize{$\pm$0.0173}             & 0.700\scriptsize{$\pm$0.0132}             & \underline{0.684}\scriptsize{$\pm$0.0187} & 0.759\scriptsize{$\pm$0.0118}             \\
 $\oursv$ (25)                                                                                                                         & 0.655\scriptsize{$\pm$0.0194}             & 0.726\scriptsize{$\pm$0.0126}             & 0.626\scriptsize{$\pm$0.0174}             & 0.706\scriptsize{$\pm$0.0130}             & 0.674\scriptsize{$\pm$0.0197}             & 0.755\scriptsize{$\pm$0.0120}             \\
 $\oursv$ (50)                                                                                                                         & \underline{0.678}\scriptsize{$\pm$0.0179} & \underline{0.743}\scriptsize{$\pm$0.0119} & \textbf{0.644}\scriptsize{$\pm$0.0178}    & \underline{0.735}\scriptsize{$\pm$0.0125} & \textbf{0.706}\scriptsize{$\pm$0.0198}    & \textbf{0.783}\scriptsize{$\pm$0.0113}    \\
\bottomrule
\end{tabular}\caption{\textbf{Ablation study of dataset size.} Linear probing results for OncoTree code ($C=46$) prediction on TCGA-OT. The best result is marked in bold and the second best is underlined.}
\label{tab:ablation_datasize_TCGA_OncoTreeCode}
\end{table}

\begin{table}[h!]
\centering
\footnotesize
\begin{tabular}{l|cc|cc|cc}
\toprule
  & \multicolumn{2}{c|}{Logistic regression} & \multicolumn{2}{c|}{SimpleShot} & \multicolumn{2}{c}{20-nearest neighbors} \\ Encoder   & Bal. acc.                                 & Weighted F1                               & Bal. acc.                                 & Weighted F1                               & Bal. acc.                                 & Weighted F1                               \\
\midrule
 $\oursv$ (12.5)                                                                                                                       & 0.520\scriptsize{$\pm$0.0111}             & 0.502\scriptsize{$\pm$0.0134}             & 0.448\scriptsize{$\pm$0.0098}             & 0.415\scriptsize{$\pm$0.0123}             & 0.511\scriptsize{$\pm$0.0112}             & 0.511\scriptsize{$\pm$0.0131}             \\
 $\oursv$ (25)                                                                                                                         & 0.526\scriptsize{$\pm$0.0109}             & 0.507\scriptsize{$\pm$0.0130}             & 0.452\scriptsize{$\pm$0.0102}             & 0.421\scriptsize{$\pm$0.0129}             & 0.515\scriptsize{$\pm$0.0110}             & 0.514\scriptsize{$\pm$0.0127}             \\
 $\oursv$ (50)                                                                                                                         & \underline{0.537}\scriptsize{$\pm$0.0102} & \underline{0.515}\scriptsize{$\pm$0.0127} & \textbf{0.468}\scriptsize{$\pm$0.0101}    & \textbf{0.435}\scriptsize{$\pm$0.0123}    & \textbf{0.537}\scriptsize{$\pm$0.0109}    & \textbf{0.532}\scriptsize{$\pm$0.0127}    \\
\bottomrule
\end{tabular}\caption{\textbf{Ablation study of dataset size.} Linear probing results for OncoTree code ($C=108$) prediction on OT108. The best result is marked in bold and the second best is underlined.}
\label{tab:ablation_datasize_OP108_OncoTreeCode}
\end{table}

\begin{table}[h!]
\centering
\footnotesize
\begin{tabular}{l|cc|cc|cc}
\toprule
  & \multicolumn{2}{c|}{Logistic regression} & \multicolumn{2}{c|}{SimpleShot} & \multicolumn{2}{c}{20-nearest neighbors} \\ Encoder   & Bal. acc.                                 & Weighted F1                               & Bal. acc.                                 & Weighted F1                               & Bal. acc.                                 & Weighted F1                               \\
\midrule
 $\oursv$ (12.5)                                                                                                                       & 0.723\scriptsize{$\pm$0.0209}             & 0.773\scriptsize{$\pm$0.0177}             & \underline{0.630}\scriptsize{$\pm$0.0216} & \underline{0.690}\scriptsize{$\pm$0.0207} & 0.708\scriptsize{$\pm$0.0211}             & 0.714\scriptsize{$\pm$0.0186}             \\
 $\oursv$ (25)                                                                                                                         & 0.721\scriptsize{$\pm$0.0204}             & 0.774\scriptsize{$\pm$0.0191}             & 0.620\scriptsize{$\pm$0.0203}             & 0.689\scriptsize{$\pm$0.0202}             & \underline{0.725}\scriptsize{$\pm$0.0208} & \underline{0.724}\scriptsize{$\pm$0.0184} \\
 $\oursv$ (50)                                                                                                                         & \textbf{0.733}\scriptsize{$\pm$0.0206}    & \underline{0.784}\scriptsize{$\pm$0.0180} & 0.618\scriptsize{$\pm$0.0207}             & 0.684\scriptsize{$\pm$0.0207}             & 0.719\scriptsize{$\pm$0.0202}             & \textbf{0.727}\scriptsize{$\pm$0.0184}    \\
\bottomrule
\end{tabular}\caption{\textbf{Ablation study of dataset size.} Linear probing results for tumor type ($C=30$) prediction on EBRAINS. The best result is marked in bold and the second best is underlined.}
\label{tab:ablation_datasize_ebrains_diagnosis}
\end{table}

\begin{table}[h!]
\centering
\footnotesize
\begin{tabular}{l|cc|cc|cc}
\toprule
  & \multicolumn{2}{c|}{Logistic regression} & \multicolumn{2}{c|}{SimpleShot} & \multicolumn{2}{c}{20-nearest neighbors} \\ Encoder   & Bal. acc.                                 & Weighted F1                               & Bal. acc.                                 & Weighted F1                               & Bal. acc.                                 & Weighted F1                               \\
\midrule
 $\oursv$ (tbase)                                                                                                                      & 0.818\scriptsize{$\pm$0.0060}             & \underline{0.873}\scriptsize{$\pm$0.0037} & 0.809\scriptsize{$\pm$0.0058}             & 0.857\scriptsize{$\pm$0.0040}             & \textbf{0.813}\scriptsize{$\pm$0.0052}    & \textbf{0.843}\scriptsize{$\pm$0.0039}    \\
 $\oursv$                                                                                                                              & \underline{0.820}\scriptsize{$\pm$0.0055} & 0.870\scriptsize{$\pm$0.0037}             & \underline{0.818}\scriptsize{$\pm$0.0053} & \underline{0.864}\scriptsize{$\pm$0.0039} & \underline{0.812}\scriptsize{$\pm$0.0050} & \underline{0.842}\scriptsize{$\pm$0.0038} \\
 $\oursv$ (base)                                                                                                                       & \textbf{0.827}\scriptsize{$\pm$0.0052}    & \textbf{0.876}\scriptsize{$\pm$0.0037}    & \textbf{0.821}\scriptsize{$\pm$0.0052}    & \textbf{0.868}\scriptsize{$\pm$0.0037}    & 0.803\scriptsize{$\pm$0.0053}             & 0.841\scriptsize{$\pm$0.0039}             \\
\bottomrule
\end{tabular}\caption{\textbf{Ablation study of model sizes.} Linear probing results for tumor subtype ($C=32$) prediction on TCGA-UT-8K. The best result is marked in bold and the second best is underlined.}
\label{tab:ablation_model_size_TCGAUniformTumor_cancer}
\end{table}

\begin{table}[h!]
\centering
\footnotesize
\begin{tabular}{l|cc|cc|cc}
\toprule
  & \multicolumn{2}{c|}{Logistic regression} & \multicolumn{2}{c|}{SimpleShot} & \multicolumn{2}{c}{20-nearest neighbors} \\ Encoder   & Bal. acc.                                 & Weighted F1                               & Bal. acc.                                 & Weighted F1                               & Bal. acc.                                 & Weighted F1                               \\
\midrule
 $\oursv$ (tbase)                                                                                                                      & \underline{0.675}\scriptsize{$\pm$0.0190} & 0.747\scriptsize{$\pm$0.0117}             & \textbf{0.643}\scriptsize{$\pm$0.0155}    & \underline{0.732}\scriptsize{$\pm$0.0127} & \underline{0.706}\scriptsize{$\pm$0.0189} & 0.771\scriptsize{$\pm$0.0114}             \\
 $\oursv$                                                                                                                              & \textbf{0.690}\scriptsize{$\pm$0.0196}    & \textbf{0.758}\scriptsize{$\pm$0.0123}    & \underline{0.639}\scriptsize{$\pm$0.0149} & \textbf{0.744}\scriptsize{$\pm$0.0124}    & \underline{0.706}\scriptsize{$\pm$0.0183} & \underline{0.775}\scriptsize{$\pm$0.0116} \\
 $\oursv$ (base)                                                                                                                       & 0.670\scriptsize{$\pm$0.0178}             & \underline{0.750}\scriptsize{$\pm$0.0119} & 0.621\scriptsize{$\pm$0.0165}             & 0.723\scriptsize{$\pm$0.0126}             & \textbf{0.709}\scriptsize{$\pm$0.0186}    & \textbf{0.779}\scriptsize{$\pm$0.0117}    \\
\bottomrule
\end{tabular}\caption{\textbf{Ablation study of model sizes.} Linear probing results for OncoTree code ($C=46$) prediction on TCGA-OT. The best result is marked in bold and the second best is underlined.}
\label{tab:ablation_model_size_TCGA_OncoTreeCode}
\end{table}

\begin{table}[h!]
\centering
\footnotesize
\begin{tabular}{l|cc|cc|cc}
\toprule
  & \multicolumn{2}{c|}{Logistic regression} & \multicolumn{2}{c|}{SimpleShot} & \multicolumn{2}{c}{20-nearest neighbors} \\ Encoder   & Bal. acc.                                 & Weighted F1                               & Bal. acc.                                 & Weighted F1                               & Bal. acc.                                 & Weighted F1                               \\
\midrule
 $\oursv$ (tbase)                                                                                                                      & \underline{0.551}\scriptsize{$\pm$0.0107} & \underline{0.529}\scriptsize{$\pm$0.0130} & \textbf{0.467}\scriptsize{$\pm$0.0102}    & \underline{0.428}\scriptsize{$\pm$0.0131} & \textbf{0.528}\scriptsize{$\pm$0.0108}    & \textbf{0.523}\scriptsize{$\pm$0.0127}    \\
 $\oursv$                                                                                                                              & \textbf{0.558}\scriptsize{$\pm$0.0104}    & \textbf{0.536}\scriptsize{$\pm$0.0127}    & \underline{0.464}\scriptsize{$\pm$0.0105} & \textbf{0.430}\scriptsize{$\pm$0.0128}    & \underline{0.524}\scriptsize{$\pm$0.0106} & \underline{0.517}\scriptsize{$\pm$0.0129} \\
 $\oursv$ (base)                                                                                                                       & 0.539\scriptsize{$\pm$0.0104}             & 0.521\scriptsize{$\pm$0.0126}             & 0.449\scriptsize{$\pm$0.0100}             & 0.416\scriptsize{$\pm$0.0128}             & 0.484\scriptsize{$\pm$0.0105}             & 0.474\scriptsize{$\pm$0.0131}             \\
\bottomrule
\end{tabular}\caption{\textbf{Ablation study of model sizes.} Linear probing results for OncoTree code ($C=108$) prediction on OT108. The best result is marked in bold and the second best is underlined.}
\label{tab:ablation_model_size_OP108_OncoTreeCode}
\end{table}

\begin{table}[h!]
\centering
\footnotesize
\begin{tabular}{l|cc|cc|cc}
\toprule
  & \multicolumn{2}{c|}{Logistic regression} & \multicolumn{2}{c|}{SimpleShot} & \multicolumn{2}{c}{20-nearest neighbors} \\ Encoder   & Bal. acc.                                 & Weighted F1                               & Bal. acc.                                 & Weighted F1                               & Bal. acc.                                 & Weighted F1                               \\
\midrule
 $\oursv$ (tbase)                                                                                                                      & \underline{0.709}\scriptsize{$\pm$0.0211} & 0.761\scriptsize{$\pm$0.0187}             & \underline{0.636}\scriptsize{$\pm$0.0202} & \underline{0.696}\scriptsize{$\pm$0.0201} & 0.693\scriptsize{$\pm$0.0219}             & \underline{0.701}\scriptsize{$\pm$0.0192} \\
 $\oursv$                                                                                                                              & \textbf{0.732}\scriptsize{$\pm$0.0208}    & \textbf{0.785}\scriptsize{$\pm$0.0175}    & \textbf{0.656}\scriptsize{$\pm$0.0187}    & \textbf{0.709}\scriptsize{$\pm$0.0199}    & \textbf{0.742}\scriptsize{$\pm$0.0192}    & \textbf{0.727}\scriptsize{$\pm$0.0182}    \\
 $\oursv$ (base)                                                                                                                       & \textbf{0.732}\scriptsize{$\pm$0.0199}    & \underline{0.781}\scriptsize{$\pm$0.0177} & 0.621\scriptsize{$\pm$0.0203}             & 0.682\scriptsize{$\pm$0.0209}             & \underline{0.698}\scriptsize{$\pm$0.0201} & 0.691\scriptsize{$\pm$0.0192}             \\
\bottomrule
\end{tabular}\caption{\textbf{Ablation study of model sizes.} Linear probing results for tumor type ($C=30$) prediction on EBRAINS. The best result is marked in bold and the second best is underlined.}
\label{tab:ablation_model_size_ebrains_diagnosis}
\end{table}
\clearpage
\begin{table}[h!]
\centering
\footnotesize
\begin{tabular}{l|cccccc}
\toprule
 Encoder           & $k$=1                                     & $k$=2                                     & $k$=4                                     & $k$=8                                     & $k$=16                                    & $k$=32                                    \\
\midrule
 Mean pool (CONCH) & 0.495\scriptsize{$\pm$0.0367}             & 0.597\scriptsize{$\pm$0.0289}             & 0.667\scriptsize{$\pm$0.0201}             & 0.711\scriptsize{$\pm$0.0156}             & 0.738\scriptsize{$\pm$0.0132}             & 0.755\scriptsize{$\pm$0.0105}             \\
 ABMIL (CONCH)     & 0.539\scriptsize{$\pm$0.0459}             & 0.662\scriptsize{$\pm$0.0275}             & 0.724\scriptsize{$\pm$0.0133}             & 0.759\scriptsize{$\pm$0.0124}             & 0.787\scriptsize{$\pm$0.0111}             & 0.805\scriptsize{$\pm$0.0109}             \\
 GigaPath          & 0.282\scriptsize{$\pm$0.0286}             & 0.381\scriptsize{$\pm$0.0282}             & 0.476\scriptsize{$\pm$0.0193}             & 0.557\scriptsize{$\pm$0.0204}             & 0.620\scriptsize{$\pm$0.0138}             & 0.659\scriptsize{$\pm$0.0112}             \\
 PRISM             & 0.469\scriptsize{$\pm$0.0380}             & 0.560\scriptsize{$\pm$0.0284}             & 0.627\scriptsize{$\pm$0.0253}             & 0.674\scriptsize{$\pm$0.0181}             & 0.704\scriptsize{$\pm$0.0151}             & 0.725\scriptsize{$\pm$0.0121}             \\
 CHIEF             & 0.255\scriptsize{$\pm$0.0292}             & 0.339\scriptsize{$\pm$0.0304}             & 0.424\scriptsize{$\pm$0.0198}             & 0.503\scriptsize{$\pm$0.0153}             & 0.564\scriptsize{$\pm$0.0135}             & 0.609\scriptsize{$\pm$0.0113}             \\
\bottomrule
\end{tabular}\caption{\textbf{Few shot experiments} with $k\in\{1, 2, 4, 8, 16, 32\}$ shots per class for tumor subtype ($C=2$) prediction on TCGA-UT-8K. Results are given in balanced accuracy of linear probing evaluation or ABMIL training, respectively. If a class contains less than $k$ samples, all samples are used. The best result is marked in bold and the second best is underlined.}
\label{tab:fewshot_log_reg_TCGAUniformTumor_cancer}
\end{table}

\begin{table}[h!]
\centering
\footnotesize
\begin{tabular}{l|cccccc}
\toprule
 Encoder           & $k$=1                                     & $k$=2                                     & $k$=4                                     & $k$=8                                     & $k$=16                                    & $k$=32                                    \\
\midrule
 Mean pool (CONCH) & 0.396\scriptsize{$\pm$0.0293}             & 0.466\scriptsize{$\pm$0.0302}             & 0.521\scriptsize{$\pm$0.0233}             & 0.555\scriptsize{$\pm$0.0217}             & 0.583\scriptsize{$\pm$0.0193}             & 0.600\scriptsize{$\pm$0.0173}             \\
 ABMIL (CONCH)     & 0.407\scriptsize{$\pm$0.0333}             & 0.505\scriptsize{$\pm$0.0257}             & 0.563\scriptsize{$\pm$0.0158}             & 0.607\scriptsize{$\pm$0.0170}             & 0.634\scriptsize{$\pm$0.0192}             & 0.659\scriptsize{$\pm$0.0138}             \\
 GigaPath          & 0.210\scriptsize{$\pm$0.0268}             & 0.289\scriptsize{$\pm$0.0294}             & 0.373\scriptsize{$\pm$0.0192}             & 0.438\scriptsize{$\pm$0.0193}             & 0.502\scriptsize{$\pm$0.0189}             & 0.537\scriptsize{$\pm$0.0171}             \\
 PRISM             & 0.441\scriptsize{$\pm$0.0282}             & 0.519\scriptsize{$\pm$0.0196}             & 0.584\scriptsize{$\pm$0.0159}             & 0.621\scriptsize{$\pm$0.0148}             & 0.645\scriptsize{$\pm$0.0110}             & 0.662\scriptsize{$\pm$0.0098}             \\
 CHIEF             & 0.219\scriptsize{$\pm$0.0248}             & 0.287\scriptsize{$\pm$0.0301}             & 0.362\scriptsize{$\pm$0.0256}             & 0.423\scriptsize{$\pm$0.0207}             & 0.476\scriptsize{$\pm$0.0177}             & 0.510\scriptsize{$\pm$0.0152}             \\
\bottomrule
\end{tabular}\caption{\textbf{Few shot experiments} with $k\in\{1, 2, 4, 8, 16, 32\}$ shots per class for OncoTree code ($C=2$) prediction on TCGA-OT. Results are given in balanced accuracy of linear probing evaluation or ABMIL training, respectively. If a class contains less than $k$ samples, all samples are used. The best result is marked in bold and the second best is underlined.}
\label{tab:fewshot_log_reg_TCGA_OncoTreeCode}
\end{table}

\begin{table}[h!]
\centering
\footnotesize
\begin{tabular}{l|cccccc}
\toprule
 Encoder           & $k$=1                                     & $k$=2                                     & $k$=4                                     & $k$=8                                     & $k$=16                                    & $k$=32                                    \\
\midrule
 Mean pool (CONCH) & 0.201\scriptsize{$\pm$0.0144}             & 0.276\scriptsize{$\pm$0.0123}             & 0.338\scriptsize{$\pm$0.0110}             & 0.392\scriptsize{$\pm$0.0094}             & 0.430\scriptsize{$\pm$0.0087}             & 0.462\scriptsize{$\pm$0.0065}             \\
 ABMIL (CONCH)     & 0.236\scriptsize{$\pm$0.0186}             & 0.352\scriptsize{$\pm$0.0161}             & 0.437\scriptsize{$\pm$0.0104}             & 0.496\scriptsize{$\pm$0.0086}             & 0.535\scriptsize{$\pm$0.0098}             & 0.553\scriptsize{$\pm$0.0065}             \\
 GigaPath          & 0.136\scriptsize{$\pm$0.0110}             & 0.202\scriptsize{$\pm$0.0121}             & 0.276\scriptsize{$\pm$0.0102}             & 0.353\scriptsize{$\pm$0.0106}             & 0.416\scriptsize{$\pm$0.0089}             & 0.452\scriptsize{$\pm$0.0054}             \\
 PRISM             & 0.271\scriptsize{$\pm$0.0161}             & 0.346\scriptsize{$\pm$0.0137}             & 0.409\scriptsize{$\pm$0.0096}             & 0.457\scriptsize{$\pm$0.0099}             & 0.489\scriptsize{$\pm$0.0086}             & 0.501\scriptsize{$\pm$0.0058}             \\
 CHIEF             & 0.126\scriptsize{$\pm$0.0111}             & 0.179\scriptsize{$\pm$0.0108}             & 0.242\scriptsize{$\pm$0.0089}             & 0.311\scriptsize{$\pm$0.0109}             & 0.368\scriptsize{$\pm$0.0072}             & 0.405\scriptsize{$\pm$0.0055}             \\
\bottomrule
\end{tabular}\caption{\textbf{Few shot experiments} with $k\in\{1, 2, 4, 8, 16, 32\}$ shots per class for OncoTree code ($C=2$) prediction on OT108. Results are given in balanced accuracy of linear probing evaluation or ABMIL training, respectively. If a class contains less than $k$ samples, all samples are used. The best result is marked in bold and the second best is underlined.}
\label{tab:fewshot_log_reg_OP108_OncoTreeCode}
\end{table}

\begin{table}[h!]
\centering
\footnotesize
\begin{tabular}{l|cccccc}
\toprule
 Encoder           & $k$=1                                     & $k$=2                                     & $k$=4                                     & $k$=8                                     & $k$=16                                    & $k$=32                                    \\
\midrule
 Mean pool (CONCH) & 0.354\scriptsize{$\pm$0.0267}             & 0.445\scriptsize{$\pm$0.0310}             & 0.526\scriptsize{$\pm$0.0212}             & 0.589\scriptsize{$\pm$0.0147}             & 0.638\scriptsize{$\pm$0.0176}             & 0.668\scriptsize{$\pm$0.0127}             \\
 ABMIL (CONCH)     & 0.388\scriptsize{$\pm$0.0417}             & 0.544\scriptsize{$\pm$0.0372}             & \underline{0.629}\scriptsize{$\pm$0.0230} & 0.671\scriptsize{$\pm$0.0195}             & 0.702\scriptsize{$\pm$0.0138}             & 0.729\scriptsize{$\pm$0.0118}             \\
 GigaPath          & 0.292\scriptsize{$\pm$0.0284}             & 0.398\scriptsize{$\pm$0.0250}             & 0.507\scriptsize{$\pm$0.0206}             & 0.589\scriptsize{$\pm$0.0190}             & 0.658\scriptsize{$\pm$0.0175}             & 0.709\scriptsize{$\pm$0.0122}             \\
 PRISM             & 0.399\scriptsize{$\pm$0.0328}             & 0.484\scriptsize{$\pm$0.0192}             & 0.548\scriptsize{$\pm$0.0225}             & 0.595\scriptsize{$\pm$0.0193}             & 0.628\scriptsize{$\pm$0.0173}             & 0.659\scriptsize{$\pm$0.0156}             \\
 CHIEF             & 0.226\scriptsize{$\pm$0.0280}             & 0.319\scriptsize{$\pm$0.0278}             & 0.418\scriptsize{$\pm$0.0237}             & 0.509\scriptsize{$\pm$0.0171}             & 0.578\scriptsize{$\pm$0.0158}             & 0.637\scriptsize{$\pm$0.0122}             \\
\bottomrule
\end{tabular}\caption{\textbf{Few shot experiments} with $k\in\{1, 2, 4, 8, 16, 32\}$ shots per class for tumor type ($C=2$) prediction on EBRAINS. Results are given in balanced accuracy of linear probing evaluation or ABMIL training, respectively. If a class contains less than $k$ samples, all samples are used. The best result is marked in bold and the second best is underlined.}
\label{tab:fewshot_log_reg_ebrains_diagnosis}
\end{table}

\begin{table}[h!]
\centering
\footnotesize
\begin{tabular}{l|cccccc}
\toprule
 Encoder           & $k$=1                                     & $k$=2                                     & $k$=4                                     & $k$=8                                     & $k$=16                                    & $k$=32                                    \\
\midrule
 Mean pool (CONCH) & 0.505\scriptsize{$\pm$0.0387}             & 0.597\scriptsize{$\pm$0.0284}             & 0.666\scriptsize{$\pm$0.0173}             & 0.711\scriptsize{$\pm$0.0186}             & 0.735\scriptsize{$\pm$0.0116}             & 0.748\scriptsize{$\pm$0.0126}             \\
 GigaPath          & 0.292\scriptsize{$\pm$0.0297}             & 0.366\scriptsize{$\pm$0.0289}             & 0.441\scriptsize{$\pm$0.0205}             & 0.505\scriptsize{$\pm$0.0187}             & 0.543\scriptsize{$\pm$0.0145}             & 0.562\scriptsize{$\pm$0.0152}             \\
 PRISM             & 0.478\scriptsize{$\pm$0.0403}             & 0.569\scriptsize{$\pm$0.0276}             & 0.641\scriptsize{$\pm$0.0191}             & 0.677\scriptsize{$\pm$0.0137}             & 0.694\scriptsize{$\pm$0.0101}             & 0.702\scriptsize{$\pm$0.0091}             \\
 CHIEF             & 0.262\scriptsize{$\pm$0.0295}             & 0.323\scriptsize{$\pm$0.0286}             & 0.391\scriptsize{$\pm$0.0220}             & 0.440\scriptsize{$\pm$0.0162}             & 0.468\scriptsize{$\pm$0.0148}             & 0.482\scriptsize{$\pm$0.0121}             \\
 $\oursv$          & \underline{0.639}\scriptsize{$\pm$0.0385} & \underline{0.727}\scriptsize{$\pm$0.0243} & \underline{0.776}\scriptsize{$\pm$0.0155} & \underline{0.798}\scriptsize{$\pm$0.0122} & \underline{0.809}\scriptsize{$\pm$0.0094} & \underline{0.816}\scriptsize{$\pm$0.0100} \\
\bottomrule
\end{tabular}\caption{\textbf{Few shot experiments} with $k\in\{1, 2, 4, 8, 16, 32\}$ shots per class for tumor subtype ($C=2$) prediction on TCGA-UT-8K. Results are given in balanced accuracy of SimpleShot evaluation. If a class contains less than $k$ samples, all samples are used. The best result is marked in bold and the second best is underlined.}
\label{tab:fewshot_prototypes_TCGAUniformTumor_cancer}
\end{table}

\begin{table}[h!]
\centering
\footnotesize
%
\caption{\textbf{Few shot experiments} with $k\in\{1, 2, 4, 8, 16, 32\}$ shots per class for OncoTree code ($C=2$) prediction on TCGA-OT. Results are given in balanced accuracy of SimpleShot evaluation. If a class contains less than $k$ samples, all samples are used. The best result is marked in bold and the second best is underlined.}
\label{tab:fewshot_prototypes_TCGA_OncoTreeCode}
\end{table}

\begin{table}[h!]
\centering
\footnotesize
%
\caption{\textbf{Few shot experiments} with $k\in\{1, 2, 4, 8, 16, 32\}$ shots per class for OncoTree code ($C=2$) prediction on OT108. Results are given in balanced accuracy of SimpleShot evaluation. If a class contains less than $k$ samples, all samples are used. The best result is marked in bold and the second best is underlined.}
\label{tab:fewshot_prototypes_OP108_OncoTreeCode}
\end{table}

\begin{table}[h!]
\centering
\footnotesize
%
\caption{\textbf{Few shot experiments} with $k\in\{1, 2, 4, 8, 16, 32\}$ shots per class for tumor type ($C=2$) prediction on EBRAINS. Results are given in balanced accuracy of SimpleShot evaluation. If a class contains less than $k$ samples, all samples are used. The best result is marked in bold and the second best is underlined.}
\label{tab:fewshot_prototypes_ebrains_diagnosis}
\end{table}
\clearpage
\begin{table}[h!]
\centering
\footnotesize
\begin{tabular}{l|ccc}
\toprule
 Encoder             & Bal. acc.                                 & Weighted F1                               & AUROC                                     \\
\midrule
 PRISM               & 0.536\scriptsize{$\pm$0.0061}             & 0.599\scriptsize{$\pm$0.0055}             & 0.966\scriptsize{$\pm$0.0010}             \\
 $\oursvcr$          & \textbf{0.761}\scriptsize{$\pm$0.0063}    & \textbf{0.798}\scriptsize{$\pm$0.0043}    & \textbf{0.989}\scriptsize{$\pm$0.0006}    \\
\bottomrule
\end{tabular}\caption{\textbf{Zero-shot results} for tumor subtype ($C=32$) prediction on TCGA-UT-8K. The best result is marked in bold and the second best is underlined.}
\label{tab:zeroshot_TCGAUniformTumor_cancer}
\end{table}

\begin{table}[h!]
\centering
\footnotesize
%
\caption{\textbf{Zero-shot results} for OncoTree code ($C=46$) prediction on TCGA-OT. The best result is marked in bold and the second best is underlined.}
\label{tab:zeroshot_TCGA_OncoTreeCode}
\end{table}

\begin{table}[h!]
\centering
\footnotesize
%
\caption{\textbf{Zero-shot results} for OncoTree code ($C=108$) prediction on OT108. The best result is marked in bold and the second best is underlined.}
\label{tab:zeroshot_OP108_OncoTreeCode}
\end{table}

\begin{table}[h!]
\centering
\footnotesize
%
\caption{\textbf{Zero-shot results} for tumor type ($C=30$) prediction on eBrains. The best result is marked in bold and the second best is underlined.}
\label{tab:zeroshot_ebrains_diagnosis}
\end{table}

\begin{table}[h!]
\centering
\footnotesize
%
\caption{\textbf{Zero-shot results} for histological pattern ($C=5$) prediction on DHMC-LUAD. The best result is marked in bold and the second best is underlined.}
\label{tab:zeroshot_dhmc_luad_label}
\end{table}

\begin{table}[h!]
\centering
\footnotesize
%
\caption{\textbf{Zero-shot results} for OncoTree code ($C=3$) prediction on TCGA-RCC. The best result is marked in bold and the second best is underlined.}
\label{tab:zeroshot_TCGA-RCC_OncoTreeCode}
\end{table}

\begin{table}[h!]
\centering
\footnotesize
%
\caption{\textbf{Zero-shot results} for OncoTree code ($C=3$) prediction on CPTAC-DHMC-RCC. The best result is marked in bold and the second best is underlined.}
\label{tab:zeroshot_cptac_dhmc_rcc_OncoTreeCode}
\end{table}

\begin{table}[h!]
\centering
\footnotesize
%
\caption{\textbf{Zero-shot results} for tumor subtype ($C=2$) prediction on TCGA-BRCA. The best result is marked in bold and the second best is underlined.}
\label{tab:zeroshot_TCGA-BRCA_subtype}
\end{table}

\begin{table}[h!]
\centering
\footnotesize
%
\caption{\textbf{Zero-shot results} for tumor subtype ($C=2$) prediction on TCGA-NSCLC. The best result is marked in bold and the second best is underlined.}
\label{tab:zeroshot_TCGA-NSCLC_subtype}
\end{table}

\begin{table}[h!]
\centering
\footnotesize
%
\caption{\textbf{Zero-shot results} for tumor subtype ($C=2$) prediction on CPTAC-NSCLC. The best result is marked in bold and the second best is underlined.}
\label{tab:zeroshot_cptac_nsclc_subtype}
\end{table}

\begin{table}[h!]
\centering
\footnotesize
\begin{tabular}{l|ccc}
\toprule
 Encoder    & Bal. acc.                                 & Weighted F1                               & AUROC                                     \\
\midrule
 PRISM      & \underline{0.584}\scriptsize{$\pm$0.0179} & \underline{0.573}\scriptsize{$\pm$0.0163} & \underline{0.576}\scriptsize{$\pm$0.0210} \\
 $\oursvcr$ & \textbf{0.693}\scriptsize{$\pm$0.0148}    & \textbf{0.633}\scriptsize{$\pm$0.0166}    & \textbf{0.762}\scriptsize{$\pm$0.0159}    \\
\bottomrule
\end{tabular}\caption{\textbf{Zero-shot results} for antibody-mediated rejection ($C=2$) prediction on Renal allograft rejection. The best result is marked in bold and the second best is underlined.}
\label{tab:zeroshot_MANTA_HE_label_amr}
\end{table}

\begin{table}[h!]
\centering
\footnotesize
%
\caption{\textbf{Zero-shot results} for cellular rejection ($C=2$) prediction on Renal allograft rejection. The best result is marked in bold and the second best is underlined.}
\label{tab:zeroshot_MANTA_HE_label_cell}
\end{table}

\caption{\textbf{Zero-shot results} for Interstitial fibrosis and tubular atrophy (IFTA) status ($C=3$) prediction on Renal allograft rejection. The best result is marked in bold and the second best is underlined.}

\begin{table}[h!]
\centering
\footnotesize
%
\caption{\textbf{Zero-shot results} for cellular rejection ($C=2$) prediction on CRANE. The best result is marked in bold and the second best is underlined.}
\label{tab:zeroshot_CRANE_cellular}
\end{table}
\clearpage
\begin{table}[h!]
\centering
\footnotesize
%
\caption{\textbf{Ablation study of vision-language design.} Linear probing results for tumor subtype ($C=32$) prediction on TCGA-UT-8K. The best result is marked in bold and the second best is underlined.}
\label{tab:ablation_vl_TCGAUniformTumor_cancer}
\end{table}

\begin{table}[h!]
\centering
\footnotesize
%
\caption{\textbf{Ablation study of vision-language design.} Linear probing results for OncoTree code ($C=46$) prediction on TCGA-OT. The best result is marked in bold and the second best is underlined.}
\label{tab:ablation_vl_TCGA_OncoTreeCode}
\end{table}

\begin{table}[h!]
\centering
\footnotesize
%
\caption{\textbf{Ablation study of vision-language design.} Linear probing results for OncoTree code ($C=108$) prediction on OT108. The best result is marked in bold and the second best is underlined.}
\label{tab:ablation_vl_OP108_OncoTreeCode}
\end{table}

\begin{table}[h!]
\centering
\footnotesize
%
\caption{\textbf{Ablation study of vision-language design.} Linear probing results for tumor type ($C=30$) prediction on EBRAINS. The best result is marked in bold and the second best is underlined.}
\label{tab:ablation_vl_ebrains_diagnosis}
\end{table}
\clearpage
\begin{table}[h]
  \centering
  %
  \caption{\textbf{Hyperparameters used in pretraining the unimodal vision model}. 4 $\times$ 80GB NVIDIA A100 GPUs were used for training. Batch size refers to the total batch size across GPUs.}
  \label{tab:hparams_ibot}
\end{table}

\begin{table}[h]
  \centering
  %
  \caption{\textbf{Hyperparameters used in visual-caption pretraining}. 8 $\times$ 80GB NVIDIA A100 GPUs were used for training. Effective batch size used for optimization is batch size $\times$ gradient accumulation steps. The maximum sequence length for captions is set to 128.}
  \label{tab:hparams_coca_caption}
\end{table}

\begin{table}[h]
  \centering
  %
  \caption{\textbf{Hyperparameters used in visual-report pretraining}. 8 $\times$ 80GB NVIDIA A100 GPUs were used for training. Effective batch size used for optimization is batch size $\times$ gradient accumulation steps. The maximum sequence length for captions is set to 128.}
  \label{tab:hparams_coca_report}
\end{table}

\begin{table}[h]
  \centering
  %
  \caption{\textbf{Hyperparameters used in CONCH v1.5 training}. 8 $\times$ 80GB NVIDIA GPUs were used for training CONCH v1.5. We initialized vision backbone with UNI \cite{chen2024towards}, while keeping the text tower configuration of CONCH v1 unchanged. Effective batch size used for optimization is batch size $\times$ gradient accumulation steps. The maximum sequence length for captions is set to 128.}
  \label{tab:hparams_coca_base}
\end{table}

\begin{table}[]
    \centering
    %
    \caption{\textbf{Prompt templates} used for zero-shot slide classification. The name of the class replaces \texttt{CLASSNAME}. See \textbf{Tables \ref{tab:tcga_nsclc_brca_prompts}-\ref{tab:heart_prompts}} for class prompts of each task.}
    \label{tab:zero_shot_prompts}
\end{table}

\begin{table*}[]
    \centering
    %
    \caption{\textbf{Class prompts for TCGA NSCLC and TCGA BRCA subtyping.}.}
    \label{tab:tcga_nsclc_brca_prompts}
\end{table*}

\begin{table*}[]
    \centering
    \footnotesize
    %
    \caption{\textbf{Class prompts for OT-108 and TCGA-OncoTree subtyping}.}
    \label{tab:ot108_tcga_prompts_1}
\end{table*}

\begin{table*}[]
    \centering
    \footnotesize
    %
    \caption{\textbf{Class prompts for OT-108 and TCGA-OncoTree subtyping}. Continued.}
    \label{tab:ot108_tcga_prompts_2}
\end{table*}

\begin{table*}[]
    \centering
    \footnotesize
    %
    \caption{\textbf{Class prompts for OT-108 and TCGA-OncoTree subtyping}. Continued.}
    \label{tab:ot108_tcga_prompts_3}
\end{table*}

\begin{table*}[]
    \centering
    \footnotesize
    %
    \caption{\textbf{Class prompts for OT-108 and TCGA-OncoTree subtyping}. Continued.}
    \label{tab:ot108_tcga_prompts_4}
\end{table*}

\begin{table*}[]
    \centering
    \footnotesize
    %
    \caption{\textbf{Class prompts for TCGA UniformTumor subtyping}.}
    \label{tab:tcga_ut_prompt_1}
\end{table*}

\begin{table*}[]
    \centering
    \footnotesize
    %
    \caption{\textbf{Class prompts for TCGA UniformTumor subtyping}. Continued.}
    \label{tab:tcga_ut_prompt_2}
\end{table*}

\begin{table*}[]
    \centering
    \footnotesize
    %
    \caption{\textbf{Class prompts for TCGA UniformTumor subtyping}. Continued.}
    \label{tab:tcga_ut_prompt_3}
\end{table*}

\begin{table*}[]
    \centering
    \footnotesize
    %
    \caption{\textbf{Class prompts for TCGA UniformTumor subtyping}. Continued.}
    \label{tab:tcga_ut_prompt_4}
\end{table*}

\begin{table*}[]
    \centering
    \footnotesize
    %
    \caption{\textbf{Class prompts for EBRAINS subtyping}.}
    \label{tab:ebrains_prompt_1}
\end{table*}

\begin{table*}[]
    \centering
    \footnotesize
    %
    \caption{\textbf{Class prompts for EBRAINS subtyping}. Continued.}
    \label{tab:ebrains_prompt_2}
\end{table*}

\begin{table*}[]
    \centering
    \footnotesize
    %
    \caption{\textbf{Class prompts for EBRAINS subtyping}. Continued.}
    \label{tab:ebrains_prompt_3}
\end{table*}

\begin{table*}[]
    \centering
    %
    \caption{\textbf{Class prompts for renal allograft rejection.}.}
    \label{tab:renal_prompts}
\end{table*}

\begin{table*}[]
    \centering
    %
    \caption{\textbf{Class prompts for cellular-mediated heart allograft rejection.}.}
    \label{tab:heart_prompts}
\end{table*}

\begin{table*}[h!]
    \centering
    \footnotesize
    %
    \caption{\textbf{PathChat prompts for synthetic caption generation} for $8192\times 8192$ pixels ROIs at $20\times$ magnification.}
    \label{tab:pathchat_prompt}
\end{table*}

\begin{table*}[h!]
    \centering
    \footnotesize
    %
    \caption{\textbf{Prompts for synthetic caption rewriting using Qwen-2-7b instruct}.}
    \label{tab:caption_rewrite}
\end{table*}

\begin{table*}[h!]
    \centering
    \footnotesize
    %
    \caption{\textbf{Prompts for rewriting parsed slide-level reports.}}
    \label{tab:report_rewrite}
\end{table*}
\clearpage
\begin{table}[h!]
\centering
\footnotesize
%
\caption{\textbf{Cross-modal retrieval results (slide-to-report)} on TCGA-Slide-Reports. The best result is marked in bold and the second best is underlined.}
\label{tab:i2t_retrieval}
\end{table}

\end{document}